\documentclass[utf8]{FrontiersinHarvard} 

\usepackage{url,hyperref,lineno,microtype,subcaption}
\usepackage[onehalfspacing]{setspace}
\usepackage{caption,booktabs}
\usepackage{makecell}
\usepackage{multirow}

\def\keyFont{\fontsize{8}{11}\helveticabold }
\def\firstAuthorLast{Boorman {et~al.}} 
\def\Authors{
P.~G.~Boorman\,$^{1,*}$,
N.~Torres-Alb\`{a}\,$^{2}$,
A.~Annuar\,$^{3}$,
S.~Marchesi\,$^{4,2}$,
R.~Pfeifle\,$^{5}$,
D.~Stern\,$^{6}$,
F.~Civano\,$^{5}$,
M.~Balokovi\'{c}\,$^{7,8}$,
J.~Buchner\,$^{9}$,
C.~Ricci\,$^{10,11}$,
D.~M.~Alexander\,$^{12}$,
W.~N.~Brandt\,$^{13,14,15}$,
M.~Brightman\,$^{1}$,
C.~T.~Chen\,$^{16,17}$,
S.~Creech\,$^{18}$,
P.~Gandhi$^{19}$,
J.~A.~Garc\'{i}a\,$^{5,1}$,
F.~Harrison\,$^{1}$,
R.~Hickox\,$^{20}$,
E.~Kammoun\,$^{21,22}$,
S.~LaMassa\,$^{23}$,
G.~Lanzuisi\,$^{4}$,
L.~Marcotulli\,$^{7,8}$,
K.~Madsen\,$^{5}$,
G.~Matt\,$^{21}$,
G.~Matzeu\,$^{24}$,
E.~Nardini\,$^{22}$,
J.~M.~Piotrowska$^{1}$,
A.~Pizzetti\,$^{2}$,
S.~Puccetti\,$^{25}$,
D.~Sicilian\,$^{26}$,
R.~Silver\,$^{5}$,
D.~J.~Walton\,$^{27}$,
D.~R.~Wilkins\,$^{28}$,
X.~Zhao\,$^{29}$,
and the \textit{HEX-P} Collaboration
}

\def\Address{\footnotesize
$^{1}$Cahill Center for Astrophysics, California Institute of Technology, 1216 East California Boulevard, Pasadena, CA 91125, USA \\
$^{2}$Department of Physics and Astronomy, Clemson University, Kinard Lab of Physics, Clemson, SC 29634, USA \\
$^{3}$Department of Applied Physics, Faculty of Science and Technology, Universiti Kebangsaan Malaysia, Selangor, Malaysia \\
$^{4}$INAF—Osservatorio di Astrofisica e Scienza dello Spazio di Bologna, Via Piero Gobetti, 93/3, I-40129, Bologna, Italy \\
$^{5}$X-ray Astrophysics Laboratory, NASA Goddard Space Flight Center, Greenbelt, MD 20771, USA \\
$^{6}$Jet Propulsion Laboratory, California Institute of Technology, Pasadena, CA 91109, USA \\
$^{7}$Yale Center for Astronomy \& Astrophysics, 219 Prospect Street, New Haven, CT 06511, USA \\
$^{8}$Department of Physics, Yale University, P.O. Box 2018120, New Haven, CT 06520, USA \\
$^{9}$Max-Planck-Institut f\"{u}r extraterrestrische Physik, Giessenbachstra{\ss}e 1, D-85748 Garching, Germany \\
$^{10}$N\'{u}cleo de Astronom \'{i}a de la Facultad de Ingenier\'{i}a, Universidad Diego Portales, Av. Ej\'{e}rcito Libertador 441, Santiago, Chile \\
$^{11}$Kavli Institute for Astronomy and Astrophysics, Peking University, Beijing 100871, China \\
$^{12}$Centre for Extragalactic Astronomy, Department of Physics, Durham University, UK \\
$^{13}$Department of Astronomy \& Astrophysics, The Pennsylvania State University, University Park, PA 16802, USA \\
$^{14}$Institute for Gravitation and the Cosmos, The Pennsylvania State University, University Park, PA 16802, USA \\
$^{15}$Department of Physics, The Pennsylvania State University, University Park, PA 16802, USA \\
$^{16}$Science and Technology Institute, Universities Space Research Association, Huntsville, AL 35805, USA \\
$^{17}$Astrophysics Office, NASA Marshall Space Flight Center, ST12, Huntsville, AL 35812, USA \\
$^{18}$Department of Physics \& Astronomy, University of Utah, 115 South 1400 East, Salt Lake City, UT 84112, USA \\
$^{19}$Department of Physics \& Astronomy, Faculty of Physical Sciences and Engineering, University of Southampton, UK \\
$^{20}$Department of Physics and Astronomy, Dartmouth College, Hanover, NH 03755 \\
$^{21}$Dipartimento di Matematica e Fisica, Universit\'{a} degli Studi Roma Tre, via della Vasca Navale 84, I-00146 Roma, Italy \\
$^{22}$INAF – Osservatorio Astrofisico di Arcetri, Largo Enrico Fermi 5, I-50125 Firenze, Italy \\
$^{23}$Space Telescope Science Institute, 3700 San Martin Drive, Baltimore, MD 21210, USA \\
$^{24}$European Space Agency (ESA), European Space Astronomy Centre (ESAC), Madrid, Spain \\
$^{25}$ASI - Agenzia Spaziale Italiana, Via del Politecnico snc, 00133 Roma, Italy \\
$^{26}$Department of Physics, University of Miami, Coral Gables, FL 33146 \\
$^{27}$Centre for Astrophysics Research, University of Hertfordshire, College Lane, Hatfield, UK \\
$^{28}$Kavli Institute for Particle Astrophysics and Cosmology, Stanford University,  Stanford, CA, USA \\
$^{29}$Harvard-Smithsonian Center for Astrophysics, 60 Garden Street, Cambridge, MA 02138, USA
}

\def\corrAuthor{Peter~G.~Boorman}

\def\corrEmail{boorman@caltech.edu}

\begin{document}
\onecolumn
\firstpage{1}

\title{\textit{The High Energy X-ray Probe (HEX-P)}:\\
The Circum-nuclear Environment of Growing Supermassive Black Holes}
\author[\firstAuthorLast ]{\Authors} 
\address{} 
\correspondance{} 

\extraAuth{}

\maketitle

\begin{abstract}

Ever since the discovery of the first Active Galactic Nuclei (AGN), substantial observational and theoretical effort has been invested into understanding how massive black holes have evolved across cosmic time. Circum-nuclear obscuration is now established as a crucial component, with almost every AGN observed known to display signatures of some level of obscuration in their X-ray spectra. But despite more than six decades of effort, substantial open questions remain: {\em How does the accretion power impact the structure of the circum-nuclear obscurer? What are the dynamical properties of the obscurer? Can dense circum-nuclear obscuration exist around intrinsically weak AGN? How many intermediate mass black holes occupy the centers of dwarf galaxies?} In this paper, we showcase a number of next-generation prospects attainable with the \textit{High Energy X-ray Probe} (\textit{HEX-P}\footnote{\href{https://hexp.org}{https://hexp.org}}) to contribute towards solving these questions in the 2030s. The uniquely broad (0.2\,--\,80\,keV) and strictly simultaneous X-ray passband of \textit{HEX-P} makes it ideally suited for studying the temporal co-evolution between the central engine and circum-nuclear obscurer. Improved sensitivities and reduced background will enable the development of spectroscopic models complemented by current and future multi-wavelength observations. We show that the angular resolution of \textit{HEX-P} both below and above 10\,keV will enable the discovery and confirmation of accreting massive black holes at both low accretion power and low black hole masses even when concealed by thick obscuration. In combination with other next-generation observations of the dusty hearts of nearby galaxies, \textit{HEX-P} will hence be pivotal in paving the way towards a complete picture of black hole growth and galaxy co-evolution.

\tiny
 \keyFont{ \section{Keywords:} X-ray, AGN, obscuration, black hole, galaxies, Compton-thick, high energy, spectral modelling} 
\end{abstract}

\section{Introduction}\label{sec:introduction}

\subsection{The Prevalence of Obscured Accretion onto Supermassive Black Holes}

It is now well established that obscuration is an omnipresent ingredient in the growth of supermassive black holes, often parameterised to have equivalent hydrogen column densities $N_{\rm H}$\,$>$\,10$^{22}$\,cm$^{-2}$ along the line-of-sight. Prime evidence arises from X-ray surveys and population synthesis studies, which have found heavily obscured AGN to dramatically dominate the AGN population at all but the strongest accretion powers, irrespective of redshift \citep{Comastri95,Gandhi03,Gilli07,Treister09,Akylas12,Ueda14,Buchner14,Buchner15,Brandt15,Aird15,Lansbury17,Ananna19,Ricci22,Ananna22}. Though some portion resides on galactic scales (e.g., \citealt{Buchner17a,Gilli22,Andonie23}), the densest {\em Compton-thick} obscuration ($N_{\rm H}$\,$>$\,1.5\,$\times$\,10$^{24}$\,cm$^{-2}$)\footnote{The Compton-thick threshold often corresponds to the inverse of the Thomson-scattering cross-section, but the actual threshold will depend on other factors such as abundances, etc. See the \texttt{MYtorus} manual (\url{http://mytorus.com/mytorus-instructions.html}) for more information.} is expected to reside on circum-nuclear parsec scales, similar to the sizes often invoked in unified schemes \citep{Antonucci93,Urry95,Netzer15,RamosAlmeida17}.

The Compton-thick fraction is often inferred to be similarly substantial to the obscured AGN population across cosmic time (see discussion in e.g., \citealt{Comastri15,Civano23}), even after considering the non-trivial dependence with the nature of the intrinsic X-ray-emitting corona and/or accretion flow (e.g., \citealt{Gandhi07,Vasudevan16,Kammoun23,Piotrowska23}). For example, the latest population synthesis models from \citet{Ananna19} constrain the abundance of Compton-thick AGN to be 50\,$\pm$9\% within $z$\,=\,0.1 and 56\,$\pm$\,9\% within $z$\,=\,1 of all AGN.

Theoretical models of supermassive black hole growth additionally suggest that enhanced circum-nuclear obscuration is intricately linked to not just supermassive black hole accretion (e.g., \citealt{Fabian99}), but also galaxy-supermassive black hole co-evolution (e.g., \citealt{AnglesAlcazar21}) and galaxy-galaxy interactions as a whole (e.g., \citealt{Springel05,Hopkins06,Pfeifle23_hexp}). Although it is still uncertain as to the exact role the dense circum-nuclear obscurer plays, some viable options include a feeding reservoir for the central black hole (e.g., \citealt{StorchiBergmann19}), or a by-product of the central engine itself (e.g., \citealt{Wada12}). Compton-thick AGN are hence pertinent targets to unveil the drivers of galaxy growth and understand the co-evolution between supermassive black holes and galaxies, as highlighted in the Astro2020 Decadal Survey\footnote{\href{https://www.nationalacademies.org/our-work/decadal-survey-on-astronomy-and-astrophysics-2020-astro2020}{https://www.nationalacademies.org/our-work/decadal-survey-on-astronomy-and-astrophysics-2020-astro2020}}.

However, Compton-thick AGN are one of the most difficult classes of AGN to detect and study \citep{Hickox18,Asmus20,Brandt22}. For energies $E$\,$<$\,10\,keV, the intrinsic X-ray flux from the corona is mostly extinguished via the photoelectric effect and only a few percent of the intrinsic flux escapes \citep{Gupta21}. Some flux survives in the form of narrow X-ray fluorescent lines at specific energies, with those arising from neutral iron K at 6.4\,keV (rest frame) typically being the strongest. The remaining AGN flux observed is dominated by X-ray photons that have undergone Compton recoil in one or multiple scatterings and escaped the obscurer, giving rise to the underlying Compton-scattered continuum. At $\sim$\,20\,--\,40\,keV, the continuum peaks into a broad Compton hump with overall shape determined by the geometry of the obscurer (e.g., \citealt{Matt00,Murphy09,Buchner19}).

\subsection{X-ray Spectroscopic Modelling of Circum-Nuclear Obscuration}

Many X-ray spectroscopic models describing the broadband X-ray emission from obscured AGN are available to date with varying geometric prescriptions for the obscurer. Such variations can be broadly separated into (1) ad-hoc (i.e. computationally-convenient) geometries, including smooth density obscurers (e.g., \texttt{etorus}; \citealt{Ikeda09}, \texttt{MYtorus}; \citealt{Murphy09}, \texttt{BNsphere}; \citealt{Brightman11a}, \texttt{RXtorus}; \citealt{Paltani17}, \texttt{borus}; \citealt{Balokovic18,Balokovic19}, \texttt{wedge}; \citealt{Buchner19}), clumpy obscurers (e.g., \texttt{Ctorus}; \citealt{Liu14}, \texttt{XCLUMPY}; \citealt{Tanimoto19}, \texttt{UXCLUMPY}; \citealt{Buchner19}), and combinations of different unique geometric components (see e.g., the polar gas simulations from \citealt{Liu19,Mckaig22} or the broadband physical model of the Circinus Galaxy in \citealt{Andonie22}) and (2) geometries that emerge from radiative simulations (e.g., \texttt{warped-disk}, \texttt{radiative-fountain}; \citealt{Buchner21}). There has also been a surge in the availability of ray tracing packages designed to enable the production of bespoke user-defined X-ray spectral models in arbitrary geometries and the inclusion of additional physical processes (e.g., \textsc{MONACO} \citealt{Odaka11,Odaka16}, \textsc{RefleX}\footnote{\href{https://www.astro.unige.ch/reflex/}{https://www.astro.unige.ch/reflex/}}; \citealt{Paltani17,Ricci23}, \textsc{XARS}\footnote{\href{https://github.com/JohannesBuchner/xars}{https://github.com/JohannesBuchner/xars}}; \citealt{Buchner19}, \textsc{SKIRT}; \citealt{VanderMeulen23}\footnote{\href{https://skirt.ugent.be/root/\_contributing.html\#ContributingRepositories}{https://skirt.ugent.be/root/\_contributing.html\#ContributingRepositories}}).

The ability for accurate and precise inference from circum-nuclear obscuration models with ever-increasing numbers of fit parameters is currently met by substantial challenges. The first is exploring the degenerate and multi-modal (i.e. non-identifiable) parameter spaces inherent to the spectral model libraries that result from ray tracing simulations. A typical model to explain the 0.2\,--\,80\,keV spectra of obscured AGN can consist of $\gtrsim$\,10 parameters describing the intrinsic X-ray spectrum, the geometric prescription of the surrounding circum-nuclear obscurer and other contaminating soft X-ray emissions. The corresponding multi-dimensional parameter spaces are very complex and do not necessarily lead to unique spectral solutions when compared with alternative geometric models of the obscurer (e.g., \citealt{Saha22,Kallova23}). As such, parameter exploration, model verification and model comparison are all non-trivial and can be exceedingly expensive to compute with increased numbers of fit parameters (e.g., \citealt{vanDyk01,Buchner14,Buchner23}). Increased complexity of obscuration models will also require more ray-tracing simulations to compute. Due to the corresponding trade-off between exploring fewer geometries versus coarser parameter grid resolution, the conventional use of multi-dimensional tables and grid interpolation to fit spectra may become obsolete entirely. A promising alternative is emulation, which has been shown to accelerate the computation time associated with radiative transfer simulations \citep{Kerzendorf21,RinoSilvestre22} and avoid the requirement for coarse gridding of parameters into multi-dimensional tables entirely \citep{Matzeu22}.

The second challenge is the observational requirement for high quality {\em broadband} spectroscopy of Compton-thick AGN to test complex physical models. Valuable insights have been attained with focusing X-ray optics $<$\,10\,keV, typically capable of isolating the Fe\,K$\alpha$ complex and underlying reflection continuum from contaminating non-AGN spectral features (e.g., \citealt{Risaliti99,Brightman11a,Lamassa17}). But without similar sensitivities $>$\,10\,keV, strong ambiguity still remains relating to the shape of the Compton hump (see discussion in \citealt{Brightman15,Lamassa19,Lamassa23}). Hard X-ray sensitivities provided by coded aperture masks have limited previous studies to higher observed X-ray fluxes (e.g., \citealt{Yaqoob12,Gandhi13,Gandhi15,Ricci17_bassV}), and reduced spectral resolution that is insufficient to strongly constrain geometrical properties of the circum-nuclear environment (see discussion in e.g., \citealt{Balokovic17,Tanimoto22}).

\textit{NuSTAR} \citep{Harrison13} provided the first focusing hard X-ray telescope in orbit, opening a new era into the pursuit and understanding of obscured accretion onto supermassive black holes. To date, \textit{NuSTAR} has provided the most sensitive insights into the X-ray obscuration of the brightest Compton-thick AGN known \citep{Puccetti14,Arevalo14,Bauer15,Puccetti16}, as well as the wider population identified previously with wide-field hard X-ray monitoring surveys (e.g., \citealt{Annuar15,Gandhi17,Marchesi17,Marchesi19a,Zhao21,Traina21,TorresAlba21,Pizzetti22,Tanimoto22,Silver22}). \textit{NuSTAR} has also enabled unambiguous Compton-thick line-of-sight column density classifications for a bulk of the Compton-thick population that had not been detected $>$\,10\,keV before \citep{Balokovic14,Gandhi14,Ptak15,Masini16,Boorman16,Annuar17,Ricci17b,Lamassa19,Kammoun20}. Broadband X-ray spectroscopy has thus proven to be a crucial tool for understanding the obscurer, provided consistent sensitivities are attainable with simultaneous soft X-ray observations across the entire passband \citep{Marchesi18,Marchesi19b}.

\subsection{Complex Structure Revealed by Variability}

There is mounting evidence that the obscurer can be clumpy rather than smooth. This was initially motivated by the infrared spectrum of AGN, which showed less prominent Silicate features than expected from smooth obscuration models \citep[e.g.,][]{Jaffe04,Nenkova08a,Elitzur06,Risaliti07,Honig07}. 
Models such as Chaotic Cold Accretion \citep{Gaspari13,Gaspari15,Gaspari20} suggests clumpy accretion from random angles onto the central 100\,pc induced by radiative cooling and turbulance (see e.g., \citealt{Rose19,Gaspari20,Maccagni21,Temi22}). Obscurer geometries arising from radiation-driven outflows also can produce dynamic, filamentary and clumpy structures (e.g., \citealt{Vollmer02,Wada12,Chan16} and references therein). Such clumpy/filamentary models of AGN obscuration predict observed changes in the line-of-sight obscuration.

In X-rays, an inhomogeneity in the circum-nuclear material can vary (i) the accretion luminosity and/or (ii) obscuration level. Such changes can be detected and disambiguated with sufficiently sensitive time-resolved X-ray spectroscopy with a wide-enough passband \citep{Ricci22_variability}. For line-of-sight column density variations $\Delta$\,$N_{\rm H}$\,$\lesssim$\,10$^{23}$\,cm$^{-2}$, the photoelectric turnover $\lesssim$\,10\,keV has been used to robustly confirm obscuration variations (e.g., \citealt{Risaliti02,Risaliti05,Markowitz14}). However, for variations $\Delta$\,$N_{\rm H}$\,$\gtrsim$\,10$^{23}\,-\,10^{24}$\,cm$^{-2}$ sensitive broadband spectroscopy is advantageous to provide constraints on the underlying absorbed spectrum $<$\,10\,keV as well as the reprocessed spectrum $>$\,10\,keV to avoid the strong degeneracy between the spectral slope, obscuration level and amount of reprocessing (e.g., \citealt{Walton14,Rivers15,Lefkir23}). Decoupling such large changes in obscuration from intrinsic flux variations exclusively in Compton-thick AGN is currently even less represented in the literature, owing in part to the observational demand for observing variations in the Compton hump that are non-trivial to disentangle $\gtrsim$\,10--20\,keV in all but the brightest targets (e.g., \citealt{Puccetti14,Marinucci16,Nardini17,Zaino20,Kayal23}).

Column density variations are expected to occur over periods of time from $\sim$1\,day up to several months, assuming a typical range of obscuring cloud filling factors, velocities and distances from the accreting black hole \citep[e.g,][]{Nenkova08a}. Tentative column density variability timescales on the order of years also exist (e.g., \citealt{Gandhi17,Masini17,Laha20,TorresAlba23}), but additional sensitive monitoring is required to quantify its prevalence in the obscured AGN population. Thus X-ray obscuration variability is a powerful tool for providing reliable constraints on the location of obscuring clouds and their distances from the accreting supermassive black hole \citep{Markowitz14,Buchner19}. By combining numerous epochs of broadband X-ray observations with physical obscuration models, the global properties of the circum-nuclear environment (such as covering factor and average column density) can be decoupled from the epoch-dependent variable components to provide the tightest constraints on obscurer properties in the heavily obscured AGN population currently known (e.g., \citealt{Ricci16,Balokovic18,Pizzetti22,Marchesi22,TorresAlba23,Kayal23}). For such observations, simultaneous observations from $\sim$\,0.2\,--\,80\,keV are essential. These are challenging to achieve, currently requiring coordination of complementary missions (e.g., \textit{XMM-Newton} and \textit{NuSTAR}), which has limited the sample size for such studies (see Section~\ref{sec:variability}).

\subsection{The Circum-Nuclear Environment at Low Accretion Power}
Volume-limited multi-wavelength surveys have revealed that the majority of supermassive black holes in the nearby Universe are underfed (e.g., \citealt{Ho97,Ho08,Baldi18,Baldi21a,Baldi21b,Williams22}). This implies that the majority of local galaxies host low-luminosity AGN, often parameterised to have bolometric luminosities $L_{{\rm bol}}$\,$\lesssim$\,$10^{42}$\,erg\,s$^{-1}$, and/or Eddington-scaled bolometric luminosities (also known as the Eddington ratio) of $\lambda_{\rm Edd}$\,=\,$L_{\rm bol}$\,/\,$L_{\rm Edd}$\,$\lesssim$\,$10^{-3}$ (e.g., \citealt{Elitzur06,Honig07,Kawamuro16}). However, our understanding of the circum-nuclear environment in AGN at low luminosities and accretion powers is currently very incomplete.

Low-luminosity AGN are known to lack an ultraviolet bump in their spectral energy distribution (e.g., \citealt{Ho99,Nemmen06,Eracleous10}) and share similar characteristics to low-luminosity/quiescent accreting stellar mass black holes (e.g., \citealt{Nagar05,Kording06,Svoboda17,FernandezOntiveros21,Moravec22}), suggesting the absence of a standard optically thick, geometrically thin accretion disk \citep{Shakura73}. At X-ray wavelengths, the absence of Fe\,K$\alpha$ emission lines and/or the Compton hump in some low-luminosity AGN also supports this notion, indicating the truncation or absence of a standard accretion disk \citep{Terashima02,GonzalezMartin09,Younes11,Ursini15,Young18,Younes19,OsorioClavijo22}. Some studies have predicted the collapse and disappearance of the broad-line region and obscuring structure in low-luminosity AGN due to insufficient radiation pressure (e.g., \citealt{Elitzur06,Honig07,Elitzur09}). Although there is some observational evidence supporting these predictions, data is often limited due to the relative difficulty of selecting and studying low-luminosity AGN relative to their more luminous counterparts \citep[e.g.,][]{Maoz05,Ho08,Trump11,HernandezGarcia16,GonzalezMartin17,Ricci17_eddrat}. Measurements of the obscuring covering factor on a source-by-source basis in large samples via X-ray spectroscopic modelling have provided additional clues supporting this notion, though with considerable uncertainties (e.g., \citealt{Brightman11b,Vasudevan13,Brightman15,Balokovic17}).

\subsection{Accreting Intermediate Mass Black Holes}
Large numbers of intermediate mass black holes with masses $M_{\rm BH}$\,$\sim$\,10$^{2}$\,--\,10$^{5}$\,M$_{\odot}$ are required to exist throughout cosmic history to give rise to the $\sim$\,10$^{9}$\,M$_{\odot}$ supermassive black holes observed within mere hundreds of million years from the Big Bang (e.g., \citealt{Banados18}) up to the present day. A large ongoing challenge however, is to observationally identify intermediate mass black holes and to understand how they were formed \citep{Greene20}. Dwarf galaxies are useful to search for intermediate mass black holes. To explain their low masses, dwarf galaxies are expected to have undergone fewer mergers than more massive galaxies which in turn restricts the availability of fuel for the central black holes to grow. Dwarf galaxies in the local Universe are hence expected to contain the seeds of the first supermassive black holes, and the dwarf galaxy black hole occupation fraction is a crucial piece of the puzzle (e.g., \citealt{Volonteri08,Volonteri10,Greene12,Reines22}).

A useful strategy is to search for intermediate mass black hole signatures during episodes of accretion in dwarf galaxy AGN. The difficulty is that any biases imposed on selecting accreting supermassive black holes in AGN are exacerbated in the case of intermediate mass black holes in low-mass galaxies. Dwarf galaxies often have high levels of star formation which can be significantly stronger than the optical emission associated with the accretion disc surrounding accreting intermediate mass black holes (e.g., \citealt{Moran14,Trump15}). Optical spectroscopy has proven a useful tool for identifying unobscured dwarf AGN signatures (e.g., \citealt{Greene04,Greene07,Reines13,Baldassare18}). However, this technique requires that host galaxy dilution be minimal and that the AGN be largely unobscured while accreting at high rates, close to the Eddington limit. X-ray observations are less affected by host galaxy contamination but soft X-rays can be readily absorbed leading to the same biases encountered for more massive black holes in AGN (e.g., \citealt{Brandt15,Hickox18}). Broadband X-ray observations including hard X-rays are hence crucial to disentangle obscured accreting massive black holes from individual host galaxy X-ray binaries that often have different predicted hard X-ray spectral shapes (see discussion in e.g., \citealt{Lehmer23}). Detailed broadband X-ray spectroscopic studies of obscured massive black holes in low-mass galaxies are currently rare though due to the requirement for sufficient sensitivities (e.g., \citealt{Ansh23,Mohanadas23}).

\subsection{The \textit{HEX-P} Perspective}

\textit{The High-Energy X-ray Probe} (\textit{HEX-P}; \citealt{Madsen23}) is a probe-class mission concept that offers sensitive coverage ($0.2-80$\,keV) of the X-ray spectrum with exceptional spectral, timing and angular capabilities. It features two high-energy telescopes (HET) that focus hard X-rays, and soft X-ray coverage with a low-energy telescope (LET).

The LET (0.2\,--\,25\,keV) consists of a segmented mirror assembly coated with Ir on monocrystalline silicon that achieves a half power diameter of 3.5'', and a low-energy DEPFET detector, of the same type as the Wide Field Imager (WFI; \citealt{Meidinger20}) onboard \textit{Athena} \citep{Nandra13}. It has 512\,$\times$\,512 pixels that cover a field of view of 11.3'\,$\times$\,11.3'. It has an effective passband of 0.2\,--\,25\,keV, and a full frame readout time of 2\,ms, which can be operated in a 128 and 64 channel window mode for higher count-rates to mitigate pile-up and achieve faster readout. Pile-up effects remain below an acceptable limit of $\sim 1\%$ for fluxes up to $\sim 100$\,mCrab in the smallest window configuration (64w). Excising the core of the PSF, a common practice in X-ray astronomy, will allow for observations of brighter sources, with a~typical loss of up to $\sim$\,60\% of the total photon counts.

The HETs (2\,--\,80\,keV) consist of two co-aligned telescopes and detector modules. The optics are made of Ni-electroformed full shell mirror substrates, leveraging the heritage of \textit{XMM-Newton}\ \citep{Jansen01}, and coated with Pt/C and W/Si multilayers for an effective passband of 2\,--\,80\,keV. The high-energy detectors are of the same type as those onboard \textit{NuSTAR}\ \citep{Harrison13}, and they consist of 16 CZT sensors per focal plane, tiled 4\,$\times$\,4, for a total of 128\,$\times$\,128 pixel spanning a field of view slightly larger than for the LET, of 13.4'\,$\times$\,13.4'.

The unique improvements yielded by \textit{HEX-P} will provide significant advancements in the study of supermassive black hole growth. Enhanced sensitivity above 10\,keV relative to \textit{NuSTAR} will enable detailed modelling constraints of the faintest Compton-thick AGN currently known (see Section~\ref{sec:llagn} and \citealt{Pfeifle23_hexp}), as well as the completion of the local AGN census which is predominantly restricted by our ability to uncover Compton-thick AGN \citep{Asmus20}. The strictly simultaneous broadband coverage will also remove any ambiguity associated with spectral component variability which can significantly affect the inference of key system parameters (see discussion in \citealt{Balokovic18,Balokovic21,TorresAlba23} and Section~\ref{sec:variability}). Lastly, the extended passband of the LET up to energies of $\sim$\,25\,keV will provide sensitive overlapping X-ray spectroscopy in all three \textit{HEX-P} cameras. The energy range $\sim$\,5\,--\,8\,keV contains the iron K lines which hold enormous diagnostic value for the structure of the circum-nuclear obscurer when combined with the underlying continuum $\gtrsim$\,10\,keV (e.g., \citealt{Balokovic18}). Having three individual instruments across this wavelength range will also provide independent verification for blue-shifted absorption features arising from outflowing material that have proven difficult to detect in heavily obscured AGN to-date (e.g., \citealt{Matzeu19}). At higher energies, the passband between $\sim$8\,--\,25\,keV encompasses the first inflection point of the Compton hump which holds exciting potential as a fingerprint-like identifier for the circum-nuclear obscurer(s) surrounding AGN (see e.g., \citealt{Buchner19,Buchner21}).

The paper is organised as follows. Section~\ref{sec:ctpop} presents the latest compilation of published Compton-thick AGN within $\sim$\,400\,Mpc confirmed in part by \textit{NuSTAR} observations. To our knowledge, this is the largest compilation of local Compton-thick AGN constructed to date which combines sources selected at a variety of different wavelengths. In Section~\ref{sec:megamasers} we present a detailed X-ray spectral analysis of local megamaser AGN, highlighting the prospects for \textit{HEX-P} and next-generation obscuration models to study the effects of radiative feedback from AGN. Section~\ref{sec:variability} showcases the future possibilities with multi-epoch studies attainable with high-sensitivity, strictly simultaneous, broadband X-ray spectroscopy. The current and future prospects behind the behaviour of dense AGN obscurers at extremely low luminosities are discussed in Section~\ref{sec:llagn} followed by the prospects for detecting faint obscured AGN in dwarf galaxies in Section~\ref{sec:dwarf}. We then provide a quantitative estimate of the volume accessible by \textit{HEX-P} for robust characterisation of Compton-thick obscuration and compare to the current state-of-the-art in Section~\ref{sec:sensitivity}. We present our summaries in Section~\ref{sec:summary}.

All the \textit{HEX-P} simulations presented in this work were produced with a set of response files that represent the observatory performance based on current best estimates (v07 - 17-04-2023; see \citealt{Madsen23}). The effective area is derived from a ray-trace of the mirror design including obscuration by all known structures. The detector responses are based on simulations performed by the respective hardware groups, with an optical blocking filter for the LET and a Be window and thermal insulation for the HET. The LET background was derived from a GEANT4 simulation \citep{Eraerds21} of the WFI instrument, and the one for the HET from a GEANT4 simulation of the \textit{NuSTAR} instrument, both positioned at L1.

\section{The Database of Compton-Thick AGN (DoCTA)}\label{sec:ctpop}

To understand the Compton-thick AGN population uncovered to-date, and as a natural starting point for our \textit{HEX-P} simulations, we construct a comprehensive list of Compton-thick AGN identified in the literature. To ensure accurate modelling of the underlying Compton scattered continuum in such sources, we limit our search to targets confirmed with spectral modelling that included \textit{NuSTAR} data. Whilst other X-ray missions such as \textit{Suzaku}/Hard X-ray Detector, \textit{Swift}/Burst Alert Telescope and \textit{INTEGRAL} instruments have provided hard X-ray spectroscopic constraints for some Compton-thick AGN (e.g., \citealt{Yaqoob12,Vasylenko13,Gandhi13,Gandhi15,Ricci15}), we limit ourselves to the requirement for \textit{NuSTAR} due to its 100-fold increase in sensitivity $>$\,10\,keV relative to previous missions \citep{Harrison13}. The resulting Database of Compton-thick AGN (DoCTA) was created as follows:

\begin{itemize}
\item[1.] \textit{Literature search:} We first identify a list of peer-reviewed publications with a NASA ADS\footnote{\href{https://ui.adsabs.harvard.edu}{https://ui.adsabs.harvard.edu}} search query for any refereed paper containing the phrase `Compton-thick' and `\textit{NuSTAR}' somewhere in its main body of text. The search returned 690 refereed publications.
\item[2.] \textit{Literature refinement:} We manually filtered through the list of publications identified in Step~1, finding $\sim$90 publications that use AGN X-ray spectroscopic modelling with \textit{NuSTAR} (often but not always complemented by soft X-ray spectroscopy from a different instrument) to constrain line-of-sight column density via obscuration models of some form.
\item[3.] \textit{Line-of-sight column density\footnote{We refer to `line-of-sight' or `angle-averaged' column densities throughout this work to distinguish between the Thomson depth of material along the line-of-sight and the angle-averaged Thomson depth of material out of the line-of-sight.}:} We sought to extract the line-of-sight column density of each system for every acceptable model fit presented per publication. Our reasoning behind this strategy was to be as complete of the literature as possible, enabling inclusion of sources that are classified as Compton-thick with a specific model setup, but not in others. A number of sources have line-of-sight column density measurement upper bounds \textit{consistent} with the Compton-thick limit, but to be conservative we only consider sources with at least one line-of-sight column density measurement lower 90\% confidence bound above the Compton-thick threshold of 1.5$\times$10$^{24}$\,cm$^{-2}$.
\item[4.] \textit{Computing unabsorbed luminosities:} To understand the fundamental demographics of the sample, we next estimated intrinsic luminosity. Given the inevitably large number of local AGN selected, a large number of redshift-independent distances are available for the sources. We cross-matched the initial set of objects from Step~3 with the NASA Extragalactic Database\footnote{\href{https://ned.ipac.caltech.edu}{https://ned.ipac.caltech.edu}} using sexagesimal coordinates to remove duplicates arising from different published identifiers. We then downloaded all redshift-independent distances per source and used the median distance where more than five distance measurements were available. To overcome possible effects from discrepant distance estimates to a given source, we additionally store the distances used by each work and manually correct intrinsic 2\,--\,10\,keV luminosities into fluxes when intrinsic fluxes are not provided. We use the median redshift-independent distance listed on NED where available, otherwise the luminosity distance is calculated assuming the cosmological parameters $H_{0}$\,=\,70.0\,km\,s$^{-1}$\,Mpc$^{-1}$, ${\rm \Omega}_{\Lambda}$\,=\,0.7 and ${\rm \Omega}_{\rm M}$\,=\,0.3.
\end{itemize}

\begin{figure*}
\begin{center}
\includegraphics[width=1.0\textwidth]{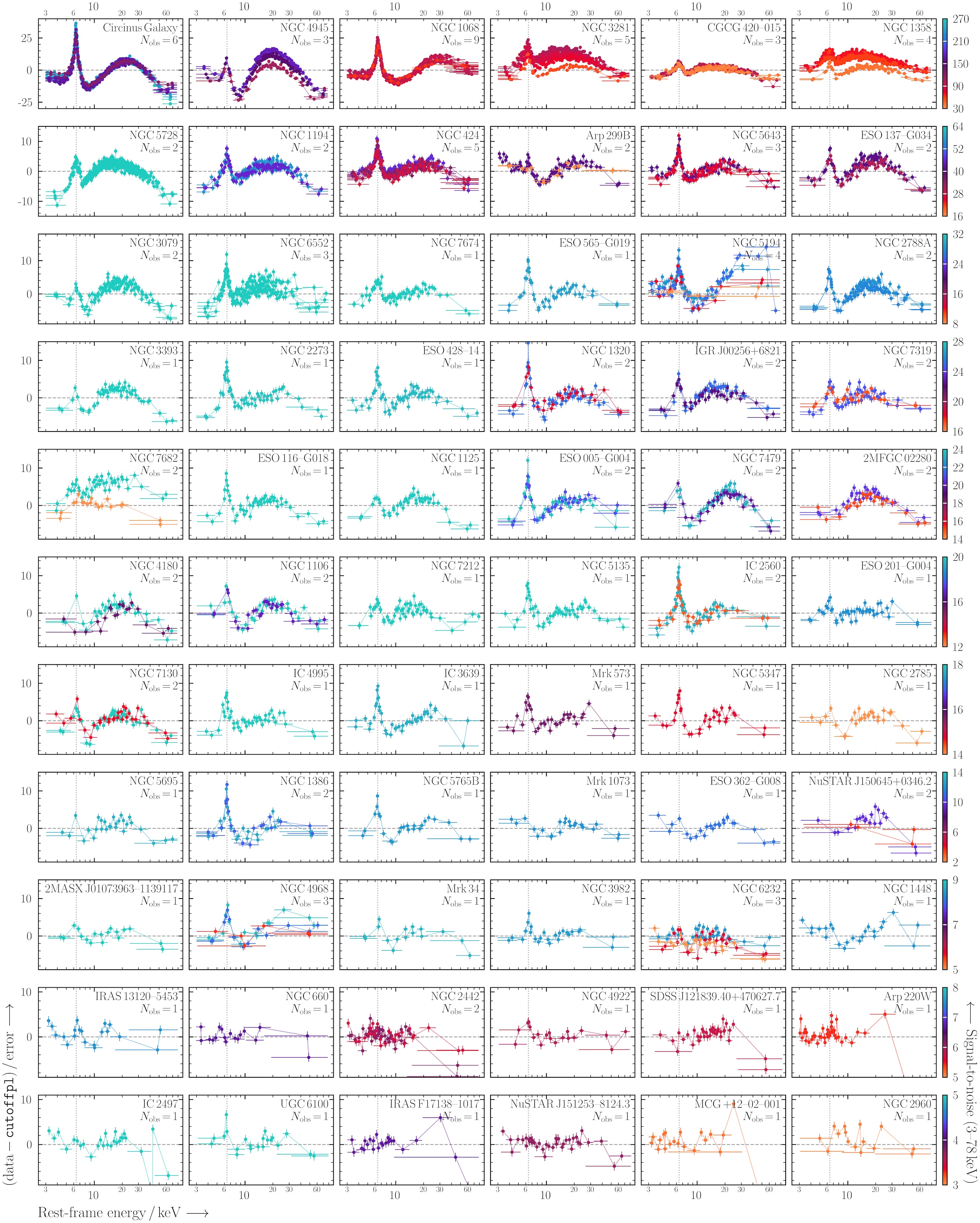}
\end{center}
\caption{Every Compton-thick AGN in DoCTA confirmed by \textit{NuSTAR}. Each panel plots the relative residual for a simple \texttt{cutoffpl} fit to each \textit{NuSTAR} spectrum with $>$\,10\,ks of data, and is coloured by the signal-to-noise ratio in the 3\,--\,78\,keV band.}\label{fig:docta_nustar}
\end{figure*}

DoCTA contains 66 Compton-thick AGN candidates, with a wide range of possible scientific applications.  The primary focus for this work is to assess the ability of current circum-nuclear obscuration models paired with modern broadband X-ray spectral sensitivity to provide unique solutions for intrinsic luminosity and line-of-sight column density, in the absence of considerable source variability (see Figure~\ref{fig:docta_nustar} \& Section~\ref{sec:variability} for more information on this assumption).

\begin{figure*}
\begin{center}
\includegraphics[width=\textwidth]{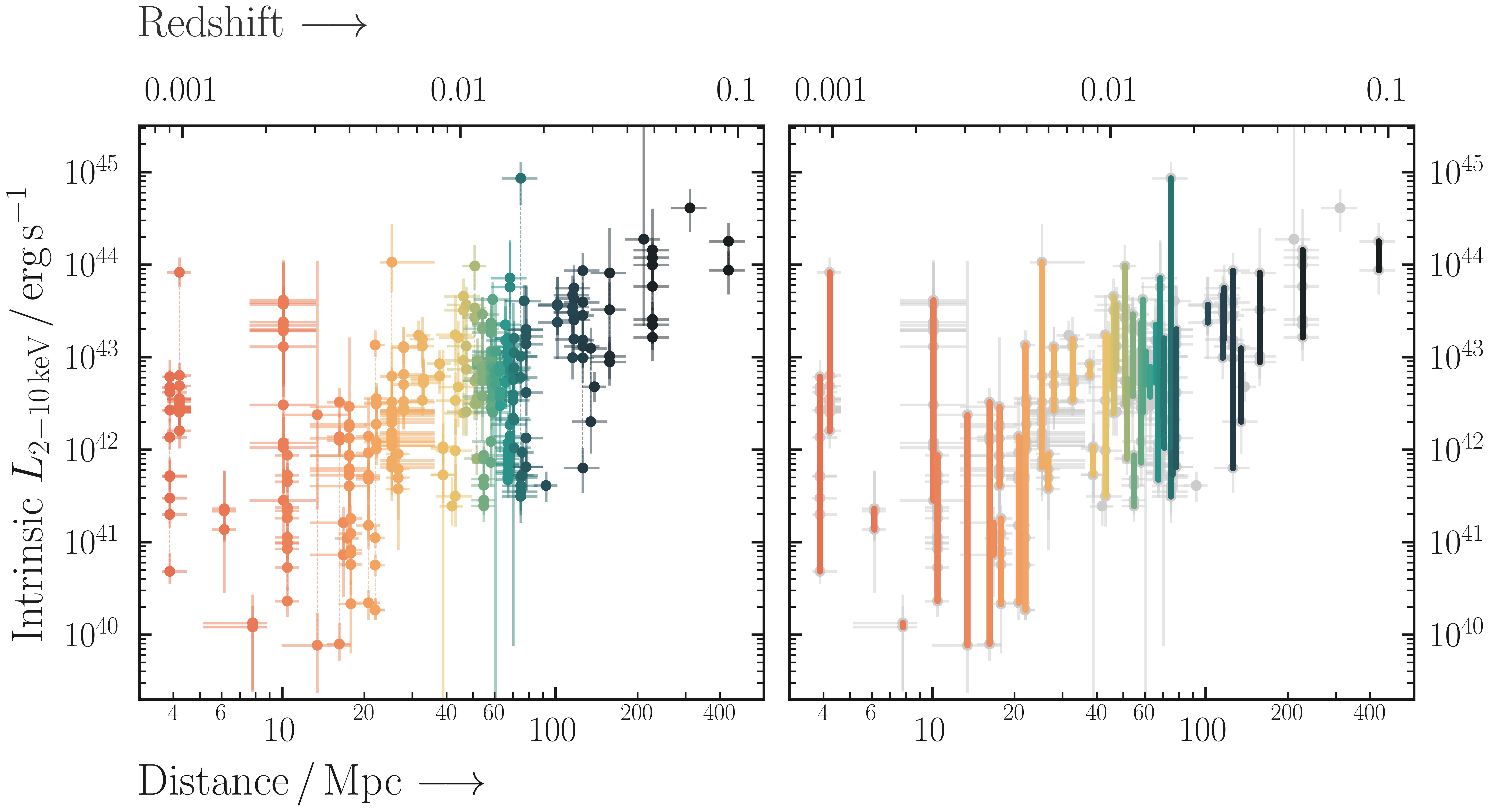}
\end{center}
\caption{\textit{(Left)} Distance vs. intrinsic (i.e. unabsorbed) 2\,--\,10\,keV luminosity reported in the literature for all local Compton-thick AGN. We only include Compton-thick classifications that included \textit{NuSTAR} data in the spectral analysis. Multiple reported luminosities are connected by a vertical dashed line. \textit{(Right)} Same as the left panel, with vertical bars indicating the range in best-fit luminosities for each source. For most sources the reported luminosities vary over one to two orders of magnitude, in some extreme cases over three orders of magnitude. The colour coding is used to distinguish between different sources, and does not correspond to a physical parameter.}\label{fig:docta_logL}
\end{figure*}

Figure~\ref{fig:docta_logL} presents the distance vs. unabsorbed 2\,--\,10\,keV X-ray luminosity for every intrinsic 2\,--\,10\,keV luminosity measurement of every source in DoCTA. The left panel shows that DoCTA is limited to $\lesssim$\,100\,Mpc for the conventional Seyfert definition with unabsorbed 2\,--\,10\,keV luminosity, $L_{2-10\,{\rm keV}}$\,$>$\,10$^{42}$\,erg\,s$^{-1}$. Beyond $\sim$\,250\,Mpc, only quasars ($L_{2-10\,{\rm keV}}$\,$>$\,10$^{44}$\,erg\,s$^{-1}$) have been found as Compton-thick. But overall there is a significant range of published intrinsic luminosities per source, with some estimates ranging over $>$\,3 orders of magnitude in the most extreme cases. This trend is shown in Figure~\ref{fig:docta_logL}, which marks the range in intrinsic luminosity reported in the spectral fits included in DoCTA. Under the assumption of negligible source variability (c.f. Figure~\ref{fig:docta_nustar}), such variation can arise from choosing different circum-nuclear obscuration models. This also leads to a significant systematic uncertainty in parameter inference (i.e. unabsorbed luminosity, but also line-of-sight column density and covering factors). It is concerning that this occurs even in the local Universe where the observing conditions are optimal (e.g., bright fluxes thanks to source proximity and spatial separation of nearby nuclear contaminants). Some sources, particularly at larger distances, appear to show little luminosity variation. However, this may be because those targets have been studied in fewer publications than famous nearby sources (e.g., the Circinus Galaxy, NGC\,4945).

\begin{figure*}
\begin{center}
\includegraphics[width=\textwidth]{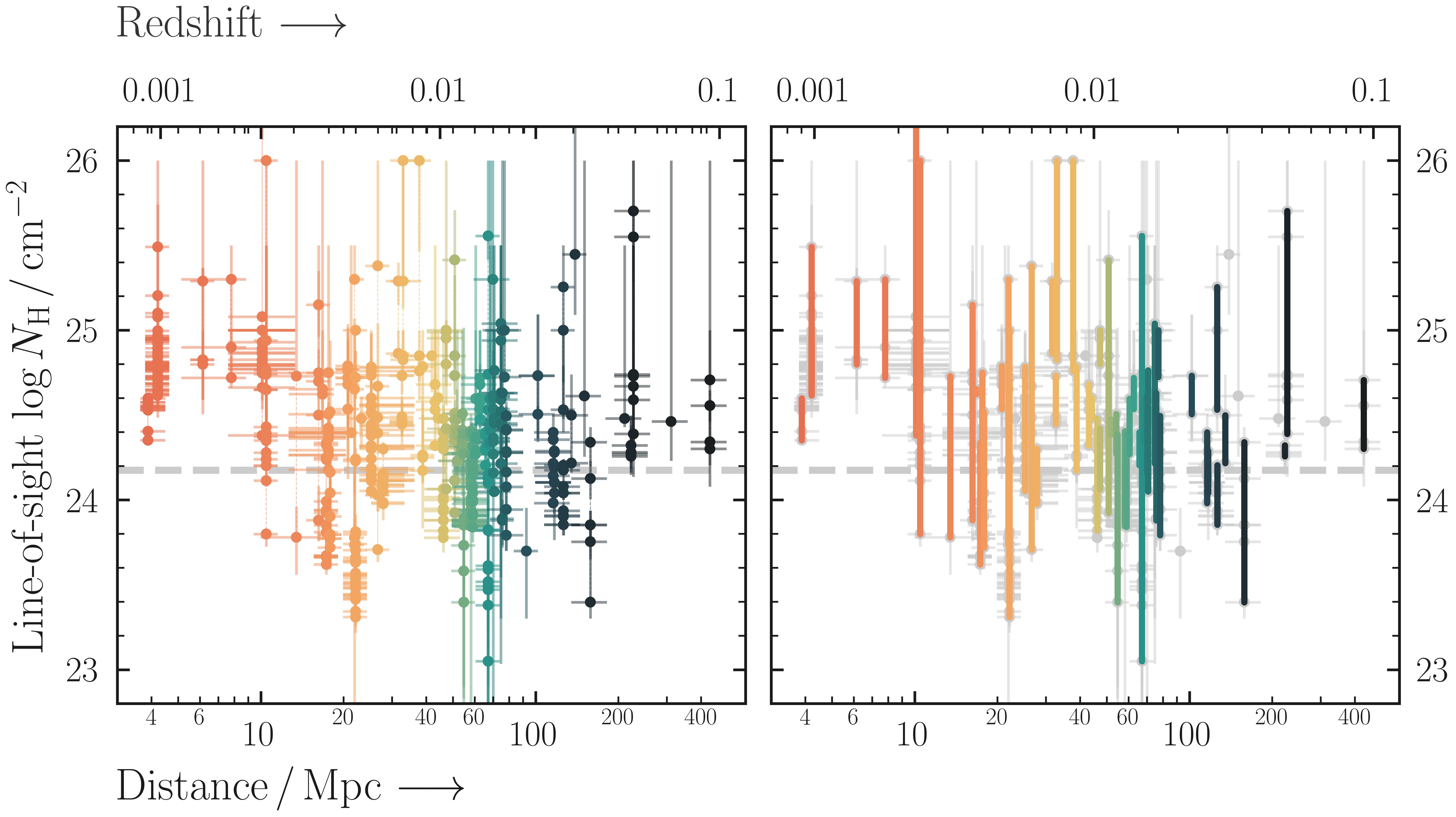}
\end{center}
\caption{Left and right panels are the same as in Figure~\ref{fig:docta_logL}, apart from the reported line-of-sight column density is shown on the vertical axis. The Compton-thick threshold adopted for this work is shown with a horizontal dashed line. The large range between reported line-of-sight log\,$N_{\rm H}$ values per source in the Compton-thick regime highlights the challenge with current models to constrain the upper boundary of column density for Compton-thick AGN. The large range is due to modelling degeneracies remaining in low signal-to-noise ratio hard X-ray data and variability in time relative to the complementary soft X-ray observations. Both issues will be directly addressed with \textit{HEX-P}.}\label{fig:docta_NH}
\end{figure*}

Figure~\ref{fig:docta_NH} presents distance vs. line-of-sight column density from the literature for every available measurement of every source in DoCTA. Similarly to Figure~\ref{fig:docta_logL}, we find a significant range of measured line-of-sight column densities. Such a range can arise from subtle differences in the physical properties of the obscurer assumed in different models. For example, line-of-sight column density is inextricably linked to predicted intrinsic luminosity since an increase in line-of-sight column density requires an increase in intrinsic luminosity to explain the additional absorption. Alternatively, parameter differences across obscuration models can arise from how the parameters are represented in the corresponding table models used in the spectral fitting (e.g., the number of parameter grid points). A primary effect of such confusion is that a significant number of sources have published measured line-of-sight column densities both above and below the Compton-thick limit (horizontal dashed line) to 90\% confidence. Such uncertainty can clearly have fundamental model-dependent effects on measurements of the Compton-thick AGN abundance, for example. A subset of the DoCTA sources are known Changing-Obscuration AGN \citep{Ricci22_variability}, in which the line-of-sight column density varies both below and above the Compton-thick threshold (e.g., NGC\,1358; \citealt{Marchesi22}). However, Compton-thick Changing-Obscuration events are currently observationally rare meaning that the wide range in line-of-sight column densities are not expected to be dominated by Changing-Look AGN.

A large number of different model prescriptions for the obscuration-based reprocessed spectrum in AGN are available in the literature today as well as different bespoke setups that incorporate those models. A prime example of the latter is the use of decoupled models, in which the global average properties of the reprocessor are decoupled from the reprocessing effects along the line-of-sight (see \citealt{Yaqoob12} for an detailed review of such techniques). On the practical side, decoupled model fitting often improves the fit due to the larger range of spectral shapes attainable. On the theoretical side, it can be interpreted as flux variability, or line-of-sight column density variations arising from a clumpy obscurer. As such, decoupled model setups can often include an overall scaling of the intrinsic continuum relative to the reprocessed one. \citet{Lamassa19} have shown the effect of manually altering the contribution from reprocessing in the broadband spectral fitting of NGC\,4968, finding that an increase in reprocessed flux corresponds to an overall decrease in intrinsic continuum flux, as expected.

Many obscured AGN are variable in hard X-rays (e.g., \citealt{TorresAlba23}), including bright Compton-thick AGN (e.g., \citealt{Puccetti14,Marinucci16,Marchesi22}). To understand the high-energy (E\,$>$\,10\,keV) spectral constraints for the DoCTA population, and to qualitatively understand the possibility of variability impacting the ranges shown in Figures~\ref{fig:docta_logL}~\&~\ref{fig:docta_NH}, we extracted all archival \textit{NuSTAR} data available per source with $>$\,10\,ks of net exposure time in both FPMA and FPMB. While intra-observation variability is not unheard of in Compton-thick AGN (e.g., \citealt{Puccetti14}), it is currently rare in the literature, so we choose to extract epoch-averaged spectra. The \textit{NuSTAR} data for both FPMA and FPMB were processed using the \textit{NuSTAR} Data Analysis Software package within \texttt{HEAsoft}. The task \texttt{nupipeline} was used to generate cleaned event files. Spectra and corresponding response files were generated using \texttt{nuproducts} with circular source regions of 20\,pixels ($\sim$49'') and background regions as large as possible on the same detector as the target. Each spectrum was then binned using the optimal binning scheme of \citet{Kaastra16}.

Figure~\ref{fig:docta_nustar} presents residuals in the form of \texttt{(data\,-\,model)\,/\,error} for every extracted \textit{NuSTAR} spectrum after fitting a simple \texttt{zcutoffpl} model in PyXspec to only the first observed spectrum per source in the 3\,--\,78\,keV band. The figure is organised vertically into bins of \textit{NuSTAR}/FPMA 3\,--\,78\,keV signal-to-noise ratio, increasing from the bottom to top row. A number of interesting features are visible from Figure~\ref{fig:docta_nustar}. Firstly there is a large diversity in shapes of the Fe\,K complex (rest frame 6.4\,keV is marked with a vertical line in each panel) and Compton hump across the sample. A number of factors can contribute to observed spectral diversity in heavily obscured AGN, whether it be from contamination in the Fe\,K band and softer energies ($E$\,$\lesssim$\,8\,keV) arising from competing spectral components (e.g., \citealt{Annuar15,Reynolds15,Farrah16,Gandhi17}) or due to the structure of the obscurer itself at $E$\,$\gtrsim$\,8\,keV (e.g., \citealt{Buchner19,Buchner21}). As noted by \citealt{Bauer15}, it is physically unlikely for a single column density obscurer to surround AGN and a plausible alternative could be a continuous distribution of obscurers with varying column densities and other geometric parameters, consistent with Galactic molecular cloud studies (e.g., \citealt{Goodman09}).

Most of the highest signal-to-noise ratio Compton-thick AGN on the top row of Figure~\ref{fig:docta_nustar} display significant spectral variability with \textit{NuSTAR} \citep{Puccetti14,Marinucci16,Marchesi22,Kayal23}. Only a small number of targets have been selected for \textit{NuSTAR} follow-up because of known variability; with sufficient sensitivity, repeated observations of others may well show that obscurer-based variability is ubiquitous amongst Compton-thick AGN. For the remainder of the DoCTA population, there are either insufficient \textit{NuSTAR} epochs to search for variability (see panels with $N_{\rm obs}$\,=\,1 in the figure) or the visual difference in the observed reflection spectra is small. Furthermore, the lowest third of DoCTA sources in terms of signal-to-noise ratio have insufficient data quality to reveal any spectral variability in detail (see Section~\ref{sec:variability} for further discussion).

\textit{HEX-P} is optimised in many ways to guide the future development of X-ray spectral models. First, the improved broadband sensitivity achieved by reducing the background level will result in a significant improvement in the observed signal-to-noise ratio for the bulk of the Compton-thick AGN population present in DoCTA. Such improvements will fundamentally decrease the number of possibilities for non-unique spectral fits in which parameter posteriors are significantly different. Second, the extended range of the LET to energies $>$\,10\,keV means that there will be three instruments in total providing sensitive spectra over the energy range corresponding to the inflection point of the Compton hump. Detailed spectral modelling of Compton hump diversity is currently an under-used resource for constraining the covering factor of material with different column densities surrounding the central engine (e.g., \citealt{Buchner19}). Lastly, the simultaneous broadband focusing capabilities of \textit{HEX-P} are a novel concept amongst previous, current and future planned X-ray missions. Broadband coverage removes any possible issues that can arise from variability or significantly mis-matched data quality in the soft and hard X-ray bands.

\section{Developing next-generation models of the circum-nuclear environment}\label{sec:megamasers}
All physical obscurer models feature multiple geometric degrees of freedom that are unique to the geometry assumed (e.g., some combination of line-of-sight column density, global obscurer column density, inclination angle, covering factors, etc.). However, the relative importance for each parameter in a given model fit is often non-trivial with many inter-parameter dependencies and multi-modal solutions to consider. Studies of the obscuration properties of AGN have shown that the covering factor is related to the Eddington-scaled accretion rate (e.g., \citealt{Fabian99,Fabian08,Ricci17_eddrat}), meaning that the geometry of the obscurer may be inherently related to the intrinsic properties of the central engine itself. An optimal sample of AGN to observe in X-rays for the development of future obscuration models would hence include (1) Compton-thick obscuration to ensure the reprocessed emission dominates the observed spectrum, (2) known inclination to remove a geometrical degree of freedom, and (3) precise measurements of black hole mass and multi-wavelength coverage to provide an independent estimate of Eddington-scaled accretion rate.

Disk megamasers satisfy all three criteria. The 22\,GHz radio emission line emitted by water vapour is produced by maser amplification\footnote{The name megamaser is assigned to sources with 22\,GHz luminosities in excess of 10\,$L_{\odot}$.}, and requires highly inclined sight lines to be detected (e.g., \citealt{Zaw20}). In agreement with the unified model of AGN \citep{Antonucci93,Urry95,Netzer15}, megamasers are thus often found in Compton-thick AGN in which highly inclined lines-of-sight lead to the highest column densities toward the central engine (e.g., \citealt{Greenhill08,Masini16,Panessa20}). Accurate Very Long Baseline Interferometry maser mapping additionally provides one of the most precise estimates of the central black hole mass currently known, enabling accurate measurements of Eddington-scaled accretion rate as long as robust bolometric luminosity estimates are available \citep{Brightman16}. We additionally note that under the unified model, the privileged inclination angles required for 22\,GHz water megamaser detection would be purely an orientation effect, with the circum-nuclear obscurer being somewhat similar in all AGN. Thus future astrophysical surveys of megamasers may be an extremely useful tool not just for studying the circum-nuclear properties of obscured AGN, but the entire AGN population.

In the following sections, we analyse a sample of confirmed Compton-thick AGN with detected water megamaser emission as a basis for developing the next generation of physically-motivated obscuration models for \textit{HEX-P}.

\subsection{\textit{NuSTAR}-confirmed Compton-thick megamasers}
As a basis for our simulations, we selected a sample of ten Compton-thick AGN in DoCTA with confirmed 22\,GHz megamaser in the literature \citep{Masini16,Panessa20}. We additionally included NGC\,2960 from \citet{Masini16} since the target was one of the lowest signal-to-noise ratio sources in their analysis, providing an interesting comparison for \textit{HEX-P}. The sample considered is shown in Table~\ref{tab:megamaser_info}. To ensure accurate spectral simulations, we then complemented the longest \textit{NuSTAR} exposure per source with the closest \textit{Chandra} observations in time available. Each \textit{Chandra} observation was reprocessed using the \texttt{chandra\_repro} command in \texttt{CIAO} \citep{Fruscione06}, before creating circular source\,$+$\,background and annular background-only regions centred on the target with each level~2 event file. Owing to the poorer angular resolution of \textit{NuSTAR} compared to \textit{Chandra}, we additionally created circular source\,$+$\,background regions for all clearly visible off-nuclear sources that were within the \textit{NuSTAR} extraction region. Spectra and response files were then produced using the \texttt{specextract} command. The breakdown of the sample in terms of source properties and X-ray observations are shown in Table~\ref{tab:megamaser_info}. The level of flux contaminating the \textit{NuSTAR} spectra from extracted off-nuclear sources was found to be negligible compared to all AGN apart from NGC\,5643. The source has a well-studied ultra luminous X-ray source that needed to be accounted for in our spectral analysis \citep{Annuar15}. We note that at a separation of $\sim$\,50'', the AGN and ultra luminous X-ray source would be easily resolved by the HETs and LET onboard \textit{HEX-P} (see Section~\ref{sec:m51}).

\begin{table}
\centering
\caption{Megamaser sample properties and X-ray observations used in this work.\label{tab:megamaser_info}}
\begin{tabular}{cccccccccc}
\toprule
     \makecell[ct]{Name \\ \,\, \\ (1)} &       \makecell[ct]{z \\ \,\, \\ (2)} &  \makecell[ct]{$D$ \\ Mpc \\ (3)} & \makecell[ct]{Type \\ \,\, \\ (4)} & \makecell[ct]{M$_{\rm BH}$ \\ 10$^{6}$\,M$_{\odot}$ \\ (5)} & \makecell[ct]{Ref. \\ \,\, \\ (6)} &  \makecell[ct]{Obs. ID \\ \textit{NuSTAR} \\ (7)} &  \makecell[ct]{T\,/\,ks \\ \textit{NuSTAR} \\ (8)} &  \makecell[ct]{Obs. ID \\ \textit{Chandra} \\ (9)} &  \makecell[ct]{T\,/\,ks \\ \textit{Chandra} \\ (10)} \\
\midrule
 IC\,2560 & 0.00976 &                       32.9 &      $i$ &                                    3.5\,$\pm$\,0.5 &    1 &                                50001039004 &                                        49.6 &                                        4908 &                                         55.7 \\
NGC\,1194 & 0.01360 &                       61.3 &      $z$ &                                       65\,$\pm$\,3 &    2 &                                60501011002 &                                        58.3 &                                       22881 &                                         21.1 \\
NGC\,1386 & 0.00290 &                       16.2 &      $i$ &                                1.2$^{+1.1}_{-0.6}$ &    3 &                                60201024002 &                                        26.4 &                                       13257 &                                         34.3 \\
NGC\,2273 & 0.00614 &                       30.3 &      $i$ &                                    7.5\,$\pm$\,0.4 &    2 &                                60001064002 &                                        23.2 &                                       19377 &                                         10.1 \\
NGC\,2960 & 0.01650 &                       92.5 &      $i$ &                                   11.6\,$\pm$\,0.5 &    2 &                                60001069002 &                                        20.7 &                                       22270 &                                         10.1 \\
NGC\,3079 & 0.00369 &                       17.6 &      $i$ &                                2.4$^{+2.4}_{-1.2}$ &    3 &                                60662004002 &                                        24.6 &                                       20947 &                                         45.1 \\
NGC\,3393 & 0.01250 &                       56.3 &      $z$ &                                       31\,$\pm$\,2 &    4 &                                60061205002 &                                        15.7 &                                       13968 &                                         28.5 \\
NGC\,5643 & 0.00400 &                       10.5 &      $i$ &                                                 -- &   -- &                                60061362006 &                                        48.1 &                                       17664 &                                         42.1 \\
NGC\,5728 & 0.00932 &                       27.9 &      $i$ &                                                 -- &   -- &                                60662002002 &                                        24.9 &                                       23254 &                                         20.1 \\
NGC\,7479 & 0.00792 &                       25.2 &      $i$ &                                                 -- &   -- &                                60061316002 &                                        23.6 &                                       10120 &                                         10.2 \\
\bottomrule
\end{tabular}
{\raggedright \textbf{Notes.} (1) -- Galaxy identifier. (2) -- Spectroscopic redshift. (3) -- Distance in Mpc. (4) Distance type; $i$\,=\,redshift-independent, $z$\,=\,redshift-dependent. (5) -- Black hole mass in 10$^{6}$\,M$_{\odot}$ dervied exclusively from megamaser emission. (6) -- Black hole mass reference; 1: \citet{Yamauchi12}, 2: \citet{Kuo11}, 3: \citet{McConnell13}, 4: \citet{Kondratko08}. (7) -- \textit{NuSTAR} observation ID. (8) -- \textit{NuSTAR}\,/\,FPMA exposure in ks. (9) \textit{Chandra} observation ID. (10) \textit{Chandra} exposure time. \par}
\end{table}

For X-ray spectral fitting, we use BXA v2.9 which connects the PyMultiNest nested sampling algorithm \citep{Feroz09,Buchner14} to the Python wrapper for the X-ray spectral fitting environment \texttt{Xspec} \citep{Arnaud96}. We chose to fit each AGN component with the \texttt{UXCLUMPY} model and its associated omni-directional Thomson scattered emission table. Since we primarily require a good enough description of the observed \textit{Chandra}\,$+$\,\textit{NuSTAR} spectra to perform \textit{HEX-P} simulations, we did not test additional physically-motivated models. \texttt{UXCLUMPY} does however include two unique geometrical parameters that describe the covering factor of material in the obscurer; \texttt{TORsigma}, the angular dispersion of the cloud distribution and \texttt{CTKcover}, the covering factor of an additional inner ring of Compton-thick clouds surrounding the AGN. Our spectral model for the AGN in \texttt{Xspec} parlance was as follows:

\begin{equation}
\label{eq:uxclumpy}
\begin{aligned}
\mathrm{AGN\,\,Model} = & \overbrace{ \mathtt{constant}}^{\text{Cross-calibration}} \times \overbrace { \mathtt{TBabs} }^{\text{Galactic absorption}} \times~ \Big[ \overbrace{\mathtt{apec}}^{\text{Thermal}} + \overbrace{ \mathtt{uxclumpy\_cutoff\_transmit}}^{\text{Transmitted emission}} \\ 
& + ~ \overbrace{ \mathtt{uxclumpy\_cutoff\_reflect}}^{\text{Reprocessed emission}} + \underbrace{ \mathtt{constant}}_{\text{Scattered fraction}} \times \underbrace{\mathtt{uxclumpy\_cutoff\_omni}}_{\text{Warm mirror emission}} \Big] 
\end{aligned}
\end{equation}

For the ultra luminous X-ray source in NGC\,5643 we additionally include the following model:
\begin{equation}
\label{eq:ulx}
\begin{aligned}
\mathrm{ULX\,\,Model} = & \overbrace{ \mathtt{constant}}^{\text{Contamination factor}} \times \overbrace { \mathtt{TBabs} }^{\text{Galactic absorption}} \times \Big[ \overbrace{\mathtt{zTBabs}}^{\text{Intrinsic absorption}} \times \overbrace{ \mathtt{zcutoffpl}}^{\text{Hard X-ray power-law}} + \overbrace{ \mathtt{diskpbb}}^{\text{Thermal emission}} \Big] 
\end{aligned}
\end{equation}

We assumed non-informative priors for line-of-sight column density, intrinsic power-law exponential cut-off, intrinsic power-law normalisation, the omni-directional scattered fraction, Compton-thick inner ring covering factor, cosine of the obscurer dispersion, thermal soft-excess temperature and its associated normalisation. For the intrinsic power-law photon index, we assumed a Gaussian prior with mean 1.8 and standard deviation 0.15 in agreement with numerous X-ray surveys (e.g., \citealt{Ricci17_bassV}). Finally for cross-calibrations between \textit{Chandra} and FPMB relative to FPMA we assumed log-Gaussian priors with mean 0 and standard deviation 0.03, consistent with the values of \citet{Madsen15}. In total, there were 11 free parameters in the \texttt{UXCLUMPY} AGN model.

We find all sources to have column densities in excess of 10$^{24}$\,cm$^{-2}$ to 90\% confidence. Interestingly, this includes NGC\,2960 for which we find a line-of-sight column density solution of $N_{\rm H}$\,$>$\,1.5\,$\times$\,10$^{24}$\,cm$^{-2}$ to 98.2\% confidence by fitting the combined \textit{Chandra} and \textit{NuSTAR} data. Previous works that analysed the \textit{NuSTAR} data alone consistently found $N_{\rm H}$\,$<$\,10$^{24}$\,cm$^{-2}$ using the \texttt{MYtorus} obscuration model \citep{Masini16}. Due to our Compton-thick solution, we include NGC\,2960 in DoCTA ex post facto. Figure~\ref{fig:docta_nustar} shows that the \textit{NuSTAR} spectra of NGC\,2960 are amongst the lowest signal-to-noise in the 3\,--\,78\,keV band of all other Compton-thick AGN studied with \textit{NuSTAR} to date. Its low signal-to-noise ratio spectrum thus makes NGC\,2960 a challenging and very conservative example to showcase the spectral constraints attainable with \textit{HEX-P}.

\subsection{Simulating the Faint Megamaser NGC\,2960}\label{sec:NGC2960}

We simulate a grid of \textit{NuSTAR} and \textit{HEX-P}/HET$\times$2\,$+$\,LET spectra to quantify the relative improvement in physical parameter inference attainable with \textit{HEX-P} observations of NGC\,2960. Whilst the Compton-thick solution for NGC\,2960 that we report here was acquired with the inclusion of \textit{Chandra} data, we restrict our simulations to purely \textit{NuSTAR} due to the relative scarcity of simultaneous observations publicly available (see Figure~\ref{fig:simultaneous}) and as an extrapolation for the discovery space in hard X-rays of new Compton-thick AGN previously missed. In total, we simulate a range of exposures between 10\,ks and 100\,ks with ten realisations per exposure. Each simulated spectrum had the same starting spectrum -- namely the maximum a posteriori spectral fit acquired with \textit{Chandra}\,$+$\,\textit{NuSTAR}. However, we additionally increased the obscurer dispersion to 60$^{\circ}$ and Compton-thick inner ring covering factor to 30\% to provide a more substantial covering factor of material to simulate.

\begin{figure*}[!ht]
\begin{center}
\includegraphics[width=0.99\textwidth]{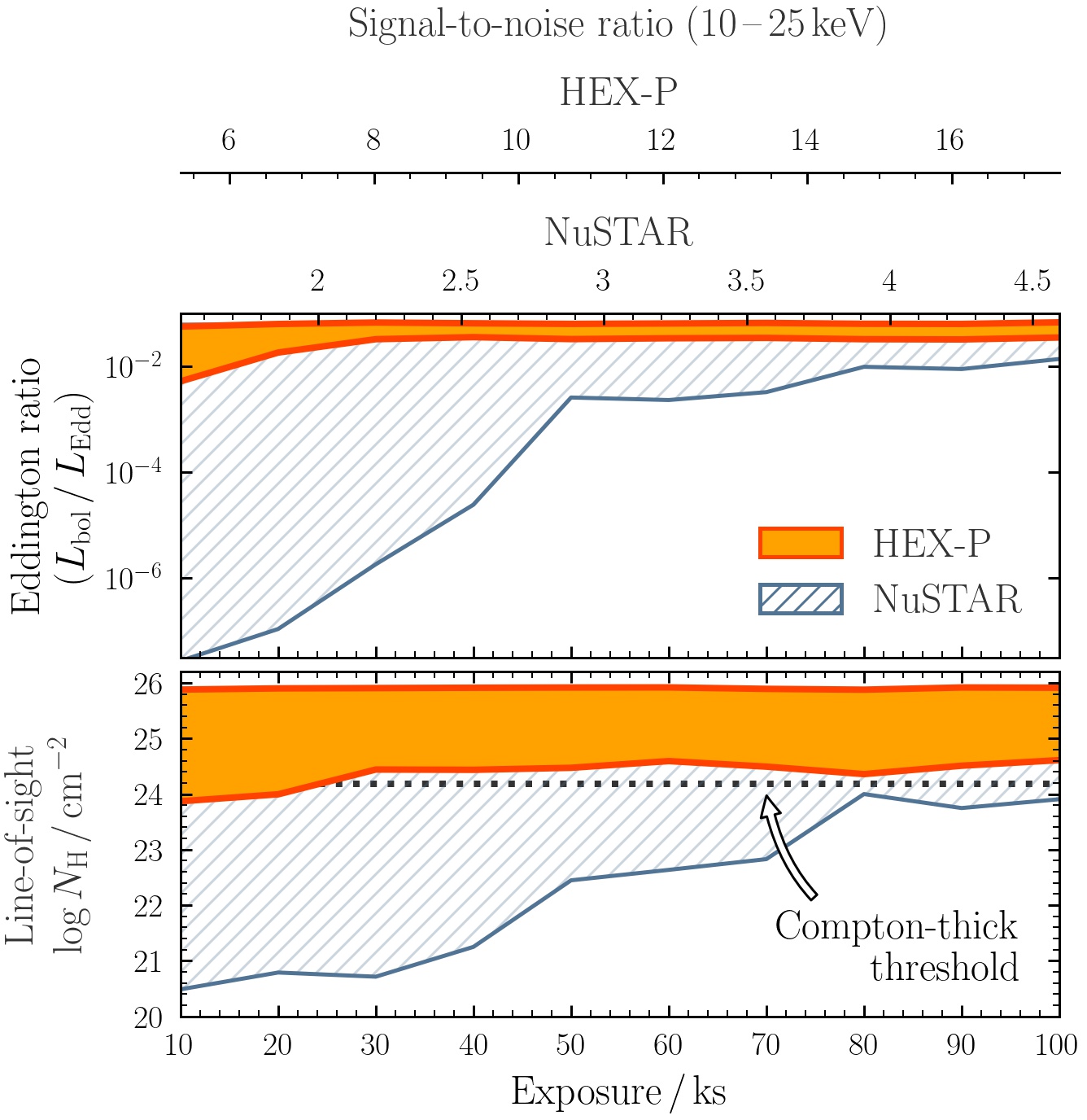}
\end{center}
\caption{Simulated parameter constraints for one of the faintest Compton-thick AGN candidates known; NGC\,2960. Bottom panel: Line-of-sight column density posterior 90\% quantile range as a function of exposure time for \textit{NuSTAR} (grey hatched region) and \textit{HEX-P} (orange filled region). \textit{HEX-P} is able to stringently confirm the target as Compton-thick (horizontal dotted line) with a modest $\sim$\,25\,ks exposure, which is not possible with \textit{NuSTAR} alone. Top panel: same as bottom, but for constrained Eddington ratio. \textit{HEX-P} is able to constrain the intrinsic luminosity, and hence accretion rate, to comparable precision of the black hole mass for exposures $\gtrsim$\,30\,ks. The top axes show the average measured signal-to-noise ratio in the 10\,--\,25\,keV band for \textit{NuSTAR} and \textit{HEX-P}, which is found to be $\sim$\,4$\times$ higher than \textit{NuSTAR} on average.}\label{fig:NGC2960}
\end{figure*}

The results of the NGC\,2960 simulation grid are shown in Figure~\ref{fig:NGC2960} in terms of 90\% posterior parameter constraints on Eddington ratio (top panel) and line-of-sight column density (bottom panel) as a function of exposure time. For Eddington ratio error propagation, we sample from the black hole mass and associated uncertainties in Table~\ref{tab:megamaser_info}. However, for the bolometric correction we use the Compton-thick bolometric correction from \citet{Brightman16} that focused on Compton-thick megamasers. To focus on the improvements attainable purely from X-ray spectral fitting as opposed to other systematics, we assume zero uncertainty on bolometric correction. We justify this choice by assuming the plethora of next-generation multi-wavelength observatories that are coming online over the next decade that will provide precise photometric data across the electromagnetic spectrum for accurate measurements of bolometric output.

Even for the low signal-to-noise ratio challenge that NGC\,2960 poses, \textit{HEX-P} is able to achieve Eddington ratio uncertainties comparable to the uncertainties on black hole mass for exposures $\gtrsim$\,30\,ks. In contrast, \textit{NuSTAR} does not reach a similar uncertainty regime for the entire range of exposures considered in the simulations. \textit{HEX-P} is additionally able to classify the target as Compton-thick to 90\% confidence for exposures $\gtrsim$\,25\,ks -- a feat that is not possible from purely \textit{NuSTAR} spectroscopy in our simulated range of exposures. We note that the remaining posterior uncertainty above the Compton-thick limit at all exposures with \textit{HEX-P} arises from the stability of the Compton-scattered component in \texttt{UXCLUMPY}. The overall shape of the reprocessed component does not change substantially for line-of-sight column densities $N_{\rm H}$\,$\gtrsim$\,5\,$\times$\,10$^{24}$\,cm$^{-2}$, such that constraining line-of-sight column densities to more than a lower limit is currently difficult. The top axes of Figure~\ref{fig:NGC2960} shows the measured signal-to-noise ratio in the 10\,--\,25\,keV band for \textit{NuSTAR} and \textit{HEX-P}. The reduced background and simultaneous coverage from three different instruments is able to boost the signal-to-noise ratio by factors of $\sim$\,4 relative to that of \textit{NuSTAR}. In the case of NGC\,2960, the boost in observed signal-to-noise ratio means that a 10\,ks exposure with \textit{HEX-P} would require $\gg$\,100\,ks of \textit{NuSTAR} exposure to reach an equivalent 10\,--\,25\,keV spectral quality. The 10\,--\,25\,keV energy band holds a plethora of information, not only regarding the line-of-sight column density but also the overall structure of the circum-nuclear obscurer (see e.g., \citealt{Buchner19,Buchner21}).

\subsection{Prospects for a New Era of Spectral Models}
\begin{figure*}[!ht]
\begin{center}
\includegraphics[width=0.99\textwidth]{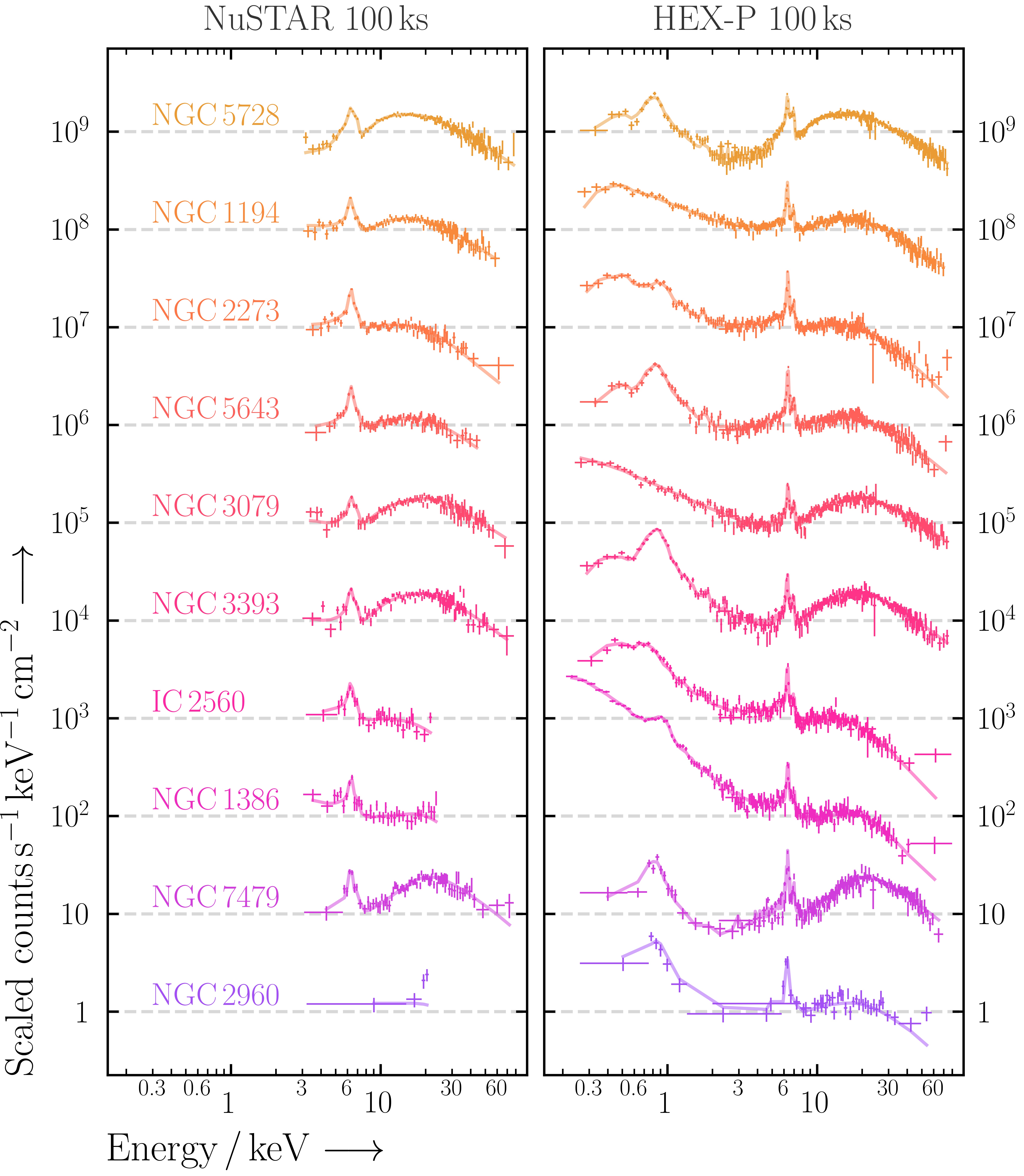}
\end{center}
\caption{Simulating the spectral prospects attainable with \textit{NuSTAR} (left) and \textit{HEX-P} (right) with 100\,ks observations of the Compton-thick megamasers considered in this work. Both panels show the folded simulated spectra from each telescope, normalised by their respective effective areas. Clearly \textit{HEX-P} offers a greatly expanded passband, with improved sensitivities up to the highest energies $\sim$\,80\,keV and much higher spectral resolution than \textit{NuSTAR}.}\label{fig:megamaser_spec}
\end{figure*}

From the signal-to-noise ratio improvements highlighted in Figure~\ref{fig:NGC2960} and the exceedingly low signal-to-noise ratio of the \textit{NuSTAR} data for NGC\,2960, it is clear that every known Compton-thick AGN will benefit greatly from modest \textit{HEX-P} observations. Next we simulate our sample of Compton-thick megamasers for 100\,ks with both \textit{NuSTAR} and \textit{HEX-P} to visually showcase the spectral improvements attainable, which will allow the development of next-generation spectral models. In Figure~\ref{fig:megamaser_spec} we present a like-for-like comparison between \textit{NuSTAR} (left column) and \textit{HEX-P} (right column), ordered from bottom to top by observed 2\,--\,10\,keV flux. A number of crucial improvements are visible.

\textit{Hard X-ray Sensitivity:} Whether it be due to the overall spectral slope, high-energy cut-off, structural properties of the obscurer or some combination of each, every AGN has a very distinctive spectral shape $>$\,10\,keV that is clearly detected with \textit{HEX-P}. In contrast, a large number of the simulated \textit{NuSTAR} spectra are not well-detected $\gtrsim$\,20\,keV. These improvements offer a number of useful avenues for model constraints and development. (1) Extending the range of detectable energies to $\gtrsim$\,50\,keV with \textit{HEX-P} will enable a dramatic reduction in confusion arising from different model components. For example, the high-energy exponential cut-off associated with the intrinsic coronal emission can be extremely difficult to disentangle from the turnover of the Compton hump (e.g., \citealt{Balokovic19,Kammoun23}). Furthermore the covering factor is very dependent on Compton hump shape, and reducing the measurement uncertainties $\gtrsim$\,20\,keV will greatly advance our ability to detect it.

\textit{Spectral Resolution:} From Figure~\ref{fig:megamaser_spec}, it is clear that the spectral resolution arising from the LET is superior to that of \textit{NuSTAR}. By combining high spectral resolution measurements of the Fe\,K region (including the Fe\,K$\alpha$ and Fe\,K$\beta$ lines) with sensitive measurements of the underlying reflection continuum up to energies $\gtrsim$\,20\,keV will enable detailed studies of fluorescence emission in heavily obscured AGN, including metallicities, dynamics and emission origins.

\textit{Simultaneous Soft X-ray Coverage:} Figure~\ref{fig:megamaser_spec} clearly shows the vast range in predicted spectral shapes from our broadband \textit{Chandra}\,$+$\,\textit{NuSTAR} fitting that are not accessible with \textit{NuSTAR}. Whilst quasi-simultaneous soft X-ray coverage is a common strategy for \textit{NuSTAR} observations, exposure times and corresponding signal-to-noise ratios are often not consistent across instruments (see Figure~\ref{fig:simultaneous}), leading to discrepant measurements of line-of-sight column density (e.g., \citealt{Marchesi18}). With \textit{HEX-P}, well-matched sensitivities with 100\% observing simultaneity will enable broadband X-ray spectral fitting effectively devoid of mismatched signal-to-noise ratio issues.

\section{Uncovering the dynamics of the thickest obscurers}\label{sec:variability}

Multi-epoch X-ray observations have proven to be a powerful tool for constraining the structure and dynamics of the obscurer. The ability to determine the amount of obscuring material along the line-of-sight as a function of time can be used to place constraints on the sizes of obscuring clumps as well as their distance from the supermassive black hole \citep[e.g.,][]{Elvis04,Risaliti09,Markowitz14}. Even for sources with sparsely-sampled light curves, changes in line-of-sight column density as a function of time can be used to gain insight into the general scales associated with the obscurer \citep[e.g.,][]{Laha20}.

Since the launch of \textit{NuSTAR}, broadband X-ray coverage has allowed the use of complex reflector models, which in turn can constrain global properties of the obscurer such as obscuring covering factor (parameterised as the fraction of sky covered when viewed from the perspective of the corona), inclination angle, and the global obscuring column density out of the line-of-sight; \citep[e.g.,][]{Balokovic18,Marchesi19a,Buchner19,Balokovic21,Zhao21}. Some works have suggested a correlation between the global properties of the obscurer and the characteristic change in line-of-sight column density as a function of timescale \citep[e.g.][]{Pizzetti22,TorresAlba23}. An advantage of multi-epoch fitting is that parameters unexpected to vary over the relatively short timescales associated with the observations (e.g., the obscurer covering factor, or global obscurer column density out of the line-of-sight) can be tied across observing epochs. Such an approach typically leads to more precise constraints on obscurer parameters since there are typically fewer regions of the parameter space compatible with multiple spectra than a single epoch-averaged spectrum (see e.g., \citealt{Balokovic18,Pizzetti22,Marchesi22,TorresAlba23}).

\textit{HEX-P}, with its capability to simultaneously observe the soft and hard X-ray bands to greatly improved sensitivity limits, will prove a key instrument for time domain studies (e.g., \citealt{Brightman23}), including the time-resolved characterisation AGN obscuration in X-rays. Non-simultaneous soft and hard band observations impose significant difficulty in disentangling intrinsic coronal luminosity variability from obscuration-related variability. For example, \citet{TorresAlba23} found that for $\sim$57\% of 12 nearby obscured AGN with confirmed long-term X-ray variability. However the two variability options could not be distinguished due to non-simultaneous soft and hard X-ray observations.

\begin{figure}
\begin{center}
\includegraphics[width=0.99\textwidth]{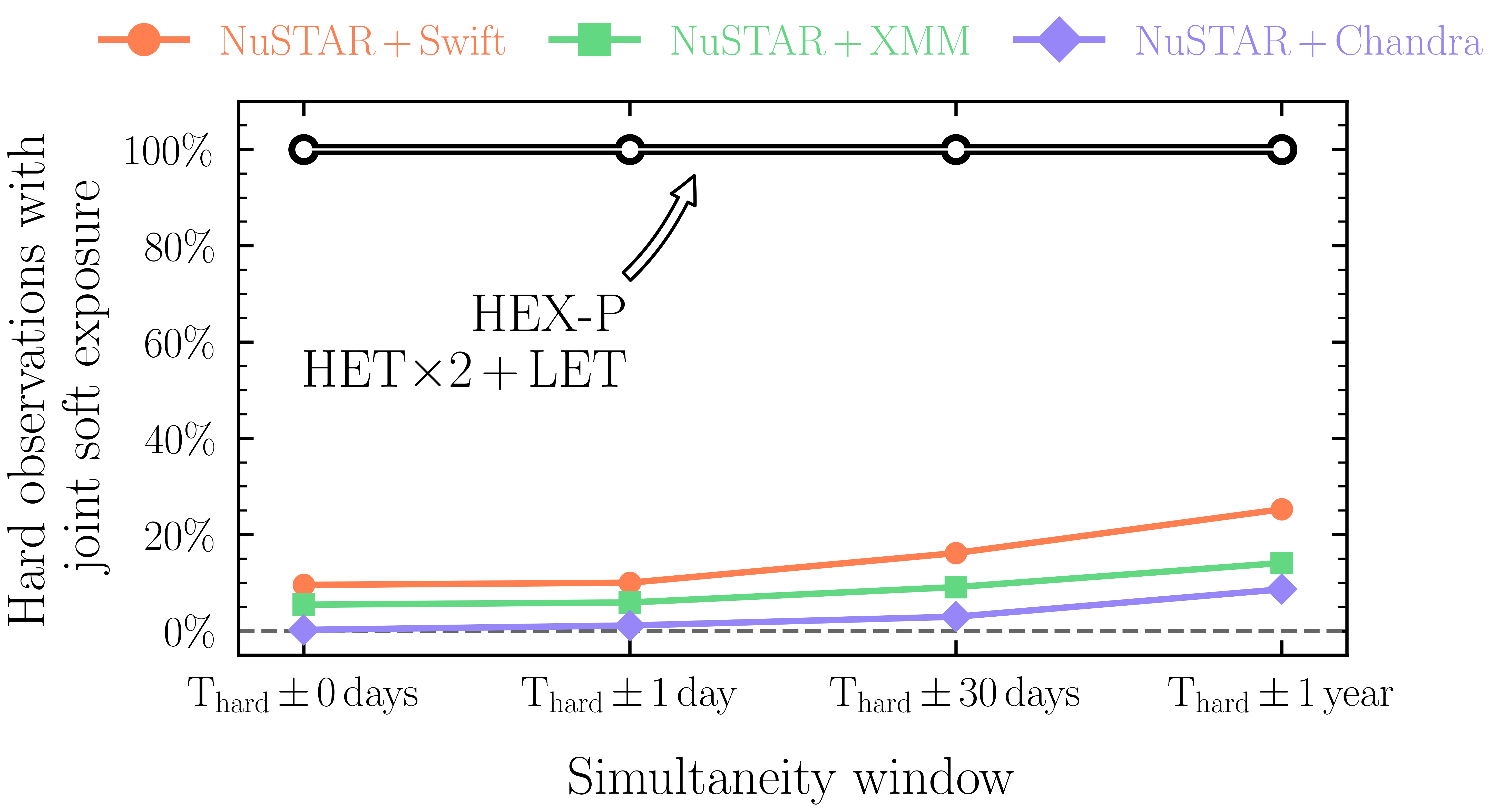}
\end{center}
\caption{The percentage of hard X-ray observations with overlapping soft X-ray exposure available for \textit{NuSTAR} vs. \textit{HEX-P}. To be conservative, in each time window (shown along the horizontal axis) we consider any hard X-ray observation with $>$\,20\% of its total exposure overlapping with a soft X-ray observation to be `joint'. Archival \textit{NuSTAR} observations were considered for any \textit{Swift}/BAT AGN from the 70-month compilation of \citet{Ricci17_bassV} obscured with line-of-sight column density $N_{\rm H}$\,$>$\,10$^{22}$\,cm$^{-2}$. Clearly, the complimentary data provided by the LET\,$+$\,HETs onboard \textit{HEX-P} will, for the very first time, provide complete simultaneous coverage across the hard and soft X-ray bands.}\label{fig:simultaneous}
\end{figure}

With the results of \citet{TorresAlba23} in mind, we sought to assess the current availability of multi-epoch broadband X-ray observations amongst the obscured AGN population. We queried the High Energy Astrophysics Science Archive Research Center\footnote{\href{https://heasarc.gsfc.nasa.gov}{https://heasarc.gsfc.nasa.gov}} for targeted \textit{NuSTAR} observations of any AGN in the 70-month BAT catalogue with line-of-sight column densities $N_{\rm H}$\,$>$\,10$^{22}$\,cm$^{-2}$ according to the X-ray spectral fitting catalogues of \citet{Ricci17_bassV}. We then searched for soft X-ray coverage from any of \textit{XMM-Newton}, \textit{Swift}/XRT, or \textit{Chandra} for each of the 372 obscured AGN with \textit{NuSTAR} observations available (439 \textit{NuSTAR} observations in total). Figure~\ref{fig:simultaneous} quantifies the frequency of joint soft\,$+$\,\textit{NuSTAR} observations in the obscured AGN sample as a function of ever-increasing hard X-ray observation simultaneity window. To be conservative, we consider any \textit{NuSTAR} observation with $>$\,20\% of its total exposure with joint soft exposure within each considered time window to be `joint'.  Even by liberally considering the non-simultaneous scenario of soft observations within one year of the \textit{NuSTAR} observations, only $\sim$\,25\% have $>$\,20\% of their exposure coincident with \textit{NuSTAR} exposure. We additionally note that having considerably different \textit{NuSTAR} and soft X-ray exposure times can give rise to dramatically different data quality across the spectral passband. Data quality mis-matches have been shown to influence measurements of obscuration parameters in heavily obscured AGN (e.g., \citealt{Marchesi18,Tanimoto22}. \textit{HEX-P} will clearly revolutionise the field, providing 100\% strictly simultaneous broadband coverage for all observations.

In the sections that follow, we quantify the advances \textit{HEX-P} will make towards multi-epoch observations of obscured AGN with detailed simulations.

\subsection{Case study I: Non-Compton-thick Obscured AGN}\label{subsec:ctn}

First we consider an obscured but not Compton-thick AGN (line-of-sight $N_{\rm H}$\,$\lesssim$\,10$^{24}$\,cm$^{-2}$) which presents both intrinsic luminosity and line-of-sight obscuration variability over three epochs of observation. To make our simulations conservative, we normalise the source flux to NGC\,835 -- the faintest AGN in the sample of \citet{TorresAlba23} with confirmed line-of-sight column density variability. We parameterise the obscurer with the \texttt{borus02} model \citep{Balokovic18} in decoupled mode, in which the obscurer properties were tuned to match the properties derived by \citet{Zhao21} for a sample of $\sim$\,100 obscured AGN. We simulate and fit with the same model to avoid any systematic uncertainties associated with the a-priori unknown obscurer that could be more dramatic for the less sensitive \textit{NuSTAR} data than \textit{HEX-P} (i.e. fewer model spectra can accommodate a given \textit{HEX-P} spectrum than \textit{NuSTAR} with reduced sensitivities and passband -- see e.g., \citealt{Saha22}). We consider three different line-of-sight column densities, namely $N_{\rm H}=1,3,6\times 10^{23}$\,cm$^{-2}$. To model additional flux variability, we include a cross-normalisation constant to the intrinsic AGN emission to simulate 50\%, 100\% and 200\% flux variability for each of the three observational epochs, respectively. We then pair line-of-sight column densities with different flux variability constant values per observational epoch such that the observed fluxes remain as similar as possible -- e.g., the lowest obscuration with the highest intrinsic luminosity, etc. We then run spectral simulations of 20\,ks exposure times with \textit{NuSTAR} and \textit{HEX-P} before re-fitting to quantitatively compare each mission's ability to disentangle the two separate forms of variability we consider.

\begin{figure*}
\begin{center}
\includegraphics[width=0.99\textwidth]{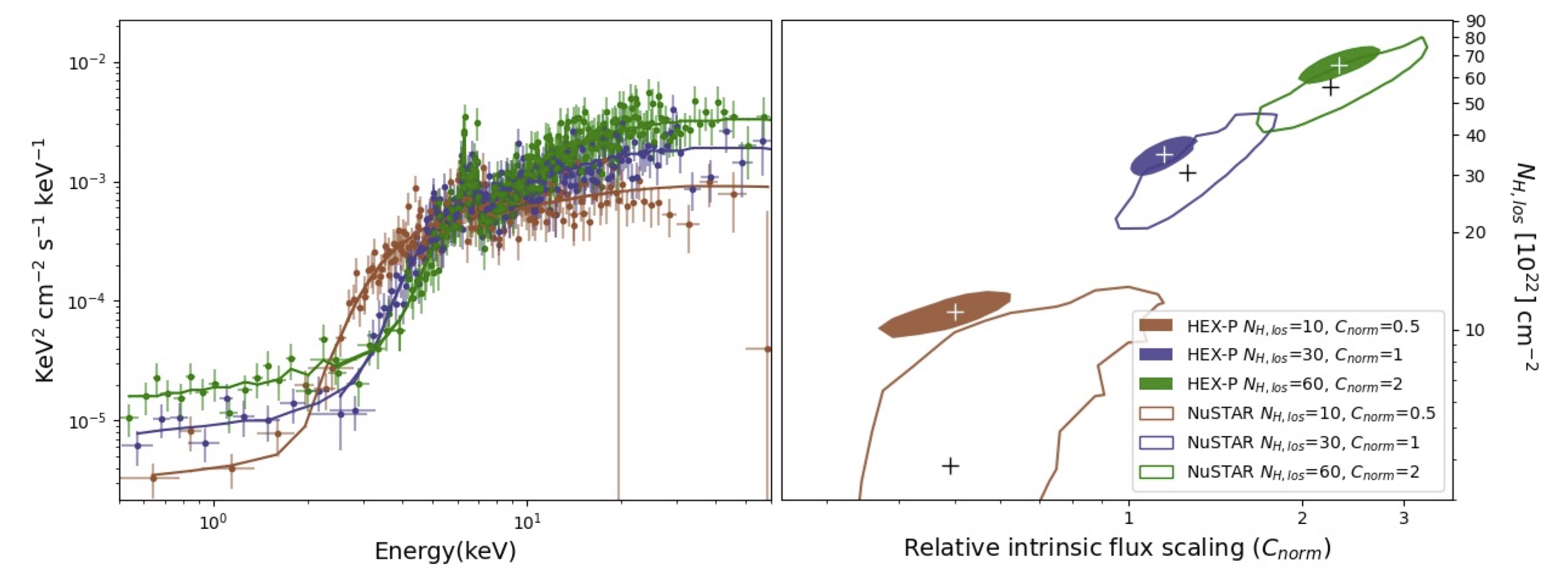}
\end{center}
\caption{\textit{Left panel:} \textit{HEX-P}/LET\,$+$\,2$\times$HET spectra simulated with 20\,ks exposures from the model described in Section~\ref{subsec:ctn}, based on NGC\,835 as a baseline. Colours correspond to the different combinations of line-of-sight column density and intrinsic flux scaling. \textit{Right panel:} Corresponding 99\% confidence contours between intrinsic flux scaling vs. line-of-sight column density derived from \textit{HEX-P} (filled) and \textit{NuSTAR} (empty). The crosses represent the best-fit values.}\label{fig:multiepoch_sims}
\end{figure*}

Figure\,\ref{fig:multiepoch_sims} shows the results of our simulations. The left panel presents the simulated \textit{HEX-P} spectra, in which the variations in column density from the photo-electric turnover and intrinsic flux from the overall normalisations are clearly visible. The right panel shows the resulting constraints in terms of intrinsic flux and line-of-sight column density for both \textit{HEX-P} and the equivalent simulated \textit{NuSTAR} spectra. Overall we find uncertainties $\sim$\,3\,--\,4 times larger with \textit{NuSTAR} than with \textit{HEX-P}. The \textit{NuSTAR} constraints can only place upper limits on the lowest line-of-sight column density scenario we considered, primarily due to its lack of simultaneous soft-band coverage encompassing the photo-electric turnover at soft energies. At higher column densities, the remaining two scenarios are consistent within 99\% confidence in terms of both line-of-sight column density constraints and intrinsic flux. In contrast, even for a short 20\,ks snapshot \textit{HEX-P} has no issue in disentangling intrinsic flux variability from line-of-sight column density variability for the full range of column densities considered. The broader passband is crucial, enabling proper characterisation of obscuration changes visible through the photoelectric turnover that are degenerate with intrinsic flux changes at harder energies.

\subsection{Case study II: Compton-thick AGN}

As discussed in Section~\ref{sec:introduction}, there have been comparatively few detailed multi-epoch broadband spectroscopic studies of Compton-thick AGN to date, and are dominated by the brightest sources known (e.g., \citealt{Puccetti14,Marinucci16}). An additional limitation with observing variability in Compton-thick AGN is that eclipsing events or intrinsic flux variations are more likely to manifest at $\gtrsim$10\,keV, where the effects of photoelectric absorption are reduced and Compton scattering dominates providing excess detectable flux (e.g., \citealt{Marinucci16,Zaino20}). The improved sensitivity at $>$10\,keV with \textit{HEX-P} will lead to new insights into broadband variability characteristics of Compton-thick AGN that have not been possible to date. To simulate the prospects attainable with \textit{HEX-P}, we use the faintest Compton-thick AGN with a published multi-epoch \textit{NuSTAR}-based campaign to date as a baseline. The nearby Seyfert~2 galaxy NGC\,1358 ($z$\,=\,0.0134) was subject to a multi-epoch monitoring campaign with \textit{NuSTAR} and \textit{XMM-Newton} between 2017\,--\,2022, and was found to be highly variable in line-of-sight column density by \citet{Marchesi22}.

For our \textit{HEX-P} simulations we use decoupled \texttt{borus02} and choose a range of line-of-sight column densities consistent with those measured for NGC\,1358 by \citet{Marchesi22} which varied above and below the Compton-thick limit in a changing-look scenario. The specific line-of-sight column densities we considered were $N_{\rm H}$\,=\,0.8,\,1.4,\,2$\times$10$^{24}$\,cm$^{-2}$. To make our simulations applicable to the wider AGN population we choose a value of $\Gamma$\,=\,1.8, consistent with the broader population of low-redshift Seyfert galaxies (e.g., \citealt{Ricci17_bassV}). The Thomson-scattered flux fraction is set to 2\%, which is again conservative considering the latest relations between scattered fraction and line-of-sight column density from \citet{Gupta21}. We additionally include a thermal \texttt{apec} component (with temperature $kT$\,=\,0.3\,keV) to model the remaining soft excess flux that the Thomson-scattered power-law does not account for. The global obscuring column density out of the line-of-sight is assumed to be $N_{\rm H}$\,=\,3.2$\times$10$^{23}$\,cm$^{-2}$ with a covering factor of 15\% within the \texttt{borus02} model.

The intrinsic 2\,--\,10\,keV luminosity of NGC\,1358 was found by \citet{Marchesi22} to be $L_{2-10\,{\rm keV}}$\,$\sim$\,6\,--\,9$\times$10$^{42}$\,erg\,s$^{-1}$ with corresponding observed 2--10\,keV flux $F_{2-10\,{\rm keV}}$\,=\,4\,--\,12$\times$10$^{-13}$\,erg\,s$^{-1}$\,cm$^{-2}$. To understand the new parameter space that \textit{HEX-P} will probe we simulate a fiducial source more than an order of magnitude fainter than NGC\,1358 with $L_{2-10\,{\rm keV}}$\,=\,5\,$\times$\,10$^{41}$\,erg\,s$^{-1}$, corresponding to observed fluxes of $F_{2-10\,{\rm keV}}$\,=\,11,\,6,\,4$\times$10$^{-14}$\,erg\,s$^{-1}$\,cm$^{-2}$ for each line-of-sight column density considered. We note that equivalent fluxes (and hence spectroscopic constraints) would be constrained for a target at ten times the distance of NGC\,1358 (i.e. $D$\,$\sim$\,500\,--\,600\,Mpc) with Seyfert-like luminosities of $L_{2-10\,{\rm keV}}$\,$\sim$\,5$\times$10$^{43}$\,erg\,s$^{-1}$. In comparison, it is currently very difficult to perform detailed X-ray spectroscopic modelling of Seyfert-luminosity Compton-thick AGN with \textit{NuSTAR} at comparable distances (e.g., \citealt{Giman23}).

We simulate one 30\,ks \textit{HEX-P} observation for each line-of-sight column density state mentioned above. The line-of-sight column density is recovered to high accuracy with relative uncertainties $\leq$\,20\%. Owing to the strong advantage of linking parameters that are not expected to vary between epochs, the global column density is precisely recovered with uncertainties $<$\,0.3\,dex, and the obscuration covering factor is correctly found to be $<$\,20\% to high confidence. Our simulations thus clearly show that a \textit{HEX-P} monitoring campaign would allow us to characterise the properties of the clumpy obscuring medium in heavily obscured AGN with unprecedented quality to far fainter flux levels than are attainable with current X-ray observatories. Such capabilities are critical for constraining the dynamics of the obscurer in the wider Compton-thick AGN population that is currently impossible.

\section{The circum-nuclear obscurer of AGN at low accretion power}\label{sec:llagn}

Our knowledge of the obscurer surrounding low accretion power AGN is currently severely incomplete. A root cause is the considerable challenge to select and classify true low accretion power AGN, especially at high line-of-sight column densities. \textit{NuSTAR} has provided an unprecedented view into the hard X-ray properties of the circum-nuclear environment of low-luminosity AGN for the first time \citep{Ursini15,Annuar17,Young18,Younes19,Annuar20,Diaz20,Balokovic21,Diaz23}. \textit{NuSTAR} has also led to the discovery and classification of a few low-luminosity Compton-thick AGN \citep[e.g.,][]{Annuar17,Brightman18,DaSilva21}, providing exciting evidence suggesting AGN can sustain a significant obscuration structure at low luminosities. However current observational studies of Compton-thick low-luminosity AGN are often hindered by the requirement for deep integration times to obtain sufficient counts for detailed X-ray spectral modelling \citep[e.g.,][]{Annuar20}.

Given the current scarcity of bona-fide low-luminosity Compton-thick AGN confirmed by broadband X-ray studies including hard X-ray observations with \textit{NuSTAR}, we sought to investigate the prospects attainable with \textit{HEX-P} for identifying, classifying and studying this elusive population in the nearby Universe. We tuned our simulations to the properties of four bona-fide low-luminosity AGN with heavy obscuration in the literature; M\,51a \citep{Brightman18}, NGC\,660 \citep{Annuar20}, NGC\,1448 \citep{Annuar17} and NGC\,2442 \citep{DaSilva21}. We note all sources have column density classifications based in part with \textit{NuSTAR}. Also, all are confirmed Compton-thick apart from NGC\,660 which has both Compton-thick and sub-Compton-thick (but still heavily obscured) solutions in \citet{Annuar20}. With this in mind, for the remainder of this Section, we refer to this sample of four sources as the low-luminosity Compton-thick AGN sample.

\subsection{Selecting and classifying low-luminosity Compton-thick AGN}\label{sec:m51}

A major challenge for studies of low-luminosity AGN is confidently associating detected sources with accretion onto a supermassive black hole, as opposed to off-nuclear accretion onto lower-mass compact objects such as ultra luminous X-ray Sources, other individual X-ray binaries or jetted emission (see \citealt{Bachetti23,Lehmer23,Connors23,Marcotulli23} for the \textit{HEX-P} perspective on ultra luminous X-ray sources, other extragalactic accreting compact objects and resolved AGN jets). A major advantage arises from spectral coverage at $\gtrsim$\,10\,keV in which the spectral curvature from accreting supermassive black holes can be dramatically different from that of lower-mass accreting compact objects. However, an additional difficulty with classification is being able to resolve emission components into individual sources to confidently ascertain the spectral parameters of the central AGN. Given the dramatic improvement in X-ray angular resolution of \textit{HEX-P} compared to both \textit{XMM-Newton} and \textit{NuSTAR}, we sought to test \textit{HEX-P}'s ability to resolve contaminating sources in the host galaxy from low-luminosity AGN.

We base our simulations on the central region of M\,51 which is known to host a Compton-thick low-luminosity AGN as well as a number of bright off-nuclear X-ray sources \citep{Brightman18}. Our main consideration here was that of `ULX-3', which is situated $\sim$\,30'' from the central AGN and is the closest spatial contamination of all four sources in the Compton-thick low-luminosity AGN sample. Though spatially resolved with \textit{Chandra}, the close separation led to strong contamination with \textit{NuSTAR} which must be accounted for by simultaneously fitting both datasets to infer spectral parameters of the central AGN and ULX-3.

To demonstrate \textit{HEX-P}’s unique capability to spatially-resolve closely separated sources in both the soft and hard X-ray energy bands, we compare current constraints from \textit{Chandra} and \textit{NuSTAR} to that of \textit{HEX-P} with simulations. We simulate \textit{HEX-P} soft ($<10$\,keV) and hard ($>10$\,keV) X-ray imaging of M\,51a and the nearby ultra luminous X-ray source \citep[ULX-3,][]{Brightman18} using the Simulated Observations of X-ray Sources \citep[\textsc{soxs}][]{Zuhone23} and Simulation of X-ray Telescopes \citep[SIXTE][]{Dauser19} software suites. We rely upon \textsc{soxs} for creating SIMulated inPUT (SIMPUT) files, which incorporates our spectral and spatial models for individual targets. For simplicity, we choose all emission to have energies $>$\,2\,keV, in order to exclude any softer extended X-ray emission from the simulation and to enable a more direct study of the resolving power of \textit{HEX-P}.

We used point source models to simulate the spatial morphology of M\,51a and ULX-3. For spectral modelling, we fit the \textit{Chandra} spectra for M\,51a and ULX-3 simultaneously with the unresolved \textit{NuSTAR} spectrum of both sources using the same \texttt{UXCLUMPY}-based model described in Section~\ref{sec:megamasers} combined with the ultra luminous X-ray source model used to fit NGC\,5643. We allow the AGN model to vary for the M\,51a \textit{Chandra} spectrum and combined \textit{NuSTAR} spectrum and vice-versa for ULX-3. The individual AGN and ULX-3 maximum a-posteriori models found with BXA were then used in conjunction with the spatial model to generate a SIMPUT file. Next, we used \textsc{sixte} to produce the telescope event files, energy-filtered imaging, and spectroscopic data products. In order to compare \textit{HEX-P}'s capabilities with current facilities, we also simulated event files for \textit{Chandra} ACIS-S (0.1--8\,keV) and \textit{NuSTAR} FPMA\,$+$\,FPMB (3\,--\,78\,keV).

\begin{figure}
\begin{center}
\includegraphics[width=0.9\textwidth]{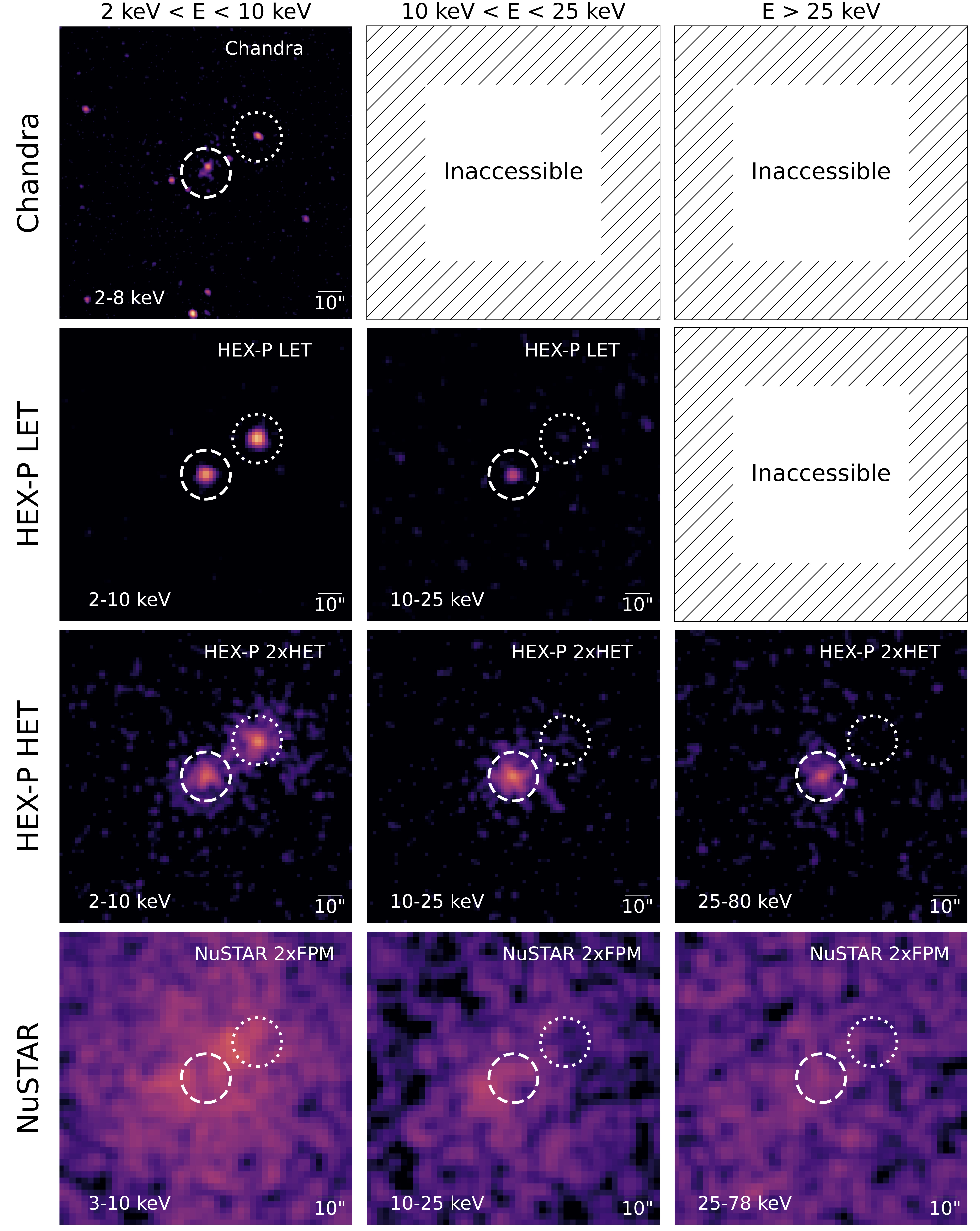}
\end{center}
\caption{Comparing the energy coverage and spatial resolution of \textit{Chandra}, \textit{NuSTAR}, and \textit{HEX-P} for the case of M\,51a and the nearby ULX-3. Each row represents \textit{Chandra}, \textit{HEX-P}\,/\,LET, \textit{HEX-P}\,/\,HET and \textit{NuSTAR}), from top to bottom, respectively. Each column represents an effective increase in observing energy passband, with left showing 2\,keV\,$<$\,E\,$<$\,10\,keV, middle 10\,keV\,$<$\,E\,$<$\,25\,keV and right $E$\,$>25$\,keV, respectively. For instruments without access to a specific energy passband, the panel is marked as `inaccessible'. Clearly, \textit{Chandra} can spatially resolve M\,51a and ULX-3, but does not have access to harder X-ray energies ($>10$\,keV). In contrast, \textit{NuSTAR} has broadband coverage ($3-78$\,keV), but lacks the angular resolution to resolve the two sources. \textit{HEX-P} will provide spatially-resolved measurements of M\,51a and ULX-3 in both the 2\,--\,25\,keV passband with LET as well as the 2\,--\,80\,keV passband with HET.
}
\label{fig:M51_sixte}
\end{figure}

We show the simulated \textit{HEX-P}\,/\,LET and HET imaging in Figure~\ref{fig:M51_sixte}, juxtaposed with the simulated \textit{Chandra} and \textit{NuSTAR} imaging of the same sources. Despite resolving the AGN and ultra luminous X-ray source well, \textit{Chandra} is only able to study the targets below 8\,keV. As discussed throughout this paper, this limitation considerably restricts the ability for physical inference of the obscurer. It is also clear from the figure that while \textit{NuSTAR} has access to soft ($<$\,10\,keV) and hard ($>$\,10\,keV) X-ray energies, it cannot spatially-resolve the X-ray sources well on these spatial scales. \textit{HEX-P} provides a unique combination of high spatial resolution {\em broadband} observations, enabling a new era of spatially-resolved closely-separated nuclear sources. These simulations showcase the power and complementary nature of \textit{HEX-P}\,/\,LET and HET imaging to provide enhanced searches of heavily obscured low-luminosity AGN in our nearby galactic neighbourhood.

\subsection{The nature of the obscurer at low accretion powers}\label{sec:llctagn_fits}
The covering factor of the obscurer at low accretion powers is currently very uncertain, due in part to the difficulty associated with selecting heavily obscured low-luminosity AGN relative to their less obscured counterparts. Here we investigate the ability of \textit{HEX-P} to study the covering factor of Compton-thick AGN via detailed spectral modelling of individual sources.

To provide a firm basis for our simulations, we begin from all archival \textit{Chandra} and \textit{NuSTAR} data available for NGC\,660, NGC\,2442 and NGC\,1448. For NGC\,2442, there were two archival \textit{NuSTAR} and two archival \textit{Chandra} observations. We extracted spectra following the same criteria as throughout this paper, before manually checking for significant variability between observations. Due to a lack of strong variability, we then co-added all \textit{NusTAR}\,/\,FPMA, \textit{NuSTAR}\,/\,FPMB and \textit{Chandra} spectra individually to provide $\sim$49.5\,ks of total \textit{Chandra} exposure and $\sim$112\,ks of \textit{NuSTAR} exposure. For NGC\,1448 and NGC\,660, we extract the \textit{Chandra} and \textit{NuSTAR} data following the methods from \citet{Annuar17} and \citet{Annuar20}, respectively. We additionally include the same absorbed power-law model components from \citet{Annuar17} for two off-nuclear contaminants inevitably included in the NGC\,1448 \textit{NuSTAR} extraction region that were resolved by \textit{Chandra}. The two contaminant model components were kept frozen to their best-fit values from \citet{Annuar17} for all fitting that involved \textit{NuSTAR}. However given the results of Section~\ref{sec:m51}, \textit{HEX-P} would easily resolve these contaminants from the AGN such that any spectral simulations and corresponding fitting of \textit{HEX-P} spectra only considered the AGN component in NGC\,1448.

We then performed spectral modelling using the \texttt{UXCLUMPY} model \citep{Buchner19} which included emission from a clumpy obscurer, as well as soft X-ray excess emission from an omni-present warm mirror as well as a thermal component with \texttt{apec}. We experimented with a number of spectral fitting setups, but due to the low signal-to-noise ratio of the observed spectra, a number of unphysical parameter constraints had to be avoided. One example is the tendency for the fit to prefer a low covering-factor obscurer (i.e. the \texttt{TORsigma} parameter in \texttt{UXCLUMPY} tended towards its minimum) in exchange for a hard X-ray photon index and overall unobscured spectrum. Similar degeneracies are well documented in the literature (e.g., \citealt{Brightman15}) and as such we opted to freeze \texttt{TORsigma} to a fiducial value of 60$^{\circ}$ for the spectral fitting of archival data.

Interestingly, the resulting parameter posteriors indicated a diversity in \texttt{CTKcover} between sources, suggesting a diversity in Compton hump shapes across the three sources fit here\footnote{The clouds that form the inner ring described by \texttt{CTKcover} in \texttt{UXCLUMPY} are assigned log\,$N_{\rm H}$ drawn from a log-normal distribution with $N_{\rm H}$\,=\,10$^{25.5\,\pm\,0.5}$\,cm$^{-2}$.}. From each acquired modal posterior model spectrum, we simulated corresponding \textit{HEX-P}\,/\,LET and \textit{HEX-P}\,/\,HET$\times$2 spectra with 100\,ks exposures before re-running the same spectral fits with \texttt{TORsigma} additionally left free to vary. Given the distribution of column densities assigned to clouds in the \texttt{UXCLUMPY} model \citep{Buchner19}, it is difficult to parametrically calculate a covering factor for a particular cloud configuration. We instead used pre-tabulated calculations of covering factor for all material with $N_{\rm H}$\,$>$\,10$^{22}$\,cm$^{-2}$ and $N_{\rm H}$\,$>$\,10$^{24}$\,cm$^{-2}$ for a two-dimensional grid of \texttt{TORsigma} and \texttt{CTKcover} values. We used grid interpolation to propagate all posterior uncertainties from \texttt{TORsigma} and \texttt{CTKcover} into posteriors for both of these column density regimes.

To investigate \textit{HEX-P}'s ability to probe precise relations between accretion power and covering factor in Compton-thick low-luminosity AGN, we required an estimate of Eddington ratio for each target. The black hole masses we used were log\,M$_{\rm BH}$\,/\,M$_{\odot}$\,=\,7.35\,$\pm$\,0.50 \citep{Annuar20}\footnote{We conservatively assumed 0.5\,dex uncertainty for NGC\,660 when propagating uncertainties into the Eddington ratio posterior.}, 7.28\,$\pm$\,0.33 \citep{Davis14} and 6.0\,$^{+0.1}_{-0.5}$ \citep{Annuar17} for NGC\,660, NGC\,2442 and NGC\,1448, respectively. We then used the bolometric correction relation determined by \citet{Nemmen14} for low-luminosity AGN to estimate the posterior distribution on the bolometric correction.

\begin{figure}
\begin{center}
\includegraphics[width=0.99\textwidth]{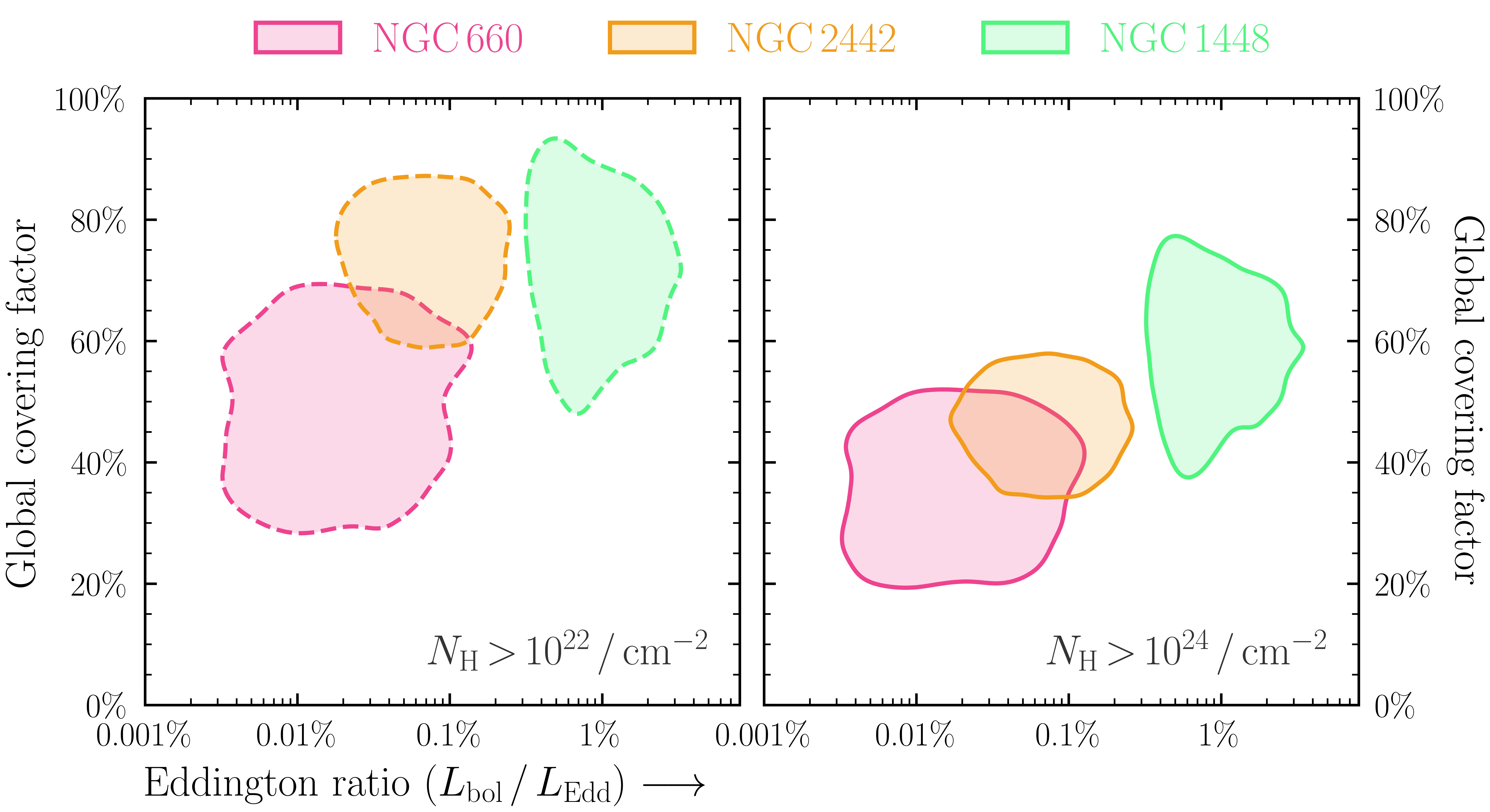}
\end{center}
\caption{\textit{HEX-P} 100\,ks simulation contours for \textit{NuSTAR}-confirmed low-luminosity Compton-thick AGN. The posterior contours show Eddington ratio vs. covering factor for material in the \texttt{UXCLUMPY} geometry with \textit{Left:} $N_{\rm H}$\,$>$\,10$^{22}$\,cm$^{-2}$ and \textit{Right:} $N_{\rm H}$\,$>$\,10$^{24}$\,cm$^{-2}$. Note due to the overall similar correlations between \texttt{TORsigma}, \texttt{CTKcover} and the global covering factors of material with $N_{\rm H}$\,$>$\,10$^{22}$\,cm$^{-2}$ and $N_{\rm H}$\,$>$\,10$^{24}$\,cm$^{-2}$ (see Section~\ref{sec:uxccov}), the propagated uncertainties in global covering factor have similar shapes. Performing similar spectral fitting to archival \textit{NuSTAR}\,$+$\,\textit{Chandra} data was insufficient to break degeneracies associated with the covering factor, photon index and line-of-sight column density -- see Section~\ref{sec:llctagn_fits} for details.}\label{fig:llctagn_uxc_cov}
\end{figure}

The corresponding two-dimensional contours shown in Figure~\ref{fig:llctagn_uxc_cov} give the posteriors on Eddington ratio vs. covering factor for material with $N_{\rm H}$\,$>$\,10$^{22}$\,cm$^{-2}$ and $N_{\rm H}$\,$>$\,10$^{24}$\,cm$^{-2}$ in the left and right panels, respectively. \textit{HEX-P} spectroscopy is thus able to constrain the covering factor to within $\lesssim$\,$\pm$\,20\% in the lowest-luminosity Compton-thick AGN currently known. Such observations are critical to understand the presence and corresponding importance of circum-nuclear obscuration in the low luminosity regime.

\section{An intermediate mass black hole confirmed with megamaser emission}\label{sec:dwarf}
To explore the parameter space attainable with \textit{HEX-P} in the search for obscured intermediate mass black holes, we start from the work of \citet{Chen17} who selected a sample of ten low-mass AGN from the 40-month \textit{NuSTAR} serendipitous survey. Of this sample, IC\,750 has a confirmed 22\,GHz water megamaser signature in the literature \citep{Zaw20} placing a tight upper bound on the central black hole mass in the intermediate mass range of $M_{\rm BH}$\,$<$\,1.4$\times$10$^{5}$\,M$_{\odot}$. As discussed in Section~\ref{sec:megamasers} megamasers are ideal targets for \textit{HEX-P} to aid the development of future circum-nuclear obscuration models, and hence IC\,750 provides an extension to the black hole mass range of known megamaser AGN. Furthermore, \citet{Chen17} performs phenomenological X-ray spectral fitting to an $\sim$30\,ks \textit{Chandra} spectrum of IC\,750, finding the source to be heavily obscured with line-of-sight $N_{\rm H}$\,$\sim$\,1.2$\times$10$^{23}$\,cm$^{-2}$. As discussed by \citet{Chen17} and throughout this work, X-ray spectral fitting to a predominantly soft-band spectrum without sensitive broadband coverage can give rise to wide systematic uncertainties on the properties of the obscurer.

We downloaded and reprocessed all archival \textit{Chandra} datasets of IC\,750 using the \texttt{chandra\_repro} command available in \texttt{CIAO} \citep{Fruscione06}. The level~2 event files were then used to create circular source\,$+$\,background and annular background-only regions centered on the target. We made sure to make the source\,$+$\,background regions small enough to remove as much contamination as possible from the $E$\,$<$\,2\,keV diffuse extended emission reported by \citet{Chen17}. The background regions were created to be as large as possible whilst avoiding off-nuclear sources and chip gaps. Source\,$+$\,background, background and response spectral files were then produced using the \texttt{specextract} command.

\begin{figure*}
\begin{center}
\includegraphics[width=\textwidth]{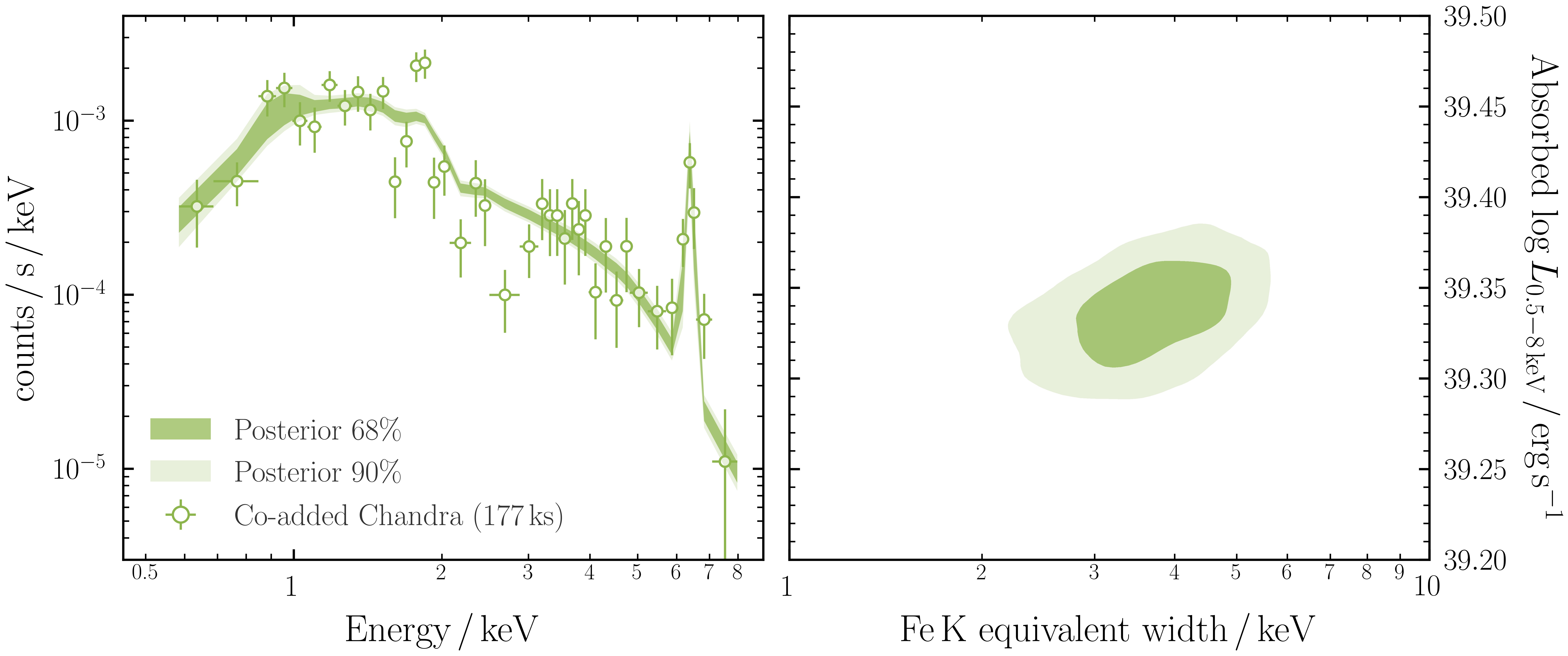}
\end{center}
\caption{\textit{(Left)} Co-added \textit{Chandra} spectrum of IC\,750, fit with a phenomenological model. \textit{(Right)} Posterior 2D contour between observed 0.5\,--\,8\,keV luminosity and Fe\,K equivalent width derived from the spectral fit shown in the left panel.}\label{fig:IC750_chandra}
\end{figure*}

After a variety of different tests of significant spectral variability, we chose to co-add the six individual X-ray spectra using the \texttt{ftool} command \texttt{addspec}. The resulting co-added spectrum contained a net exposure of 177\,ks with 441 source counts detected in the 0.5\,--\,8\,keV band (see the left panel of Figure~\ref{fig:IC750_chandra}). As an initial assessment of the spectrum, we fit a similar model to \citet{Chen17}; namely an absorbed power-law with an additional thermal component provided by \texttt{apec} and a narrow Gaussian line to represent the Fe\,K complex. The corresponding folded spectrum with shaded posterior and two-dimensional posterior between observed 0.5\,--\,8\,keV luminosity and Fe\,K equivalent width are shown in the left and right panels of Figure~\ref{fig:IC750_chandra}, respectively. Despite the observed (i.e. absorption-uncorrected) 0.5\,--\,8\,keV luminosity being consistent with ultra luminous X-ray sources (e.g., \citealt{Earnshaw19}), the observed Fe\,K equivalent width is enormous with values $>$2\,keV at $>$99\% confidence. Such large equivalent widths are rare but not unheard of in the Compton-thick AGN population \citep{Levenson02,Boorman18}, and strongly indicative of a deeply buried accreting massive black hole.

\subsubsection{Inferring the obscuration geometry with \textit{HEX-P}}
To provide a basis for \textit{HEX-P} simulations, we next turned to physically-motivated spectral models of obscuration. To test the distinguishing power of broadband X-ray spectroscopy, we fit the existing co-added \textit{Chandra} spectrum with two distinctive physical X-ray obscuration models: \texttt{BNsphere} \citep{Brightman11a}, representing a spherical distribution of matter and \texttt{UXCLUMPY} \citep{Buchner19}, representing a clumpy distribution of matter. Despite using the \texttt{BNsphere} model, we include a soft X-ray excess component in both model fits that is often attributed to a small fraction of intrinsic X-ray flux escaping through less-obscured sight-lines. Though unlikely in a spherical model (i.e. there are no less-obscured sight-lines when surrounded by fully-covering obscuration), by allowing a small amount of flux to escape we approximately reproduce a leaky-sphere geometry akin to that described in \citet{Greenwell22}. Some contribution to the soft X-ray excess in obscured AGN may still come from larger scales than the circum-nuclear obscurer (see discussion in \citealt{Gupta21}), such that our approximation is justified. Both physically-motivated obscurer model setups are assumed to be coupled (i.e. the line-of-sight obscuration is assumed to be the same as the global obscuration level out of the line-of-sight).

\begin{figure*}
\begin{center}
\includegraphics[width=\textwidth]{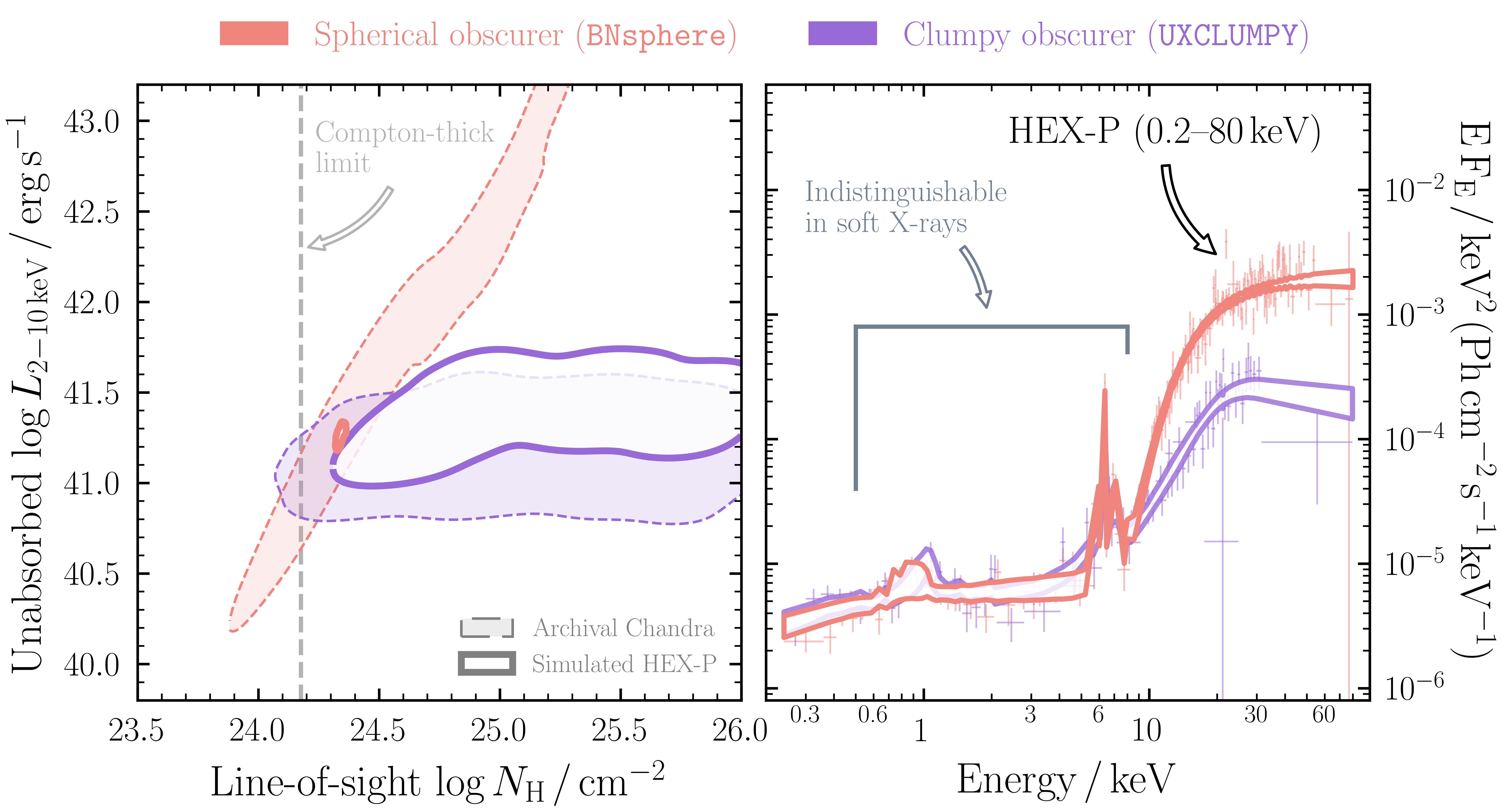}
\end{center}
\caption{\textit{(Left)} Posterior 2D contours between intrinsic 2\,--\,10\,keV luminosity and line-of-sight column density derived with the \texttt{BNsphere} and \texttt{UXCLUMPY} physically-motivated obscuration models. The posterior contours derived from fitting the archival \textit{Chandra} data are shown with light shading bounded by a dashed line. The constraints from fitting simulated \textit{HEX-P} spectra are shown with transparent white contours bounded by thick lines. \textit{HEX-P} can easily constrain the target to be Compton-thick with high confidence as well as precisely measure the intrinsic luminosity of the buried X-ray corona. \textit{(Right)} Simulated \textit{HEX-P} spectra with posterior range derived from spectral fitting. The soft X-ray band covered by archival \textit{Chandra} spectroscopy is insufficient to distinguish spectral models. In contrast, the broad passband of \textit{HEX-P} encompasses stark spectral differences in the Compton hump, enabling geometrical constraints on the obscurer.}\label{fig:IC750_hexp}
\end{figure*}

The line-of-sight column density vs. intrinsic luminosity posterior distributions from the spectral fits to the existing \textit{Chandra} data are shown in the left panel of Figure~\ref{fig:IC750_hexp} with light shading, encompassed by a dashed line. \texttt{BNsphere} and \texttt{UXCLUMPY} find IC\,750 to have line-of-sight column density in excess of the Compton-thick limit to $\sim$85\% and $\sim$99\% probability\footnote{We note that the \texttt{BNsphere} posterior gives a $>$99\% probability for a line-of-sight column density in excess of 10$^{24}$\,cm$^{-2}$.}, making IC\,750 a firm Compton-thick accreting intermediate mass black hole candidate.

However as shown in the right panel of Figure~\ref{fig:IC750_hexp}, the existing soft band \textit{Chandra} spectral coverage is incapable of distinguishing the two obscuration models. This means that the true range of the intrinsic luminosity posterior we have derived from the spectral fitting spans $\sim$\,4\,dex, and the properties of the obscurer cannot be constrained (see e.g., \citealt{Lamassa17,Lamassa19} for more discussion on this). We next simulate \textit{HEX-P}/LET and HET$\times$2 spectra from the modal posterior parameter values derived by fitting the \textit{Chandra} data, using exactly the same exposure as the co-added \textit{Chandra} spectrum for both models and re-fit with BXA. The resulting posterior contours and simulated spectra unfolded with the best-fitting model posteriors are shown with transparent white contours encompassed by thick borders in the left and right panels of Figure~\ref{fig:IC750_hexp}, respectively. Clearly the broadband coverage provided by \textit{HEX-P} is able to dramatically decrease the luminosity degeneracy of the \texttt{BNsphere} model fit, as well as definitively constrain the line-of-sight column density as Compton-thick. The dramatic difference between the two model fits in the right panel at energies $>$\,10\,keV is crucial, since \textit{HEX-P} can not only precisely measure intrinsic luminosities and line-of-sight column densities but also distinguish between obscuration geometries.

\textit{HEX-P} is hence well poised to find, classify and study the obscured accretion onto intermediate mass black holes. Gaining access to spectroscopic information of sufficient sensitivity at $>$\,10\,keV is additionally crucial to decipher genuine intermediate mass black hole accretion from other lower mass compact object accretion in the host galaxy.

\section{The Discovery Space for Compton-thick AGN}\label{sec:sensitivity}

\begin{figure*}
\begin{center}
\includegraphics[width=0.99\textwidth]{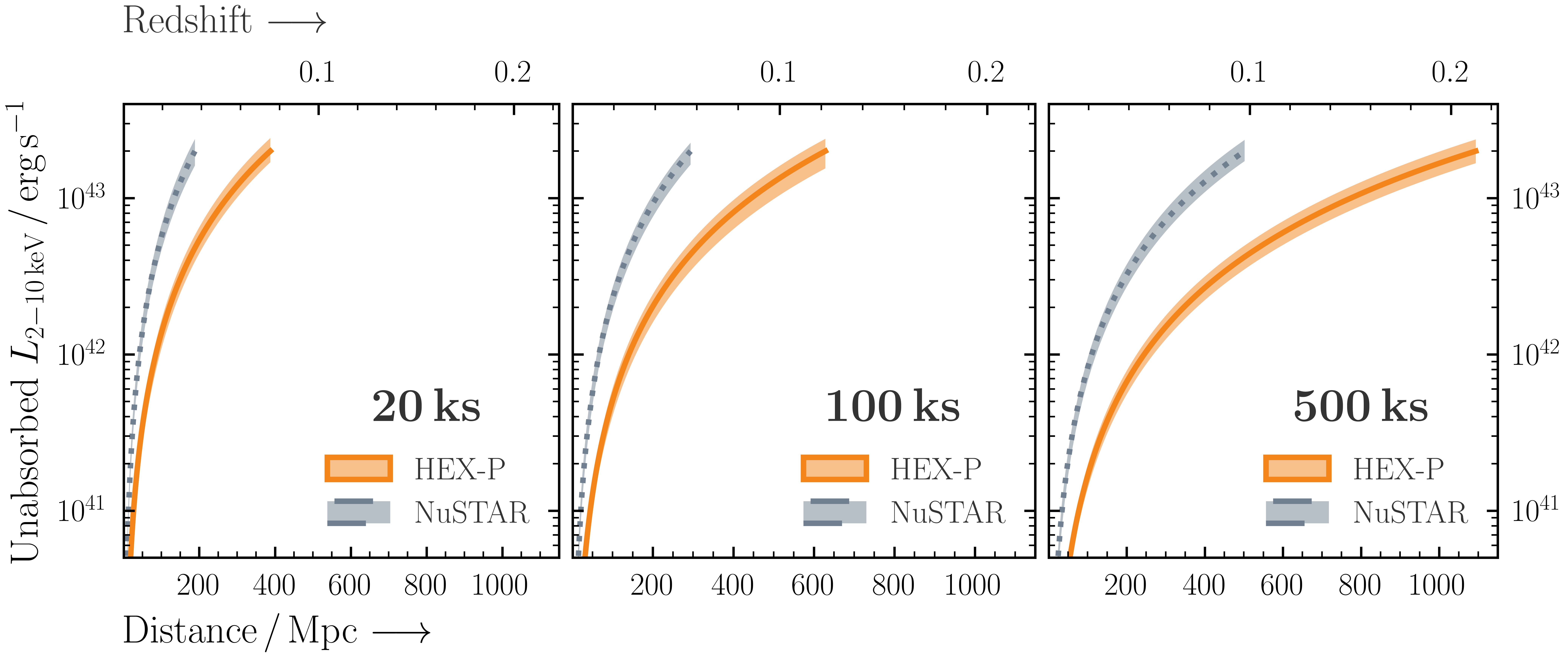}
\end{center}
\caption{Sensitivity curves for the discovery and detailed study of Compton-thick AGN in hard X-rays. Left, center and right panels correspond to exposure times of 20\,ks, 100\,ks and 500\,ks, respectively. Across all panels, each curve shows the combination of distance (horizontal axis) and unabsorbed X-ray luminosity in the 2\,--\,10\,keV band (vertical axis) that gives an observed signal-to-noise ratio of 8 in the hard 10\,--\,25\,keV band. The simulations detailed in Figure~\ref{fig:NGC2960} show that this requirement on signal-to-noise is sufficient to correctly classify an AGN as Compton-thick to 90\% confidence. By increasing the volume for discovery and detailed study by an order of magnitude relative to \textit{NuSTAR}, \textit{HEX-P} will open up a new era in the pursuit of a complete AGN census.}\label{fig:sensitivity}
\end{figure*}

The Compton-thick population remains extremely elusive and uncertain, not just in X-ray surveys of the distant Universe but even in our nearest cosmic volumes $<$\,100\,Mpc \citep{Ricci15,Asmus20,TorresAlba21}. A big open question is the volume for which we can reliably classify AGN as Compton-thick and whether the spectral data is sufficient to constrain the morphology of the circum-nuclear obscurer. Here, we demonstrate the much larger search volume accessible with \textit{HEX-P} than \textit{NuSTAR} for the detailed study of Compton-thick AGN. 

We build upon the simulations of NGC\,2960 detailed in Section~\ref{sec:NGC2960} and shown in Figure~\ref{fig:NGC2960}, and characterise the luminosity and distance space that can be accessed with a given exposure time. We created a large grid of spectral simulations for a Compton-thick AGN \texttt{UXCLUMPY} model with line-of-sight column density $N_{\rm H}$\,=\,1.5\,$\times$\,10$^{24}$\,cm$^{-2}$ and a $\sim$\,50\% covering factor of material with $N_{\rm H}$\,$>$\,10$^{22}$\,cm$^{-2}$. We simulate spectra with 20\,ks, 100\,ks and 500\,ks exposures for a grid of distance and unabsorbed 2\,--\,10\,keV luminosities. For each grid point, we generate ten realisations to estimate the scatter arising from statistical fluctuations. For each exposure time and each luminosity, we identify the distance out to which line-of-sight column densities $N_{\rm H}$\,$<$\,1.5\,$\times$\,10$^{24}$\,cm$^{-2}$ can be ruled out to 90\% significance based on the 10\,--\,25\,keV signal-to-noise ratio (as per Figure~\ref{fig:NGC2960}).

\textit{HEX-P} will double the distance out to which Compton-thick AGN can be securely discovered, over what was possible with \textit{NuSTAR}. This is demonstrated in Figure~\ref{fig:sensitivity}, in three separate panels for each exposure considered. Each panel plots the unabsorbed 2\,--\,10\,keV luminosity as a function of distance. We note that for Compton-thick column densities, the corresponding observed 2\,--\,10\,keV luminosities would be $\sim$\,1\,--\,2 dex lower, dependent upon the choice of spectral model. We choose a fiducial unabsorbed 2\,--\,10\,keV luminosity of 2\,$\times$\,10$^{43}$\,erg\,s$^{-1}$ to define the limiting distance for Compton-thick AGN classifications in our simulations. Figure~\ref{fig:sensitivity} shows that irrespective of exposure time, \textit{HEX-P} can provide detailed spectral classifications of Compton-thick AGN to $\gtrsim$\,2$\times$ the distance of \textit{NuSTAR}, increasing the available volume and number of target sources by $\sim$\,8\,--\,10$\times$. Even for the shortest exposure considered of 20\,ks, \textit{HEX-P} would provide full characterisation of the Compton-thick AGN population with unabsorbed 2\,--\,10\,keV luminosities $L_{2-10\,{\rm keV}}$\,$>$\,10$^{42}$\,erg\,s$^{-1}$ within 100\,Mpc. This encompasses the current volume cut of the highly-complete multi-wavelength selected Local AGN Survey which predicts 362$^{+145}_{-116}$ AGN within the volume with 61 AGN previously unidentified (LASr; \citealt{Asmus20}).

For very long exposures $\gtrsim$\,300\,ks, the effective volume for sensitive \textit{HEX-P} spectral modelling of the Compton-thick population extends to $\sim$\,1\,Gpc for Seyfert-luminosity AGN with unabsorbed 2\,--\,10\,keV luminosities $L_{2-10\,{\rm keV}}$\,$\gtrsim$\,10$^{43}$\,erg\,s$^{-1}$. The corresponding large accessible volume for detailed spectral modelling is testament to the vast jump in sensitivity provided by \textit{HEX-P}. This will be pivotal in probing the dusty hearts of galaxies such as mergers, Ultra/Luminous Infrared Galaxies and Compact Obscured Nuclei that have been suggested to host deeply buried Compton-thick AGN (e.g., \citealt{Iwasawa11,TorresAlba18,Aalto19,Ricci21,Falstad21,Pfeifle23_hexp}. Together with the planned \textit{HEX-P} extragalactic surveys \citep{Civano23}, and a wide array of multi-wavelength facilities, detailed studies of Compton-thick AGN will pave the way towards a complete census of black hole growth across cosmic time.

\section{Summary}\label{sec:summary}
In this paper, we detail many aspects of heavily obscured AGN studies performed in the literature to-date with an extrapolation to the prospects attainable with the next-generation \textit{HEX-P} concept. We begin by compiling an up-to-date and highly complete catalogue of Compton-thick AGN confirmed with spectroscopic fitting that incorporated \textit{NuSTAR} (The Database of Compton-thick AGN, DoCTA; see Section~\ref{sec:ctpop}). We show that there is an enormous range of measured line-of-sight column density and intrinsic luminosity for the targets, highlighting the need for improved broadband spectroscopy paired with next-generation multi-wavelength models of the circum-nuclear obscurer.

The key findings from our \textit{HEX-P} simulations are as follows:

\begin{itemize}
    \item \textit{The development of future AGN models with \textit{HEX-P}:} with the findings of DoCTA in mind, we present an analysis of archival \textit{NuSTAR}\,$+$\,\textit{Chandra} data for ten Compton-thick megamaser AGN. Megamasers have some of the best constrained black hole masses with known inclinations that would help create physically-motivated models of the obscurer that depend on accretion power. We showcase \textit{HEX-P} simulations for the faintest megamaser in our sample, finding that an exposure of 25\,ks suffices to classify the target as Compton-thick to better than 90\% confidence and to measure the intrinsic accretion rate precisely.
    \item \textit{Strictly simultaneous broadband X-ray spectroscopy:} \textit{HEX-P} will provide highly sensitive and strictly simultaneous X-ray spectroscopic observations in the 0.2\,--\,80\,keV passband for the first time. We show that future \textit{HEX-P} monitoring campaigns of heavily obscured AGN will disentangle obscuration-based variations from intrinsic flux variations to much greater precision than is possible with current instruments.
    \item \textit{The nature of the circum-nuclear environment at low accretion power:} It is currently extremely difficult to constrain the geometry of circum-nuclear obscuration at low intrinsic AGN power. We show that the enhanced sensitivities and angular resolution of \textit{HEX-P} are essential for (1) disentangling true low-luminosity AGN from off-nuclear compact objects and (2) constraining the covering factor of the obscurer in sources with $L_{\rm bol}$\,$\lesssim$\,10$^{42}$\,erg\,s$^{-1}$.
    \item \textit{The obscured growth of intermediate mass black holes:} Current estimates of the black hole occupation fraction rely on the detection of dwarf AGN that are biased towards unobscured sources. We show detailed \textit{HEX-P} simulations of one candidate Compton-thick intermediate mass AGN in the literature. We show that sensitive broadband spectroscopy from \textit{HEX-P} is sufficient to not only constrain the line-of-sight column density into the Compton-thick regime, but to differentiate between alternative physical prescriptions for the geometry of the obscurer.
    \item \textit{A new discovery space for Compton-thick AGN:} We determine the accessible volume for accurate characterisation of Compton-thick AGN. We find the improved sensitivities provided by \textit{HEX-P} will more than double the distance and increase the accessible volume by up to an order of magnitude relative to \textit{NuSTAR}.
\end{itemize}

\newpage

\section*{Appendix}

\subsection*{A.~1. Tabulated covering factors in \texttt{UXCLUMPY}}\label{sec:uxccov}

\begin{figure}[!ht]
\begin{center}
\includegraphics[width=0.9\textwidth]{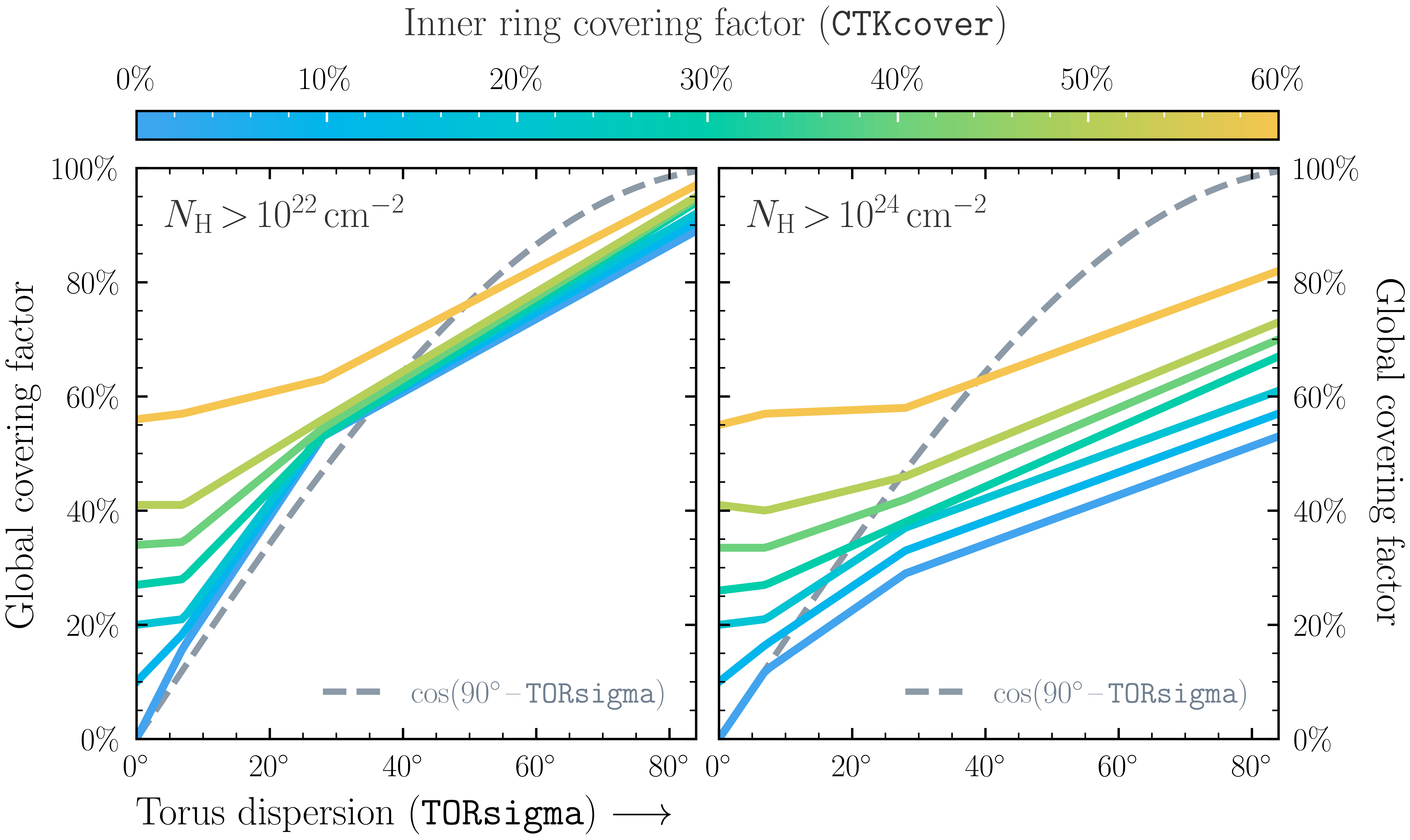}
\end{center}
\caption{Relation between the obscurer dispersion (\texttt{TORsigma}) and the global covering factor of material with $N_{\rm H}$\,$>$\,10$^{22}$\,cm$^{-2}$ (left) and $N_{\rm H}$\,$>$\,10$^{24}$\,cm$^{-2}$ (right) in \texttt{UXCLUMPY}. The colourbar shows the corresponding covering factor of the Compton-thick inner ring (\texttt{CTKcover}). The corresponding relation assuming \texttt{TORsigma} to be a strict boundary to the obscurer opening is shown with a grey dashed line.}\label{fig:uxc_tor_vs_cov}
\end{figure}

\begin{figure}[!ht]
\begin{center}
\includegraphics[width=0.9\textwidth]{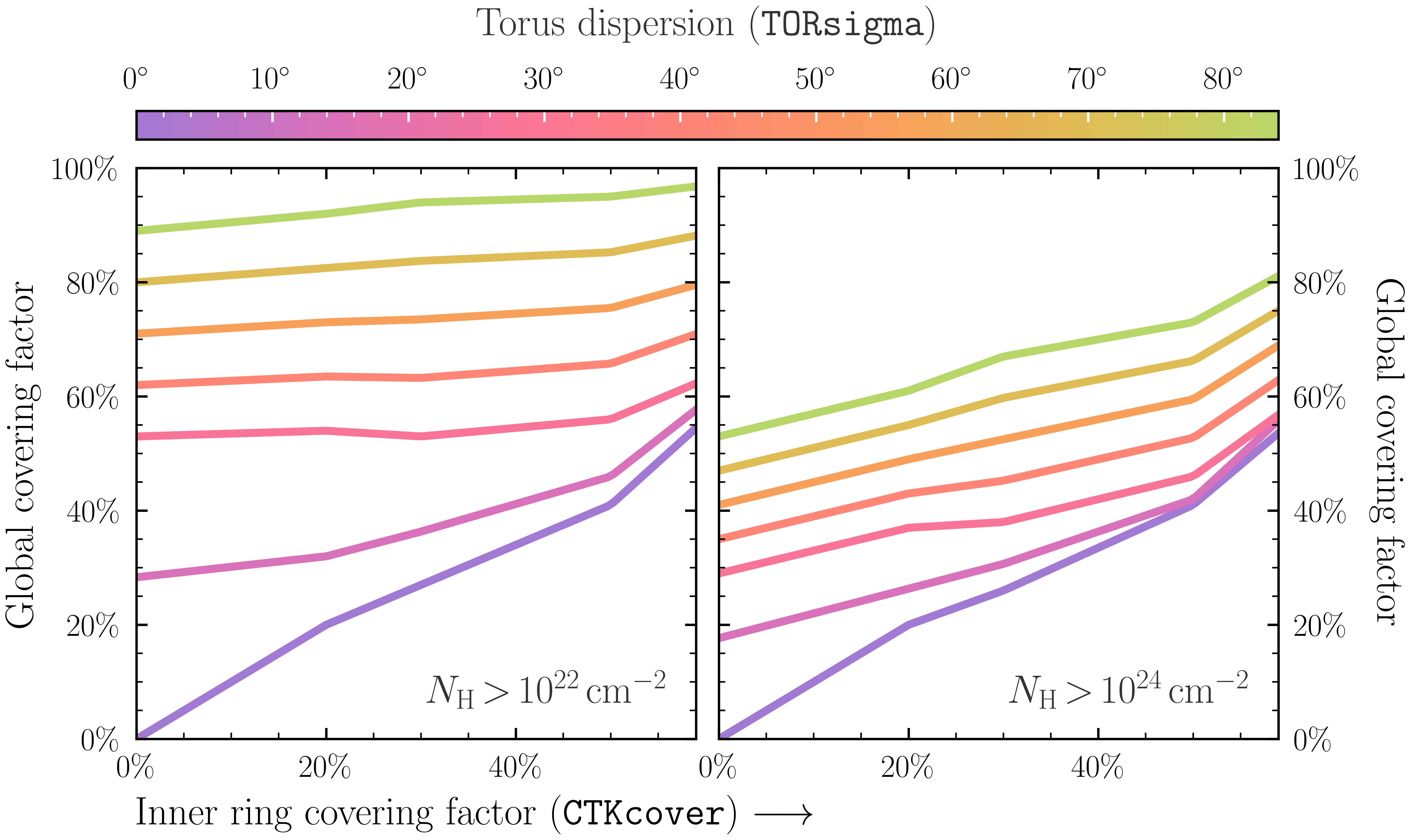}
\end{center}
\caption{Same as in Figure~\ref{fig:uxc_tor_vs_cov} with the horizontal axis and colourbar parameters swapped.}\label{fig:uxc_ctk_vs_cov}
\end{figure}

In Table~\ref{tab:uxc_cov}, we provide the tabulated covering factors in \texttt{UXCLUMPY} for all material with $N_{\rm H}$\,$>$\,10$^{22}$\,cm$^{-2}$ and $N_{\rm H}$\,$>$\,10$^{24}$\,cm$^{-2}$ as a function of \texttt{TORsigma} and \texttt{CTKcover}. We note that for each covering factor column density boundary to apply, the \texttt{nH} parameter in \texttt{UXCLUMPY} must be above the corresponding column density boundary. The covering factors plotted as a function of \texttt{TORsigma} and \texttt{CTKcover} are shown in Figures~\ref{fig:uxc_tor_vs_cov} and~\ref{fig:uxc_ctk_vs_cov}, coloured by the alternative parameter respectively.

\begin{table}[!ht]
\centering
\caption{Tabulated \texttt{UXCLUMPY} covering factors as a function of \texttt{TORsigma} and \texttt{CTKcover}.}\label{tab:uxc_cov}
\begin{tabular}{cccc}
\toprule
     \makecell[ct]{\texttt{TORsigma} \\ (1)} & \makecell[ct]{\texttt{CTKcover} \\ (2)} & \makecell[ct]{$C(N_{\rm H}>10^{22}\,{\rm cm}^{-2})$ \\ (3)} & \makecell[ct]{$C(N_{\rm H}>10^{24}\,{\rm cm}^{-2})$ \\ (4)} \\
\midrule
\multirow{ 5}{*}{84$^{\circ}$} & 0.0 & 0.89 & 0.53 \\
 & 0.2 & 0.92 & 0.61 \\
 & 0.3 & 0.94 & 0.67 \\
 & 0.5 & 0.95 & 0.73 \\
 & 0.6 & 0.97 & 0.82 \\
 \midrule
\multirow{ 5}{*}{28$^{\circ}$} & 0.0 & 0.53 & 0.29 \\
 & 0.2 & 0.54 & 0.37 \\
 & 0.3 & 0.53 & 0.38 \\
 & 0.5 & 0.56 & 0.46 \\
 & 0.6 & 0.63 & 0.58 \\
 \midrule
\multirow{ 5}{*}{7$^{\circ}$}  & 0.0 & 0.16 & 0.12 \\
  & 0.2 & 0.21 & 0.21 \\
  & 0.3 & 0.28 & 0.27 \\
  & 0.5 & 0.41 & 0.40 \\
  & 0.6 & 0.57 & 0.57 \\
  \midrule
\multirow{ 5}{*}{0$^{\circ}$}  & 0.0 & 0.00 & 0.00 \\
  & 0.2 & 0.20 & 0.20 \\
  & 0.3 & 0.27 & 0.26 \\
  & 0.5 & 0.41 & 0.41 \\
  & 0.6 & 0.56 & 0.55 \\
\bottomrule
\end{tabular}

{\raggedright \textbf{Notes.} (1) -- Obscurer dispersion in \texttt{UXCLUMPY}. (2) -- Compton-thick inner ring covering factor in \texttt{UXCLUMPY}. (3) -- Corresponding covering factor of material with $N_{\rm H}$\,$>$\,10$^{22}$\,cm$^{-2}$. (4) -- Corresponding covering factor of material with $N_{\rm H}$\,$>$\,10$^{24}$\,cm$^{-2}$. The covering factors shown in columns (3) and (4) are only valid if the corresponding \texttt{nH} parameter is above $10^{22}$\,cm$^{-2}$ and $10^{24}$\,cm$^{-2}$, respectively.\par}
\end{table}

\section*{Conflict of Interest Statement}

The authors declare that the research was conducted in the absence of any commercial or financial relationships that could be construed as a potential conflict of interest.




\section*{Acknowledgements}
P.G.B. would like to thank Abhijeet Borkar for useful comments that helped to improve the manuscript. N.T.A. acknowledges funding from NASA under contracts 80NSSC19K0531, 80NSSC20K0045, and 80NSSC20K834. J.M.P. acknowledges support from NASA grants 80NSSC21K1567 and 80NSSC22K1120. M.B acknowledges support from the YCAA Prize Postdoctoral Fellowship. C.R. acknowledges support from the Fondecyt Regular grant 1230345 and ANID BASAL project FB210003. 

This work made use of data from the \textit{NuSTAR} mission, a project led by the California Institute of Technology, managed by the Jet Propulsion Laboratory, and funded by the National Aeronautics and Space Administration. We thank the \textit{NuSTAR} Operations, Software and Calibration teams for support with the execution and analysis of these observations. This research has made use of the \textit{NuSTAR} Data Analysis Software (NuSTARDAS) jointly developed by the ASI Science Data Center (ASDC, Italy) and the California Institute of Technology (USA).

This research has made use of data obtained from the Chandra Data Archive and the Chandra Source Catalog, and software provided by the Chandra X-ray Center (CXC) in the application packages CIAO and Sherpa.

This research has made use of the NASA/IPAC Extragalactic Database (NED), which is operated by the Jet Propulsion Laboratory, California Institute of Technology, under contract with the National Aeronautics and Space Administration.

This research has made use of NASA's Astrophysics Data System Bibliographic Services.

This work made use of Astropy:\footnote{http://www.astropy.org} a community-developed core Python package and an ecosystem of tools and resources for astronomy \citep{astropy:2013, astropy:2018, astropy:2022}.

This paper made extensive use of the matplotlib \citep{Hunter07}, Scipy \citep{2020SciPy-NMeth} and Pandas \citep{mckinney-proc-scipy-2010} Python packages.



\bibliographystyle{Frontiers-Harvard} 
\bibliography{biblio}

\begin{thebibliography}{231}
\providecommand{\natexlab}[1]{#1}
\expandafter\ifx\csname urlstyle\endcsname\relax
  \providecommand{\doi}[1]{doi:\discretionary{}{}{}#1}\else
  \providecommand{\doi}{doi:\discretionary{}{}{}\begingroup
  \urlstyle{rm}\Url}\fi
\providecommand{\selectlanguage}[1]{\relax}
\providecommand{\bibAnnoteFile}[1]{%
  \IfFileExists{#1}{\begin{quotation}\noindent\textsc{Key:} #1\\
  \textsc{Annotation:}\ \input{#1}\end{quotation}}{}}
\providecommand{\bibAnnote}[2]{%
  \begin{quotation}\noindent\textsc{Key:} #1\\
  \textsc{Annotation:}\ #2\end{quotation}}

\bibitem[{{Aalto} et~al.(2019){Aalto}, {Muller}, {K{\"o}nig}, {Falstad},
  {Mangum}, {Sakamoto} et~al.}]{Aalto19}
{Aalto}, S., {Muller}, S., {K{\"o}nig}, S., {Falstad}, N., {Mangum}, J.,
  {Sakamoto}, K., et~al. (2019).
\newblock {The hidden heart of the luminous infrared galaxy IC 860. I. A
  molecular inflow feeding opaque, extreme nuclear activity}.
\newblock \emph{\aap} 627, A147.
\newblock \doi{10.1051/0004-6361/201935480}
\bibAnnoteFile{Aalto19}

\bibitem[{{Aird} et~al.(2015){Aird}, {Coil}, {Georgakakis}, {Nandra}, {Barro},
  and {P{\'e}rez-Gonz{\'a}lez}}]{Aird15}
{Aird}, J., {Coil}, A.~L., {Georgakakis}, A., {Nandra}, K., {Barro}, G., and
  {P{\'e}rez-Gonz{\'a}lez}, P.~G. (2015).
\newblock {The evolution of the X-ray luminosity functions of unabsorbed and
  absorbed AGNs out to z{\ensuremath{\sim}} 5}.
\newblock \emph{\mnras} 451, 1892--1927.
\newblock \doi{10.1093/mnras/stv1062}
\bibAnnoteFile{Aird15}

\bibitem[{{Akylas} et~al.(2012){Akylas}, {Georgakakis}, {Georgantopoulos},
  {Brightman}, and {Nandra}}]{Akylas12}
{Akylas}, A., {Georgakakis}, A., {Georgantopoulos}, I., {Brightman}, M., and
  {Nandra}, K. (2012).
\newblock {Constraining the fraction of Compton-thick AGN in the Universe by
  modelling the diffuse X-ray background spectrum}.
\newblock \emph{\aap} 546, A98.
\newblock \doi{10.1051/0004-6361/201219387}
\bibAnnoteFile{Akylas12}

\bibitem[{{Ananna} et~al.(2019){Ananna}, {Treister}, {Urry}, {Ricci},
  {Kirkpatrick}, {LaMassa} et~al.}]{Ananna19}
{Ananna}, T.~T., {Treister}, E., {Urry}, C.~M., {Ricci}, C., {Kirkpatrick}, A.,
  {LaMassa}, S., et~al. (2019).
\newblock {The Accretion History of AGNs. I. Supermassive Black Hole Population
  Synthesis Model}.
\newblock \emph{\apj} 871, 240.
\newblock \doi{10.3847/1538-4357/aafb77}
\bibAnnoteFile{Ananna19}

\bibitem[{{Ananna} et~al.(2022){Ananna}, {Urry}, {Ricci}, {Natarajan},
  {Hickox}, {Trakhtenbrot} et~al.}]{Ananna22}
{Ananna}, T.~T., {Urry}, C.~M., {Ricci}, C., {Natarajan}, P., {Hickox}, R.~C.,
  {Trakhtenbrot}, B., et~al. (2022).
\newblock {Probing the Structure and Evolution of BASS Active Galactic Nuclei
  through Eddington Ratios}.
\newblock \emph{\apjl} 939, L13.
\newblock \doi{10.3847/2041-8213/ac9979}
\bibAnnoteFile{Ananna22}

\bibitem[{{Andonie} et~al.(2023){Andonie}, {Alexander}, {Greenwell}, {Puglisi},
  {Laloux}, {Alonso-Tetilla} et~al.}]{Andonie23}
{Andonie}, C., {Alexander}, D.~M., {Greenwell}, C., {Puglisi}, A., {Laloux},
  B., {Alonso-Tetilla}, A.~V., et~al. (2023).
\newblock {Obscuration beyond the nucleus: infrared quasars can be buried in
  extreme compact starbursts}.
\newblock \emph{\mnras} \doi{10.1093/mnrasl/slad144}
\bibAnnoteFile{Andonie23}

\bibitem[{{Andonie} et~al.(2022){Andonie}, {Ricci}, {Paltani}, {Ar{\'e}valo},
  {Treister}, {Bauer} et~al.}]{Andonie22}
{Andonie}, C., {Ricci}, C., {Paltani}, S., {Ar{\'e}valo}, P., {Treister}, E.,
  {Bauer}, F., et~al. (2022).
\newblock {A multiwavelength-motivated X-ray model for the Circinus Galaxy}.
\newblock \emph{\mnras} 511, 5768--5781.
\newblock \doi{10.1093/mnras/stac403}
\bibAnnoteFile{Andonie22}

\bibitem[{{Angl{\'e}s-Alc{\'a}zar} et~al.(2021){Angl{\'e}s-Alc{\'a}zar},
  {Quataert}, {Hopkins}, {Somerville}, {Hayward}, {Faucher-Gigu{\`e}re}
  et~al.}]{AnglesAlcazar21}
{Angl{\'e}s-Alc{\'a}zar}, D., {Quataert}, E., {Hopkins}, P.~F., {Somerville},
  R.~S., {Hayward}, C.~C., {Faucher-Gigu{\`e}re}, C.-A., et~al. (2021).
\newblock {Cosmological Simulations of Quasar Fueling to Subparsec Scales Using
  Lagrangian Hyper-refinement}.
\newblock \emph{\apj} 917, 53.
\newblock \doi{10.3847/1538-4357/ac09e8}
\bibAnnoteFile{AnglesAlcazar21}

\bibitem[{{Annuar} et~al.(2017){Annuar}, {Alexander}, {Gandhi}, {Lansbury},
  {Asmus}, {Ballantyne} et~al.}]{Annuar17}
{Annuar}, A., {Alexander}, D.~M., {Gandhi}, P., {Lansbury}, G.~B., {Asmus}, D.,
  {Ballantyne}, D.~R., et~al. (2017).
\newblock {A New Compton-thick AGN in our Cosmic Backyard: Unveiling the Buried
  Nucleus in NGC 1448 with NuSTAR}.
\newblock \emph{\apj} 836, 165.
\newblock \doi{10.3847/1538-4357/836/2/165}
\bibAnnoteFile{Annuar17}

\bibitem[{Annuar et~al.(2020)Annuar, Gandhi, Alexander, Anathpindika, Asmus,
  Ballantyne et~al.}]{Annuar20}
Annuar, A., Gandhi, P., Alexander, D.~M., Anathpindika, S., Asmus, D.,
  Ballantyne, D.~R., et~al. (2020).
\newblock A deep x-ray view of the bare agn ark120. v. spin determination from
  disc-comptonisation efficiency method.
\newblock \emph{Monthly Notices of the Royal Astronomical Society} 492,
  5481--5500.
\newblock \doi{10.1093/mnras/staa205}
\bibAnnoteFile{Annuar20}

\bibitem[{{Annuar} et~al.(2015){Annuar}, {Gandhi}, {Alexander}, {Lansbury},
  {Ar{\'e}valo}, {Ballantyne} et~al.}]{Annuar15}
{Annuar}, A., {Gandhi}, P., {Alexander}, D.~M., {Lansbury}, G.~B.,
  {Ar{\'e}valo}, P., {Ballantyne}, D.~R., et~al. (2015).
\newblock {NuSTAR Observations of the Compton-thick Active Galactic Nucleus and
  Ultraluminous X-Ray Source Candidate in NGC 5643}.
\newblock \emph{\apj} 815, 36.
\newblock \doi{10.1088/0004-637X/815/1/36}
\bibAnnoteFile{Annuar15}

\bibitem[{{Ansh} et~al.(2023){Ansh}, {Chen}, {Brandt}, {Hood}, {Kammoun},
  {Lansbury} et~al.}]{Ansh23}
{Ansh}, S., {Chen}, C.-T.~J., {Brandt}, W.~N., {Hood}, C.~E., {Kammoun}, E.~S.,
  {Lansbury}, G., et~al. (2023).
\newblock {NuSTAR Observations of a Heavily X-Ray-obscured AGN in the Dwarf
  Galaxy J144013+024744}.
\newblock \emph{\apj} 942, 82.
\newblock \doi{10.3847/1538-4357/ac9382}
\bibAnnoteFile{Ansh23}

\bibitem[{{Antonucci}(1993)}]{Antonucci93}
{Antonucci}, R. (1993).
\newblock {Unified models for active galactic nuclei and quasars.}
\newblock \emph{\araa} 31, 473--521.
\newblock \doi{10.1146/annurev.aa.31.090193.002353}
\bibAnnoteFile{Antonucci93}

\bibitem[{{Ar{\'e}valo} et~al.(2014){Ar{\'e}valo}, {Bauer}, {Puccetti},
  {Walton}, {Koss}, {Boggs} et~al.}]{Arevalo14}
{Ar{\'e}valo}, P., {Bauer}, F.~E., {Puccetti}, S., {Walton}, D.~J., {Koss}, M.,
  {Boggs}, S.~E., et~al. (2014).
\newblock {The 2-79 keV X-Ray Spectrum of the Circinus Galaxy with NuSTAR,
  XMM-Newton, and Chandra: A Fully Compton-thick Active Galactic Nucleus}.
\newblock \emph{\apj} 791, 81.
\newblock \doi{10.1088/0004-637X/791/2/81}
\bibAnnoteFile{Arevalo14}

\bibitem[{{Arnaud}(1996)}]{Arnaud96}
{Arnaud}, K.~A. (1996).
\newblock {XSPEC: The First Ten Years}.
\newblock In \emph{Astronomical Data Analysis Software and Systems V}, eds.
  G.~H. {Jacoby} and J.~{Barnes}. vol. 101 of \emph{Astronomical Society of the
  Pacific Conference Series}, 17
\bibAnnoteFile{Arnaud96}

\bibitem[{{Asmus} et~al.(2020){Asmus}, {Greenwell}, {Gandhi}, {Boorman},
  {Aird}, {Alexander} et~al.}]{Asmus20}
{Asmus}, D., {Greenwell}, C.~L., {Gandhi}, P., {Boorman}, P.~G., {Aird}, J.,
  {Alexander}, D.~M., et~al. (2020).
\newblock {Local AGN survey (LASr): I. Galaxy sample, infrared colour
  selection, and predictions for AGN within 100 Mpc}.
\newblock \emph{\mnras} 494, 1784--1816.
\newblock \doi{10.1093/mnras/staa766}
\bibAnnoteFile{Asmus20}

\bibitem[{{Astropy Collaboration} et~al.(2022){Astropy Collaboration},
  {Price-Whelan}, {Lim}, {Earl}, {Starkman}, {Bradley} et~al.}]{astropy:2022}
{Astropy Collaboration}, {Price-Whelan}, A.~M., {Lim}, P.~L., {Earl}, N.,
  {Starkman}, N., {Bradley}, L., et~al. (2022).
\newblock {The Astropy Project: Sustaining and Growing a Community-oriented
  Open-source Project and the Latest Major Release (v5.0) of the Core Package}.
\newblock \emph{\apj} 935, 167.
\newblock \doi{10.3847/1538-4357/ac7c74}
\bibAnnoteFile{astropy:2022}

\bibitem[{{Astropy Collaboration} et~al.(2018){Astropy Collaboration},
  {Price-Whelan}, {Sip{\H{o}}cz}, {G{\"u}nther}, {Lim}, {Crawford}
  et~al.}]{astropy:2018}
{Astropy Collaboration}, {Price-Whelan}, A.~M., {Sip{\H{o}}cz}, B.~M.,
  {G{\"u}nther}, H.~M., {Lim}, P.~L., {Crawford}, S.~M., et~al. (2018).
\newblock {The Astropy Project: Building an Open-science Project and Status of
  the v2.0 Core Package}.
\newblock \emph{\aj} 156, 123.
\newblock \doi{10.3847/1538-3881/aabc4f}
\bibAnnoteFile{astropy:2018}

\bibitem[{{Astropy Collaboration} et~al.(2013){Astropy Collaboration},
  {Robitaille}, {Tollerud}, {Greenfield}, {Droettboom}, {Bray}
  et~al.}]{astropy:2013}
{Astropy Collaboration}, {Robitaille}, T.~P., {Tollerud}, E.~J., {Greenfield},
  P., {Droettboom}, M., {Bray}, E., et~al. (2013).
\newblock {Astropy: A community Python package for astronomy}.
\newblock \emph{\aap} 558, A33.
\newblock \doi{10.1051/0004-6361/201322068}
\bibAnnoteFile{astropy:2013}

\bibitem[{{Ba{\~n}ados} et~al.(2018){Ba{\~n}ados}, {Venemans}, {Mazzucchelli},
  {Farina}, {Walter}, {Wang} et~al.}]{Banados18}
{Ba{\~n}ados}, E., {Venemans}, B.~P., {Mazzucchelli}, C., {Farina}, E.~P.,
  {Walter}, F., {Wang}, F., et~al. (2018).
\newblock {An 800-million-solar-mass black hole in a significantly neutral
  Universe at a redshift of 7.5}.
\newblock \emph{\nat} 553, 473--476.
\newblock \doi{10.1038/nature25180}
\bibAnnoteFile{Banados18}

\bibitem[{{Bachetti} et~al.(2023){Bachetti}, {Garc\'{i}a}, {Grefenstette},
  {Stern}, {Bachetti}, and {The HEX-P Team}}]{Bachetti23}
{Bachetti}, M., {Garc\'{i}a}, J., {Grefenstette}, B., {Stern}, D., {Bachetti},
  M., and {The HEX-P Team} (2023).
\newblock {submitted}.
\newblock \emph{Frontiers in Astronomy and Space Sciences}
\bibAnnoteFile{Bachetti23}

\bibitem[{{Baldassare} et~al.(2018){Baldassare}, {Geha}, and
  {Greene}}]{Baldassare18}
{Baldassare}, V.~F., {Geha}, M., and {Greene}, J. (2018).
\newblock {Identifying AGNs in Low-mass Galaxies via Long-term Optical
  Variability}.
\newblock \emph{\apj} 868, 152.
\newblock \doi{10.3847/1538-4357/aae6cf}
\bibAnnoteFile{Baldassare18}

\bibitem[{{Baldi} et~al.(2021{\natexlab{a}}){Baldi}, {Williams}, {Beswick},
  {McHardy}, {Dullo}, {Knapen} et~al.}]{Baldi21b}
{Baldi}, R.~D., {Williams}, D.~R.~A., {Beswick}, R.~J., {McHardy}, I., {Dullo},
  B.~T., {Knapen}, J.~H., et~al. (2021{\natexlab{a}}).
\newblock {LeMMINGs III. The e-MERLIN legacy survey of the Palomar sample:
  exploring the origin of nuclear radio emission in active and inactive
  galaxies through the [O III] - radio connection}.
\newblock \emph{\mnras} 508, 2019--2038.
\newblock \doi{10.1093/mnras/stab2613}
\bibAnnoteFile{Baldi21b}

\bibitem[{{Baldi} et~al.(2018){Baldi}, {Williams}, {McHardy}, {Beswick},
  {Argo}, {Dullo} et~al.}]{Baldi18}
{Baldi}, R.~D., {Williams}, D.~R.~A., {McHardy}, I.~M., {Beswick}, R.~J.,
  {Argo}, M.~K., {Dullo}, B.~T., et~al. (2018).
\newblock {LeMMINGs - I. The eMERLIN legacy survey of nearby galaxies. 1.5-GHz
  parsec-scale radio structures and cores}.
\newblock \emph{\mnras} 476, 3478--3522.
\newblock \doi{10.1093/mnras/sty342}
\bibAnnoteFile{Baldi18}

\bibitem[{{Baldi} et~al.(2021{\natexlab{b}}){Baldi}, {Williams}, {McHardy},
  {Beswick}, {Brinks}, {Dullo} et~al.}]{Baldi21a}
{Baldi}, R.~D., {Williams}, D.~R.~A., {McHardy}, I.~M., {Beswick}, R.~J.,
  {Brinks}, E., {Dullo}, B.~T., et~al. (2021{\natexlab{b}}).
\newblock {LeMMINGs - II. The e-MERLIN legacy survey of nearby galaxies. The
  deepest radio view of the Palomar sample on parsec scale}.
\newblock \emph{\mnras} 500, 4749--4767.
\newblock \doi{10.1093/mnras/staa3519}
\bibAnnoteFile{Baldi21a}

\bibitem[{{Balokovi\' c}(2017)}]{Balokovic17}
{Balokovi\' c}, M. (2017).
\newblock \emph{{Unveiling the Structure of Active Galactic Nuclei with Hard
  X-ray Spectroscopy}}.
\newblock Ph.D. thesis, California Institute of Technology
\bibAnnoteFile{Balokovic17}

\bibitem[{{Balokovi{\'c}} et~al.(2018){Balokovi{\'c}}, {Brightman}, {Harrison},
  {Comastri}, {Ricci}, {Buchner} et~al.}]{Balokovic18}
{Balokovi{\'c}}, M., {Brightman}, M., {Harrison}, F.~A., {Comastri}, A.,
  {Ricci}, C., {Buchner}, J., et~al. (2018).
\newblock {New Spectral Model for Constraining Torus Covering Factors from
  Broadband X-Ray Spectra of Active Galactic Nuclei}.
\newblock \emph{\apj} 854, 42.
\newblock \doi{10.3847/1538-4357/aaa7eb}
\bibAnnoteFile{Balokovic18}

\bibitem[{{Balokovi{\'c}} et~al.(2021){Balokovi{\'c}}, {Cabral}, {Brenneman},
  and {Urry}}]{Balokovic21}
{Balokovi{\'c}}, M., {Cabral}, S.~E., {Brenneman}, L., and {Urry}, C.~M.
  (2021).
\newblock {Properties of the Obscuring Torus in NGC 1052 from Multiepoch
  Broadband X-Ray Spectroscopy}.
\newblock \emph{\apj} 916, 90.
\newblock \doi{10.3847/1538-4357/abff4d}
\bibAnnoteFile{Balokovic21}

\bibitem[{{Balokovi{\'c}} et~al.(2014){Balokovi{\'c}}, {Comastri}, {Harrison},
  {Alexander}, {Ballantyne}, {Bauer} et~al.}]{Balokovic14}
{Balokovi{\'c}}, M., {Comastri}, A., {Harrison}, F.~A., {Alexander}, D.~M.,
  {Ballantyne}, D.~R., {Bauer}, F.~E., et~al. (2014).
\newblock {The NuSTAR View of Nearby Compton-thick Active Galactic Nuclei: The
  Cases of NGC 424, NGC 1320, and IC 2560}.
\newblock \emph{\apj} 794, 111.
\newblock \doi{10.1088/0004-637X/794/2/111}
\bibAnnoteFile{Balokovic14}

\bibitem[{{Balokovi{\'c}} et~al.(2019){Balokovi{\'c}}, {Garc{\'\i}a}, and
  {Cabral}}]{Balokovic19}
{Balokovi{\'c}}, M., {Garc{\'\i}a}, J.~A., and {Cabral}, S.~E. (2019).
\newblock {New Tools for Self-consistent Modeling of the AGN Torus and Corona}.
\newblock \emph{Research Notes of the American Astronomical Society} 3, 173.
\newblock \doi{10.3847/2515-5172/ab578e}
\bibAnnoteFile{Balokovic19}

\bibitem[{{Bauer} et~al.(2015){Bauer}, {Ar{\'e}valo}, {Walton}, {Koss},
  {Puccetti}, {Gandhi} et~al.}]{Bauer15}
{Bauer}, F.~E., {Ar{\'e}valo}, P., {Walton}, D.~J., {Koss}, M.~J., {Puccetti},
  S., {Gandhi}, P., et~al. (2015).
\newblock {NuSTAR Spectroscopy of Multi-component X-Ray Reflection from NGC
  1068}.
\newblock \emph{\apj} 812, 116.
\newblock \doi{10.1088/0004-637X/812/2/116}
\bibAnnoteFile{Bauer15}

\bibitem[{{Boorman} et~al.(2016){Boorman}, {Gandhi}, {Alexander}, {Annuar},
  {Ballantyne}, {Bauer} et~al.}]{Boorman16}
{Boorman}, P.~G., {Gandhi}, P., {Alexander}, D.~M., {Annuar}, A., {Ballantyne},
  D.~R., {Bauer}, F., et~al. (2016).
\newblock {IC 3639{\textemdash}a New Bona Fide Compton-Thick AGN Unveiled by
  NuSTAR}.
\newblock \emph{\apj} 833, 245.
\newblock \doi{10.3847/1538-4357/833/2/245}
\bibAnnoteFile{Boorman16}

\bibitem[{{Boorman} et~al.(2018){Boorman}, {Gandhi}, {Balokovi{\'c}},
  {Brightman}, {Harrison}, {Ricci} et~al.}]{Boorman18}
{Boorman}, P.~G., {Gandhi}, P., {Balokovi{\'c}}, M., {Brightman}, M.,
  {Harrison}, F., {Ricci}, C., et~al. (2018).
\newblock {An Iwasawa-Taniguchi effect for Compton-thick active galactic
  nuclei}.
\newblock \emph{\mnras} 477, 3775--3790.
\newblock \doi{10.1093/mnras/sty861}
\bibAnnoteFile{Boorman18}

\bibitem[{{Brandt} and {Alexander}(2015)}]{Brandt15}
{Brandt}, W.~N. and {Alexander}, D.~M. (2015).
\newblock {Cosmic X-ray surveys of distant active galaxies. The demographics,
  physics, and ecology of growing supermassive black holes}.
\newblock \emph{\aapr} 23, 1.
\newblock \doi{10.1007/s00159-014-0081-z}
\bibAnnoteFile{Brandt15}

\bibitem[{{Brandt} and {Yang}(2022)}]{Brandt22}
{Brandt}, W.~N. and {Yang}, G. (2022).
\newblock {Surveys of the Cosmic X-Ray Background}.
\newblock In \emph{Handbook of X-ray and Gamma-ray Astrophysics}. 78.
\newblock \doi{10.1007/978-981-16-4544-0_130-1}
\bibAnnoteFile{Brandt22}

\bibitem[{{Brightman} et~al.(2015){Brightman}, {Balokovi{\'c}}, {Stern},
  {Ar{\'e}valo}, {Ballantyne}, {Bauer} et~al.}]{Brightman15}
{Brightman}, M., {Balokovi{\'c}}, M., {Stern}, D., {Ar{\'e}valo}, P.,
  {Ballantyne}, D.~R., {Bauer}, F.~E., et~al. (2015).
\newblock {Determining the Covering Factor of Compton-thick Active Galactic
  Nuclei with NuSTAR}.
\newblock \emph{\apj} 805, 41.
\newblock \doi{10.1088/0004-637X/805/1/41}
\bibAnnoteFile{Brightman15}

\bibitem[{{Brightman} et~al.(2023){Brightman}, {Margutti}, {Polzin}, {Jaodand},
  and {The HEX-P Team}}]{Brightman23}
{Brightman}, M., {Margutti}, R., {Polzin}, A., {Jaodand}, A., and {The HEX-P
  Team} (2023).
\newblock {submitted}.
\newblock \emph{Frontiers in Astronomy and Space Sciences}
\bibAnnoteFile{Brightman23}

\bibitem[{{Brightman} et~al.(2016){Brightman}, {Masini}, {Ballantyne},
  {Balokovi{\'c}}, {Brandt}, {Chen} et~al.}]{Brightman16}
{Brightman}, M., {Masini}, A., {Ballantyne}, D.~R., {Balokovi{\'c}}, M.,
  {Brandt}, W.~N., {Chen}, C.~T., et~al. (2016).
\newblock {A Growth-rate Indicator for Compton-thick Active Galactic Nuclei}.
\newblock \emph{\apj} 826, 93.
\newblock \doi{10.3847/0004-637X/826/1/93}
\bibAnnoteFile{Brightman16}

\bibitem[{Brightman et~al.(2018)Brightman, Masini, Ballantyne, Baloković,
  Brandt, Chen et~al.}]{Brightman18}
Brightman, M., Masini, A., Ballantyne, D.~R., Baloković, M., Brandt, W.~N.,
  Chen, C.-T.~J., et~al. (2018).
\newblock Bat agn spectroscopic survey. vii. the covering factor of dust and
  gas in swift/bat active galactic nuclei.
\newblock \emph{The Astrophysical Journal} 867, 110.
\newblock \doi{10.3847/1538-4357/aae4da}
\bibAnnoteFile{Brightman18}

\bibitem[{{Brightman} and {Nandra}(2011{\natexlab{a}})}]{Brightman11a}
{Brightman}, M. and {Nandra}, K. (2011{\natexlab{a}}).
\newblock {An XMM-Newton spectral survey of 12 {\ensuremath{\mu}}m selected
  galaxies - I. X-ray data}.
\newblock \emph{\mnras} 413, 1206--1235.
\newblock \doi{10.1111/j.1365-2966.2011.18207.x}
\bibAnnoteFile{Brightman11a}

\bibitem[{{Brightman} and {Nandra}(2011{\natexlab{b}})}]{Brightman11b}
{Brightman}, M. and {Nandra}, K. (2011{\natexlab{b}}).
\newblock {An XMM-Newton spectral survey of 12 {\ensuremath{\mu}}m selected
  galaxies - II. Implications for AGN selection and unification}.
\newblock \emph{\mnras} 414, 3084--3104.
\newblock \doi{10.1111/j.1365-2966.2011.18612.x}
\bibAnnoteFile{Brightman11b}

\bibitem[{{Buchner} and {Boorman}(2023)}]{Buchner23}
{Buchner}, J. and {Boorman}, P. (2023).
\newblock {Statistical Aspects of X-ray Spectral Analysis}.
\newblock \emph{arXiv e-prints} ,
  arXiv:2309.05705\doi{10.48550/arXiv.2309.05705}
\bibAnnoteFile{Buchner23}

\bibitem[{{Buchner} et~al.(2021){Buchner}, {Brightman}, {Balokovi{\'c}},
  {Wada}, {Bauer}, and {Nandra}}]{Buchner21}
{Buchner}, J., {Brightman}, M., {Balokovi{\'c}}, M., {Wada}, K., {Bauer},
  F.~E., and {Nandra}, K. (2021).
\newblock {Physically motivated X-ray obscurer models}.
\newblock \emph{\aap} 651, A58.
\newblock \doi{10.1051/0004-6361/201834963}
\bibAnnoteFile{Buchner21}

\bibitem[{{Buchner} et~al.(2019){Buchner}, {Brightman}, {Nandra}, {Nikutta},
  and {Bauer}}]{Buchner19}
{Buchner}, J., {Brightman}, M., {Nandra}, K., {Nikutta}, R., and {Bauer}, F.~E.
  (2019).
\newblock {X-ray spectral and eclipsing model of the clumpy obscurer in active
  galactic nuclei}.
\newblock \emph{\aap} 629, A16.
\newblock \doi{10.1051/0004-6361/201834771}
\bibAnnoteFile{Buchner19}

\bibitem[{{Buchner} et~al.(2015){Buchner}, {Georgakakis}, {Nandra},
  {Brightman}, {Menzel}, {Liu} et~al.}]{Buchner15}
{Buchner}, J., {Georgakakis}, A., {Nandra}, K., {Brightman}, M., {Menzel},
  M.-L., {Liu}, Z., et~al. (2015).
\newblock {Obscuration-dependent Evolution of Active Galactic Nuclei}.
\newblock \emph{\apj} 802, 89.
\newblock \doi{10.1088/0004-637X/802/2/89}
\bibAnnoteFile{Buchner15}

\bibitem[{{Buchner} et~al.(2014){Buchner}, {Georgakakis}, {Nandra}, {Hsu},
  {Rangel}, {Brightman} et~al.}]{Buchner14}
{Buchner}, J., {Georgakakis}, A., {Nandra}, K., {Hsu}, L., {Rangel}, C.,
  {Brightman}, M., et~al. (2014).
\newblock {X-ray spectral modelling of the AGN obscuring region in the CDFS:
  Bayesian model selection and catalogue}.
\newblock \emph{\aap} 564, A125.
\newblock \doi{10.1051/0004-6361/201322971}
\bibAnnoteFile{Buchner14}

\bibitem[{{Buchner} et~al.(2017){Buchner}, {Schulze}, and {Bauer}}]{Buchner17a}
{Buchner}, J., {Schulze}, S., and {Bauer}, F.~E. (2017).
\newblock {Galaxy gas as obscurer - I. GRBs x-ray galaxies and find an
  N$_{H}$$^{3}${\ensuremath{\propto}} M\_\{star\} relation}.
\newblock \emph{\mnras} 464, 4545--4566.
\newblock \doi{10.1093/mnras/stw2423}
\bibAnnoteFile{Buchner17a}

\bibitem[{{Chan} and {Krolik}(2016)}]{Chan16}
{Chan}, C.-H. and {Krolik}, J.~H. (2016).
\newblock {Radiation-driven Outflows from and Radiative Support in Dusty Tori
  of Active Galactic Nuclei}.
\newblock \emph{\apj} 825, 67.
\newblock \doi{10.3847/0004-637X/825/1/67}
\bibAnnoteFile{Chan16}

\bibitem[{{Chen} et~al.(2017){Chen}, {Brandt}, {Reines}, {Lansbury}, {Stern},
  {Alexander} et~al.}]{Chen17}
{Chen}, C. T.~J., {Brandt}, W.~N., {Reines}, A.~E., {Lansbury}, G., {Stern},
  D., {Alexander}, D.~M., et~al. (2017).
\newblock {Hard X-Ray-selected AGNs in Low-mass Galaxies from the NuSTAR
  Serendipitous Survey}.
\newblock \emph{\apj} 837, 48.
\newblock \doi{10.3847/1538-4357/aa5d5b}
\bibAnnoteFile{Chen17}

\bibitem[{{Civano} et~al.(2023){Civano}, {Zhao}, {Boorman}, {Marchesi},
  {Ananna}, {Creech} et~al.}]{Civano23}
{Civano}, F., {Zhao}, X., {Boorman}, P., {Marchesi}, S., {Ananna}, T.,
  {Creech}, S., et~al. (2023).
\newblock {in prep.}
\newblock \emph{Frontiers in Astronomy and Space Sciences}
\bibAnnoteFile{Civano23}

\bibitem[{{Comastri} et~al.(2015){Comastri}, {Gilli}, {Marconi}, {Risaliti},
  and {Salvati}}]{Comastri15}
{Comastri}, A., {Gilli}, R., {Marconi}, A., {Risaliti}, G., and {Salvati}, M.
  (2015).
\newblock {Mass without radiation: Heavily obscured AGNs, the X-ray background,
  and the black hole mass density}.
\newblock \emph{\aap} 574, L10.
\newblock \doi{10.1051/0004-6361/201425496}
\bibAnnoteFile{Comastri15}

\bibitem[{{Comastri} et~al.(1995){Comastri}, {Setti}, {Zamorani}, and
  {Hasinger}}]{Comastri95}
{Comastri}, A., {Setti}, G., {Zamorani}, G., and {Hasinger}, G. (1995).
\newblock {The contribution of AGNs to the X-ray background.}
\newblock \emph{\aap} 296, 1.
\newblock \doi{10.48550/arXiv.astro-ph/9409067}
\bibAnnoteFile{Comastri95}

\bibitem[{{Connors} et~al.(2023){Connors}, {Tomsick}, {Draghis}, {Coughenour},
  {Shaw}, {Garc\'ia} et~al.}]{Connors23}
{Connors}, R.~M.~T., {Tomsick}, J.~A., {Draghis}, P., {Coughenour}, B., {Shaw},
  A., {Garc\'ia}, J.~A., et~al. (2023).
\newblock {submitted}.
\newblock \emph{Frontiers in Astronomy and Space Sciences}
\bibAnnoteFile{Connors23}

\bibitem[{Da~Silva et~al.(2021)Da~Silva, Ricci, Sani, Koss, Trakhtenbrot,
  Lamperti et~al.}]{DaSilva21}
Da~Silva, R.~L., Ricci, C., Sani, E., Koss, M.~J., Trakhtenbrot, B., Lamperti,
  I., et~al. (2021).
\newblock A catalog of low-mass black holes in active galactic nuclei. iv. the
  catalog content and properties, and constraints on the host galaxies.
\newblock \emph{The Astrophysical Journal} 911, 150.
\newblock \doi{10.3847/1538-4357/abec74}
\bibAnnoteFile{DaSilva21}

\bibitem[{{Dauser} et~al.(2019){Dauser}, {Falkner}, {Lorenz}, {Kirsch},
  {Peille}, {Cucchetti} et~al.}]{Dauser19}
{Dauser}, T., {Falkner}, S., {Lorenz}, M., {Kirsch}, C., {Peille}, P.,
  {Cucchetti}, E., et~al. (2019).
\newblock {SIXTE: a generic X-ray instrument simulation toolkit}.
\newblock \emph{\aap} 630, A66.
\newblock \doi{10.1051/0004-6361/201935978}
\bibAnnoteFile{Dauser19}

\bibitem[{{Davis} et~al.(2014){Davis}, {Berrier}, {Johns}, {Shields},
  {Hartley}, {Kennefick} et~al.}]{Davis14}
{Davis}, B.~L., {Berrier}, J.~C., {Johns}, L., {Shields}, D.~W., {Hartley},
  M.~T., {Kennefick}, D., et~al. (2014).
\newblock {The Black Hole Mass Function Derived from Local Spiral Galaxies}.
\newblock \emph{\apj} 789, 124.
\newblock \doi{10.1088/0004-637X/789/2/124}
\bibAnnoteFile{Davis14}

\bibitem[{{Diaz} et~al.(2020){Diaz}, {Ar{\'e}valo},
  {Hern{\'a}ndez-Garc{\'\i}a}, {Bassani}, {Malizia}, {Gonz{\'a}lez-Mart{\'\i}n}
  et~al.}]{Diaz20}
{Diaz}, Y., {Ar{\'e}valo}, P., {Hern{\'a}ndez-Garc{\'\i}a}, L., {Bassani}, L.,
  {Malizia}, A., {Gonz{\'a}lez-Mart{\'\i}n}, O., et~al. (2020).
\newblock {Constraining X-ray reflection in the low-luminosity AGN NGC 3718
  using NuSTAR and XMM-Newton}.
\newblock \emph{\mnras} 496, 5399--5413.
\newblock \doi{10.1093/mnras/staa1762}
\bibAnnoteFile{Diaz20}

\bibitem[{{Diaz} et~al.(2023){Diaz}, {Hern{\`a}ndez-Garc{\'\i}a},
  {Ar{\'e}valo}, {L{\'o}pez-Navas}, {Ricci}, {Koss} et~al.}]{Diaz23}
{Diaz}, Y., {Hern{\`a}ndez-Garc{\'\i}a}, L., {Ar{\'e}valo}, P.,
  {L{\'o}pez-Navas}, E., {Ricci}, C., {Koss}, M., et~al. (2023).
\newblock {Constraining the X-ray reflection in low accretion-rate active
  galactic nuclei using XMM-Newton, NuSTAR, and Swift}.
\newblock \emph{\aap} 669, A114.
\newblock \doi{10.1051/0004-6361/202244678}
\bibAnnoteFile{Diaz23}

\bibitem[{{Earnshaw} et~al.(2019){Earnshaw}, {Roberts}, {Middleton}, {Walton},
  and {Mateos}}]{Earnshaw19}
{Earnshaw}, H.~P., {Roberts}, T.~P., {Middleton}, M.~J., {Walton}, D.~J., and
  {Mateos}, S. (2019).
\newblock {A new, clean catalogue of extragalactic non-nuclear X-ray sources in
  nearby galaxies}.
\newblock \emph{\mnras} 483, 5554--5573.
\newblock \doi{10.1093/mnras/sty3403}
\bibAnnoteFile{Earnshaw19}

\bibitem[{Elitzur(2006)}]{Elitzur06}
Elitzur, M. (2006).
\newblock The physics and diagnostic power of agn infrared emission.
\newblock \emph{New Astronomy Reviews} 50, 728--733.
\newblock \doi{10.1016/j.newar.2006.06.034}
\bibAnnoteFile{Elitzur06}

\bibitem[{Elitzur and Ho(2009)}]{Elitzur09}
Elitzur, M. and Ho, L.~C. (2009).
\newblock The disk-wind connection in agn: Us 322, an interesting case.
\newblock \emph{The Astrophysical Journal} 701, L91--L94.
\newblock \doi{10.1088/0004-637X/701/2/L91}
\bibAnnoteFile{Elitzur09}

\bibitem[{{Elvis} et~al.(2004){Elvis}, {Risaliti}, {Nicastro}, {Miller},
  {Fiore}, and {Puccetti}}]{Elvis04}
{Elvis}, M., {Risaliti}, G., {Nicastro}, F., {Miller}, J.~M., {Fiore}, F., and
  {Puccetti}, S. (2004).
\newblock {An Unveiling Event in the Type 2 Active Galactic Nucleus NGC 4388:A
  Challenge for a Parsec-Scale Absorber}.
\newblock \emph{\apjl} 615, L25--L28.
\newblock \doi{10.1086/424380}
\bibAnnoteFile{Elvis04}

\bibitem[{Eracleous et~al.(2010)Eracleous, Hwang, and Flohic}]{Eracleous10}
Eracleous, M., Hwang, J.~A., and Flohic, H. M. L.~G. (2010).
\newblock The narrow-line region of narrow-line and broad-line seyfert 1
  galaxies.
\newblock \emph{The Astrophysical Journal} 711, 796--811.
\newblock \doi{10.1088/0004-637X/711/2/796}
\bibAnnoteFile{Eracleous10}

\bibitem[{{Eraerds} et~al.(2021){Eraerds}, {Antonelli}, {Davis}, {Hall},
  {Hetherington}, {Holland} et~al.}]{Eraerds21}
{Eraerds}, T., {Antonelli}, V., {Davis}, C., {Hall}, D., {Hetherington}, O.,
  {Holland}, A., et~al. (2021).
\newblock {Enhanced simulations on the Athena/Wide Field Imager instrumental
  background}.
\newblock \emph{Journal of Astronomical Telescopes, Instruments, and Systems}
  7, 034001.
\newblock \doi{10.1117/1.JATIS.7.3.034001}
\bibAnnoteFile{Eraerds21}

\bibitem[{{Fabian}(1999)}]{Fabian99}
{Fabian}, A.~C. (1999).
\newblock {The obscured growth of massive black holes}.
\newblock \emph{\mnras} 308, L39--L43.
\newblock \doi{10.1046/j.1365-8711.1999.03017.x}
\bibAnnoteFile{Fabian99}

\bibitem[{{Fabian} et~al.(2008){Fabian}, {Vasudevan}, and {Gandhi}}]{Fabian08}
{Fabian}, A.~C., {Vasudevan}, R.~V., and {Gandhi}, P. (2008).
\newblock {The effect of radiation pressure on dusty absorbing gas around
  active galactic nuclei}.
\newblock \emph{\mnras} 385, L43--L47.
\newblock \doi{10.1111/j.1745-3933.2008.00430.x}
\bibAnnoteFile{Fabian08}

\bibitem[{{Falstad} et~al.(2021){Falstad}, {Aalto}, {K{\"o}nig}, {Onishi},
  {Muller}, {Gorski} et~al.}]{Falstad21}
{Falstad}, N., {Aalto}, S., {K{\"o}nig}, S., {Onishi}, K., {Muller}, S.,
  {Gorski}, M., et~al. (2021).
\newblock {CON-quest. Searching for the most obscured galaxy nuclei}.
\newblock \emph{\aap} 649, A105.
\newblock \doi{10.1051/0004-6361/202039291}
\bibAnnoteFile{Falstad21}

\bibitem[{{Farrah} et~al.(2016){Farrah}, {Balokovi{\'c}}, {Stern}, {Harris},
  {Kunimoto}, {Walton} et~al.}]{Farrah16}
{Farrah}, D., {Balokovi{\'c}}, M., {Stern}, D., {Harris}, K., {Kunimoto}, M.,
  {Walton}, D.~J., et~al. (2016).
\newblock {The Geometry of the Infrared and X-Ray Obscurer in a Dusty
  Hyperluminous Quasar}.
\newblock \emph{\apj} 831, 76.
\newblock \doi{10.3847/0004-637X/831/1/76}
\bibAnnoteFile{Farrah16}

\bibitem[{{Fern{\'a}ndez-Ontiveros} and
  {Mu{\~n}oz-Darias}(2021)}]{FernandezOntiveros21}
{Fern{\'a}ndez-Ontiveros}, J.~A. and {Mu{\~n}oz-Darias}, T. (2021).
\newblock {X-ray binary accretion states in active galactic nuclei? Sensing the
  accretion disc of supermassive black holes with mid-infrared nebular lines}.
\newblock \emph{\mnras} 504, 5726--5740.
\newblock \doi{10.1093/mnras/stab1108}
\bibAnnoteFile{FernandezOntiveros21}

\bibitem[{{Feroz} et~al.(2009){Feroz}, {Hobson}, and {Bridges}}]{Feroz09}
{Feroz}, F., {Hobson}, M.~P., and {Bridges}, M. (2009).
\newblock {MULTINEST: an efficient and robust Bayesian inference tool for
  cosmology and particle physics}.
\newblock \emph{\mnras} 398, 1601--1614.
\newblock \doi{10.1111/j.1365-2966.2009.14548.x}
\bibAnnoteFile{Feroz09}

\bibitem[{{Fruscione} et~al.(2006){Fruscione}, {McDowell}, {Allen},
  {Brickhouse}, {Burke}, {Davis} et~al.}]{Fruscione06}
{Fruscione}, A., {McDowell}, J.~C., {Allen}, G.~E., {Brickhouse}, N.~S.,
  {Burke}, D.~J., {Davis}, J.~E., et~al. (2006).
\newblock {CIAO: Chandra's data analysis system}.
\newblock In \emph{Society of Photo-Optical Instrumentation Engineers (SPIE)
  Conference Series}, eds. D.~R. {Silva} and R.~E. {Doxsey}. vol. 6270 of
  \emph{Society of Photo-Optical Instrumentation Engineers (SPIE) Conference
  Series}, 62701V.
\newblock \doi{10.1117/12.671760}
\bibAnnoteFile{Fruscione06}

\bibitem[{{Gandhi} et~al.(2017){Gandhi}, {Annuar}, {Lansbury}, {Stern},
  {Alexander}, {Bauer} et~al.}]{Gandhi17}
{Gandhi}, P., {Annuar}, A., {Lansbury}, G.~B., {Stern}, D., {Alexander}, D.~M.,
  {Bauer}, F.~E., et~al. (2017).
\newblock {The weak Fe fluorescence line and long-term X-ray evolution of the
  Compton-thick active galactic nucleus in NGC 7674}.
\newblock \emph{\mnras} 467, 4606--4621.
\newblock \doi{10.1093/mnras/stx357}
\bibAnnoteFile{Gandhi17}

\bibitem[{{Gandhi} and {Fabian}(2003)}]{Gandhi03}
{Gandhi}, P. and {Fabian}, A.~C. (2003).
\newblock {X-ray background synthesis: the infrared connection}.
\newblock \emph{\mnras} 339, 1095--1102.
\newblock \doi{10.1046/j.1365-8711.2003.06259.x}
\bibAnnoteFile{Gandhi03}

\bibitem[{{Gandhi} et~al.(2007){Gandhi}, {Fabian}, {Suebsuwong}, {Malzac},
  {Miniutti}, and {Wilman}}]{Gandhi07}
{Gandhi}, P., {Fabian}, A.~C., {Suebsuwong}, T., {Malzac}, J., {Miniutti}, G.,
  and {Wilman}, R.~J. (2007).
\newblock {Constraints on light bending and reflection from the hard X-ray
  background}.
\newblock \emph{\mnras} 382, 1005--1018.
\newblock \doi{10.1111/j.1365-2966.2007.12462.x}
\bibAnnoteFile{Gandhi07}

\bibitem[{{Gandhi} et~al.(2014){Gandhi}, {Lansbury}, {Alexander}, {Stern},
  {Ar{\'e}valo}, {Ballantyne} et~al.}]{Gandhi14}
{Gandhi}, P., {Lansbury}, G.~B., {Alexander}, D.~M., {Stern}, D.,
  {Ar{\'e}valo}, P., {Ballantyne}, D.~R., et~al. (2014).
\newblock {NuSTAR Unveils a Compton-thick Type 2 Quasar in Mrk 34}.
\newblock \emph{\apj} 792, 117.
\newblock \doi{10.1088/0004-637X/792/2/117}
\bibAnnoteFile{Gandhi14}

\bibitem[{{Gandhi} et~al.(2013){Gandhi}, {Terashima}, {Yamada}, {Mushotzky},
  {Ueda}, {Baumgartner} et~al.}]{Gandhi13}
{Gandhi}, P., {Terashima}, Y., {Yamada}, S., {Mushotzky}, R.~F., {Ueda}, Y.,
  {Baumgartner}, W.~H., et~al. (2013).
\newblock {Reflection-dominated Nuclear X-Ray Emission in the Early-type Galaxy
  ESO 565-G019}.
\newblock \emph{\apj} 773, 51.
\newblock \doi{10.1088/0004-637X/773/1/51}
\bibAnnoteFile{Gandhi13}

\bibitem[{{Gandhi} et~al.(2015){Gandhi}, {Yamada}, {Ricci}, {Asmus},
  {Mushotzky}, {Ueda} et~al.}]{Gandhi15}
{Gandhi}, P., {Yamada}, S., {Ricci}, C., {Asmus}, D., {Mushotzky}, R.~F.,
  {Ueda}, Y., et~al. (2015).
\newblock {A Compton-thick AGN in the barred spiral galaxy NGC 4785}.
\newblock \emph{\mnras} 449, 1845--1855.
\newblock \doi{10.1093/mnras/stv344}
\bibAnnoteFile{Gandhi15}

\bibitem[{{Gaspari} et~al.(2015){Gaspari}, {Brighenti}, and {Temi}}]{Gaspari15}
{Gaspari}, M., {Brighenti}, F., and {Temi}, P. (2015).
\newblock {Chaotic cold accretion on to black holes in rotating atmospheres}.
\newblock \emph{\aap} 579, A62.
\newblock \doi{10.1051/0004-6361/201526151}
\bibAnnoteFile{Gaspari15}

\bibitem[{{Gaspari} et~al.(2013){Gaspari}, {Ruszkowski}, and {Oh}}]{Gaspari13}
{Gaspari}, M., {Ruszkowski}, M., and {Oh}, S.~P. (2013).
\newblock {Chaotic cold accretion on to black holes}.
\newblock \emph{\mnras} 432, 3401--3422.
\newblock \doi{10.1093/mnras/stt692}
\bibAnnoteFile{Gaspari13}

\bibitem[{{Gaspari} et~al.(2020){Gaspari}, {Tombesi}, and {Cappi}}]{Gaspari20}
{Gaspari}, M., {Tombesi}, F., and {Cappi}, M. (2020).
\newblock {Linking macro-, meso- and microscales in multiphase AGN feeding and
  feedback}.
\newblock \emph{Nature Astronomy} 4, 10--13.
\newblock \doi{10.1038/s41550-019-0970-1}
\bibAnnoteFile{Gaspari20}

\bibitem[{{Gilli} et~al.(2007){Gilli}, {Comastri}, and {Hasinger}}]{Gilli07}
{Gilli}, R., {Comastri}, A., and {Hasinger}, G. (2007).
\newblock {The synthesis of the cosmic X-ray background in the Chandra and
  XMM-Newton era}.
\newblock \emph{\aap} 463, 79--96.
\newblock \doi{10.1051/0004-6361:20066334}
\bibAnnoteFile{Gilli07}

\bibitem[{{Gilli} et~al.(2022){Gilli}, {Norman}, {Calura}, {Vito}, {Decarli},
  {Marchesi} et~al.}]{Gilli22}
{Gilli}, R., {Norman}, C., {Calura}, F., {Vito}, F., {Decarli}, R., {Marchesi},
  S., et~al. (2022).
\newblock {Supermassive black holes at high redshift are expected to be
  obscured by their massive host galaxies' interstellar medium}.
\newblock \emph{\aap} 666, A17.
\newblock \doi{10.1051/0004-6361/202243708}
\bibAnnoteFile{Gilli22}

\bibitem[{{Giman} et~al.(2023){Giman}, {Boorman}, {Harrison}, {Stern}, and
  Balokovi\'{c}}]{Giman23}
{Giman}, A., {Boorman}, P., {Harrison}, F., {Stern}, D., and Balokovi\'{c}, M.
  (2023).
\newblock {in prep.}
\newblock \emph{\apj}
\bibAnnoteFile{Giman23}

\bibitem[{González-Martín et~al.(2017)González-Martín, Hernández-García,
  Masegosa, Marquez, and Esquej}]{GonzalezMartin17}
González-Martín, O., Hernández-García, L., Masegosa, J., Marquez, I., and
  Esquej, P. (2017).
\newblock A universal scaling for the energetics of relativistic jets from
  black hole systems.
\newblock \emph{The Astrophysical Journal} 835, 16.
\newblock \doi{10.3847/1538-4357/835/1/16}
\bibAnnoteFile{GonzalezMartin17}

\bibitem[{González-Martín et~al.(2009)González-Martín, Masegosa, Márquez,
  Guerrero, and Dultzin-Hacyan}]{GonzalezMartin09}
González-Martín, O., Masegosa, J., Márquez, I., Guerrero, M.~A., and
  Dultzin-Hacyan, D. (2009).
\newblock Unification of x-ray winds in seyfert galaxies: from ultra-fast
  outflows to warm absorbers.
\newblock \emph{Monthly Notices of the Royal Astronomical Society} 397,
  L79--L83.
\newblock \doi{10.1111/j.1745-3933.2009.00757.x}
\bibAnnoteFile{GonzalezMartin09}

\bibitem[{{Goodman} et~al.(2009){Goodman}, {Pineda}, and {Schnee}}]{Goodman09}
{Goodman}, A.~A., {Pineda}, J.~E., and {Schnee}, S.~L. (2009).
\newblock {The ``True'' Column Density Distribution in Star-Forming Molecular
  Clouds}.
\newblock \emph{\apj} 692, 91--103.
\newblock \doi{10.1088/0004-637X/692/1/91}
\bibAnnoteFile{Goodman09}

\bibitem[{{Greene}(2012)}]{Greene12}
{Greene}, J.~E. (2012).
\newblock {Low-mass black holes as the remnants of primordial black hole
  formation}.
\newblock \emph{Nature Communications} 3, 1304.
\newblock \doi{10.1038/ncomms2314}
\bibAnnoteFile{Greene12}

\bibitem[{{Greene} and {Ho}(2004)}]{Greene04}
{Greene}, J.~E. and {Ho}, L.~C. (2004).
\newblock {Active Galactic Nuclei with Candidate Intermediate-Mass Black
  Holes}.
\newblock \emph{\apj} 610, 722--736.
\newblock \doi{10.1086/421719}
\bibAnnoteFile{Greene04}

\bibitem[{{Greene} and {Ho}(2007)}]{Greene07}
{Greene}, J.~E. and {Ho}, L.~C. (2007).
\newblock {A New Sample of Low-Mass Black Holes in Active Galaxies}.
\newblock \emph{\apj} 670, 92--104.
\newblock \doi{10.1086/522082}
\bibAnnoteFile{Greene07}

\bibitem[{{Greene} et~al.(2020){Greene}, {Strader}, and {Ho}}]{Greene20}
{Greene}, J.~E., {Strader}, J., and {Ho}, L.~C. (2020).
\newblock {Intermediate-Mass Black Holes}.
\newblock \emph{\araa} 58, 257--312.
\newblock \doi{10.1146/annurev-astro-032620-021835}
\bibAnnoteFile{Greene20}

\bibitem[{{Greenhill} et~al.(2008){Greenhill}, {Tilak}, and
  {Madejski}}]{Greenhill08}
{Greenhill}, L.~J., {Tilak}, A., and {Madejski}, G. (2008).
\newblock {Prevalence of High X-Ray Obscuring Columns among AGNs that Host
  H$_{2}$O Masers}.
\newblock \emph{\apjl} 686, L13.
\newblock \doi{10.1086/592782}
\bibAnnoteFile{Greenhill08}

\bibitem[{{Greenwell} et~al.(2022){Greenwell}, {Gandhi}, {Lansbury}, {Boorman},
  {Mainieri}, and {Stern}}]{Greenwell22}
{Greenwell}, C., {Gandhi}, P., {Lansbury}, G., {Boorman}, P., {Mainieri}, V.,
  and {Stern}, D. (2022).
\newblock {XMM and NuSTAR Observations of an Optically Quiescent Quasar}.
\newblock \emph{\apjl} 934, L34.
\newblock \doi{10.3847/2041-8213/ac83a0}
\bibAnnoteFile{Greenwell22}

\bibitem[{{Gupta} et~al.(2021){Gupta}, {Ricci}, {Tortosa}, {Ueda}, {Kawamuro},
  {Koss} et~al.}]{Gupta21}
{Gupta}, K.~K., {Ricci}, C., {Tortosa}, A., {Ueda}, Y., {Kawamuro}, T., {Koss},
  M., et~al. (2021).
\newblock {BAT AGN Spectroscopic Survey XXVII: scattered X-Ray radiation in
  obscured active galactic nuclei}.
\newblock \emph{\mnras} 504, 428--443.
\newblock \doi{10.1093/mnras/stab839}
\bibAnnoteFile{Gupta21}

\bibitem[{{Harrison} et~al.(2013){Harrison}, {Craig}, {Christensen}, {Hailey},
  {Zhang}, {Boggs} et~al.}]{Harrison13}
{Harrison}, F.~A., {Craig}, W.~W., {Christensen}, F.~E., {Hailey}, C.~J.,
  {Zhang}, W.~W., {Boggs}, S.~E., et~al. (2013).
\newblock {The Nuclear Spectroscopic Telescope Array (NuSTAR) High-energy X-Ray
  Mission}.
\newblock \emph{\apj} 770, 103.
\newblock \doi{10.1088/0004-637X/770/2/103}
\bibAnnoteFile{Harrison13}

\bibitem[{Hernández-García et~al.(2016)Hernández-García, Masegosa,
  González-Martín, and Márquez}]{HernandezGarcia16}
Hernández-García, L., Masegosa, J., González-Martín, O., and Márquez, I.
  (2016).
\newblock Subarcsecond imaging of the water maser in ngc 1068 with kvn and
  vera.
\newblock \emph{The Astrophysical Journal} 824, L27.
\newblock \doi{10.3847/2041-8205/824/2/L27}
\bibAnnoteFile{HernandezGarcia16}

\bibitem[{{Hickox} and {Alexander}(2018)}]{Hickox18}
{Hickox}, R.~C. and {Alexander}, D.~M. (2018).
\newblock {Obscured Active Galactic Nuclei}.
\newblock \emph{\araa} 56, 625--671.
\newblock \doi{10.1146/annurev-astro-081817-051803}
\bibAnnoteFile{Hickox18}

\bibitem[{Ho(1997)}]{Ho97}
Ho, L.~C. (1997).
\newblock Nuclear activity in nearby galaxies.
\newblock \emph{Revista Mexicana de Astronomía y Astrofísica} 6, 55--63
\bibAnnoteFile{Ho97}

\bibitem[{Ho(1999)}]{Ho99}
Ho, L.~C. (1999).
\newblock Narrow-line seyfert 1 galaxies and their place in the universe.
\newblock \emph{The Astrophysical Journal Supplement Series} 120, 73--112.
\newblock \doi{10.1086/313181}
\bibAnnoteFile{Ho99}

\bibitem[{Ho(2008)}]{Ho08}
Ho, L.~C. (2008).
\newblock Narrow-line seyfert 1 galaxies.
\newblock \emph{Annual Review of Astronomy and Astrophysics} 46, 475--539.
\newblock \doi{10.1146/annurev.astro.45.051806.110546}
\bibAnnoteFile{Ho08}

\bibitem[{{Hopkins} et~al.(2006){Hopkins}, {Hernquist}, {Cox}, {Di Matteo},
  {Robertson}, and {Springel}}]{Hopkins06}
{Hopkins}, P.~F., {Hernquist}, L., {Cox}, T.~J., {Di Matteo}, T., {Robertson},
  B., and {Springel}, V. (2006).
\newblock {A Unified, Merger-driven Model of the Origin of Starbursts, Quasars,
  the Cosmic X-Ray Background, Supermassive Black Holes, and Galaxy Spheroids}.
\newblock \emph{\apjs} 163, 1--49.
\newblock \doi{10.1086/499298}
\bibAnnoteFile{Hopkins06}

\bibitem[{Hunter(2007)}]{Hunter07}
Hunter, J.~D. (2007).
\newblock Matplotlib: A 2d graphics environment.
\newblock \emph{Computing in Science \& Engineering} 9, 90--95.
\newblock \doi{10.1109/MCSE.2007.55}
\bibAnnoteFile{Hunter07}

\bibitem[{Hönig and Beckert(2007)}]{Honig07}
Hönig, S.~F. and Beckert, T. (2007).
\newblock The unified model of active galactic nuclei.
\newblock \emph{The Astrophysical Journal} 746, 214--234.
\newblock \doi{10.1088/0004-637X/746/2/214}
\bibAnnoteFile{Honig07}

\bibitem[{{Ikeda} et~al.(2009){Ikeda}, {Awaki}, and {Terashima}}]{Ikeda09}
{Ikeda}, S., {Awaki}, H., and {Terashima}, Y. (2009).
\newblock {Study on X-Ray Spectra of Obscured Active Galactic Nuclei Based on
  Monte Carlo Simulation{\textemdash}An Interpretation of Observed Wide-Band
  Spectra}.
\newblock \emph{\apj} 692, 608--617.
\newblock \doi{10.1088/0004-637X/692/1/608}
\bibAnnoteFile{Ikeda09}

\bibitem[{{Iwasawa} et~al.(2011){Iwasawa}, {Sanders}, {Teng}, {U}, {Armus},
  {Evans} et~al.}]{Iwasawa11}
{Iwasawa}, K., {Sanders}, D.~B., {Teng}, S.~H., {U}, V., {Armus}, L., {Evans},
  A.~S., et~al. (2011).
\newblock {C-GOALS: Chandra observations of a complete sample of luminous
  infrared galaxies from the IRAS Revised Bright Galaxy Survey}.
\newblock \emph{\aap} 529, A106.
\newblock \doi{10.1051/0004-6361/201015264}
\bibAnnoteFile{Iwasawa11}

\bibitem[{{Jaffe} et~al.(2004){Jaffe}, {Meisenheimer}, {R{\"o}ttgering},
  {Leinert}, {Richichi}, {Chesneau} et~al.}]{Jaffe04}
{Jaffe}, W., {Meisenheimer}, K., {R{\"o}ttgering}, H.~J.~A., {Leinert}, C.,
  {Richichi}, A., {Chesneau}, O., et~al. (2004).
\newblock {The central dusty torus in the active nucleus of NGC 1068}.
\newblock \emph{\nat} 429, 47--49.
\newblock \doi{10.1038/nature02531}
\bibAnnoteFile{Jaffe04}

\bibitem[{{Jansen} et~al.(2001){Jansen}, {Lumb}, {Altieri}, {Clavel}, {Ehle},
  {Erd} et~al.}]{Jansen01}
{Jansen}, F., {Lumb}, D., {Altieri}, B., {Clavel}, J., {Ehle}, M., {Erd}, C.,
  et~al. (2001).
\newblock {XMM-Newton observatory. I. The spacecraft and operations}.
\newblock \emph{\aap} 365, L1--L6.
\newblock \doi{10.1051/0004-6361:20000036}
\bibAnnoteFile{Jansen01}

\bibitem[{{Kaastra} and {Bleeker}(2016)}]{Kaastra16}
{Kaastra}, J.~S. and {Bleeker}, J.~A.~M. (2016).
\newblock {Optimal binning of X-ray spectra and response matrix design}.
\newblock \emph{\aap} 587, A151.
\newblock \doi{10.1051/0004-6361/201527395}
\bibAnnoteFile{Kaastra16}

\bibitem[{{Kallov\'{a}} et~al.(2023){Kallov\'{a}}, {Boorman}, and
  {Ricci}}]{Kallova23}
{Kallov\'{a}}, K., {Boorman}, P., and {Ricci}, C. (2023).
\newblock {in prep.}
\newblock \emph{\apj}
\bibAnnoteFile{Kallova23}

\bibitem[{{Kammoun} et~al.(2023){Kammoun}, {Lohfink}, {Masterson}, {Wilkins},
  {Zhao}, {Balokovi\'{c}} et~al.}]{Kammoun23}
{Kammoun}, E., {Lohfink}, A.~M., {Masterson}, M., {Wilkins}, D.~R., {Zhao}, X.,
  {Balokovi\'{c}}, M., et~al. (2023).
\newblock {submitted}.
\newblock \emph{Frontiers in Astronomy and Space Sciences}
\bibAnnoteFile{Kammoun23}

\bibitem[{{Kammoun} et~al.(2020){Kammoun}, {Miller}, {Koss}, {Oh}, {Zoghbi},
  {Mushotzky} et~al.}]{Kammoun20}
{Kammoun}, E.~S., {Miller}, J.~M., {Koss}, M., {Oh}, K., {Zoghbi}, A.,
  {Mushotzky}, R.~F., et~al. (2020).
\newblock {A Hard Look at Local, Optically Selected, Obscured Seyfert
  Galaxies}.
\newblock \emph{\apj} 901, 161.
\newblock \doi{10.3847/1538-4357/abb29f}
\bibAnnoteFile{Kammoun20}

\bibitem[{Kawamuro et~al.(2016)Kawamuro, Ueda, Tazaki, Terashima, Mushotzky,
  Mori et~al.}]{Kawamuro16}
Kawamuro, T., Ueda, Y., Tazaki, F., Terashima, Y., Mushotzky, R., Mori, M.,
  et~al. (2016).
\newblock Low-luminosity active galactic nuclei as observed with suzaku.
\newblock \emph{The Astrophysical Journal} 831, 37.
\newblock \doi{10.3847/0004-637X/831/1/37}
\bibAnnoteFile{Kawamuro16}

\bibitem[{{Kayal} et~al.(2023){Kayal}, {Singh}, {Ricci}, {Mithun}, {Vadawale},
  {Dewangan} et~al.}]{Kayal23}
{Kayal}, A., {Singh}, V., {Ricci}, C., {Mithun}, N.~P.~S., {Vadawale}, S.,
  {Dewangan}, G., et~al. (2023).
\newblock {Multi-epoch hard X-ray view of Compton-thick AGN Circinus Galaxy}.
\newblock \emph{\mnras} 522, 4098--4115.
\newblock \doi{10.1093/mnras/stad1216}
\bibAnnoteFile{Kayal23}

\bibitem[{{Kerzendorf} et~al.(2021){Kerzendorf}, {Vogl}, {Buchner}, {Contardo},
  {Williamson}, and {van der Smagt}}]{Kerzendorf21}
{Kerzendorf}, W.~E., {Vogl}, C., {Buchner}, J., {Contardo}, G., {Williamson},
  M., and {van der Smagt}, P. (2021).
\newblock {Dalek: A Deep Learning Emulator for TARDIS}.
\newblock \emph{\apjl} 910, L23.
\newblock \doi{10.3847/2041-8213/abeb1b}
\bibAnnoteFile{Kerzendorf21}

\bibitem[{{Kondratko} et~al.(2008){Kondratko}, {Greenhill}, and
  {Moran}}]{Kondratko08}
{Kondratko}, P.~T., {Greenhill}, L.~J., and {Moran}, J.~M. (2008).
\newblock {The Parsec-Scale Accretion Disk in NGC 3393}.
\newblock \emph{\apj} 678, 87--95.
\newblock \doi{10.1086/586879}
\bibAnnoteFile{Kondratko08}

\bibitem[{{K{\"o}rding} et~al.(2006){K{\"o}rding}, {Jester}, and
  {Fender}}]{Kording06}
{K{\"o}rding}, E.~G., {Jester}, S., and {Fender}, R. (2006).
\newblock {Accretion states and radio loudness in active galactic nuclei:
  analogies with X-ray binaries}.
\newblock \emph{\mnras} 372, 1366--1378.
\newblock \doi{10.1111/j.1365-2966.2006.10954.x}
\bibAnnoteFile{Kording06}

\bibitem[{{Kuo} et~al.(2011){Kuo}, {Braatz}, {Condon}, {Impellizzeri}, {Lo},
  {Zaw} et~al.}]{Kuo11}
{Kuo}, C.~Y., {Braatz}, J.~A., {Condon}, J.~J., {Impellizzeri}, C.~M.~V., {Lo},
  K.~Y., {Zaw}, I., et~al. (2011).
\newblock {The Megamaser Cosmology Project. III. Accurate Masses of Seven
  Supermassive Black Holes in Active Galaxies with Circumnuclear Megamaser
  Disks}.
\newblock \emph{\apj} 727, 20.
\newblock \doi{10.1088/0004-637X/727/1/20}
\bibAnnoteFile{Kuo11}

\bibitem[{{Laha} et~al.(2020){Laha}, {Markowitz}, {Krumpe}, {Nikutta},
  {Rothschild}, and {Saha}}]{Laha20}
{Laha}, S., {Markowitz}, A.~G., {Krumpe}, M., {Nikutta}, R., {Rothschild}, R.,
  and {Saha}, T. (2020).
\newblock {The Variable and Non-variable X-Ray Absorbers in Compton-thin Type
  II Active Galactic Nuclei}.
\newblock \emph{\apj} 897, 66.
\newblock \doi{10.3847/1538-4357/ab92ab}
\bibAnnoteFile{Laha20}

\bibitem[{{LaMassa} et~al.(2019){LaMassa}, {Yaqoob}, {Boorman}, {Tzanavaris},
  {Levenson}, {Gandhi} et~al.}]{Lamassa19}
{LaMassa}, S.~M., {Yaqoob}, T., {Boorman}, P.~G., {Tzanavaris}, P., {Levenson},
  N.~A., {Gandhi}, P., et~al. (2019).
\newblock {NuSTAR Uncovers an Extremely Local Compton-thick AGN in NGC 4968}.
\newblock \emph{\apj} 887, 173.
\newblock \doi{10.3847/1538-4357/ab552c}
\bibAnnoteFile{Lamassa19}

\bibitem[{{LaMassa} et~al.(2017){LaMassa}, {Yaqoob}, {Levenson}, {Boorman},
  {Heckman}, {Gandhi} et~al.}]{Lamassa17}
{LaMassa}, S.~M., {Yaqoob}, T., {Levenson}, N.~A., {Boorman}, P., {Heckman},
  T.~M., {Gandhi}, P., et~al. (2017).
\newblock {Chandra Reveals Heavy Obscuration and Circumnuclear Star Formation
  in Seyfert 2 Galaxy NGC 4968}.
\newblock \emph{\apj} 835, 91.
\newblock \doi{10.3847/1538-4357/835/1/91}
\bibAnnoteFile{Lamassa17}

\bibitem[{{LaMassa} et~al.(2023){LaMassa}, {Yaqoob}, {Tzanavaris}, {Gandhi},
  {Heckman}, {Lansbury} et~al.}]{Lamassa23}
{LaMassa}, S.~M., {Yaqoob}, T., {Tzanavaris}, P., {Gandhi}, P., {Heckman}, T.,
  {Lansbury}, G., et~al. (2023).
\newblock {The Complex X-Ray Obscuration Environment in the Radio-loud Type 2
  Quasar 3C 223}.
\newblock \emph{\apj} 944, 152.
\newblock \doi{10.3847/1538-4357/acb3bb}
\bibAnnoteFile{Lamassa23}

\bibitem[{{Lansbury} et~al.(2017){Lansbury}, {Stern}, {Aird}, {Alexander},
  {Fuentes}, {Harrison} et~al.}]{Lansbury17}
{Lansbury}, G.~B., {Stern}, D., {Aird}, J., {Alexander}, D.~M., {Fuentes}, C.,
  {Harrison}, F.~A., et~al. (2017).
\newblock {The NuSTAR Serendipitous Survey: The 40-month Catalog and the
  Properties of the Distant High-energy X-Ray Source Population}.
\newblock \emph{\apj} 836, 99.
\newblock \doi{10.3847/1538-4357/836/1/99}
\bibAnnoteFile{Lansbury17}

\bibitem[{{Lefkir} et~al.(2023){Lefkir}, {Kammoun}, {Barret}, {Boorman},
  {Matzeu}, {Miller} et~al.}]{Lefkir23}
{Lefkir}, M., {Kammoun}, E., {Barret}, D., {Boorman}, P., {Matzeu}, G.,
  {Miller}, J.~M., et~al. (2023).
\newblock {A hard look at the X-ray spectral variability of NGC 7582}.
\newblock \emph{\mnras} 522, 1169--1182.
\newblock \doi{10.1093/mnras/stad995}
\bibAnnoteFile{Lefkir23}

\bibitem[{{Lehmer} et~al.(2023){Lehmer}, {Garofali}, {Binder}, {Fornasini},
  {Vulic}, {Zezas} et~al.}]{Lehmer23}
{Lehmer}, B.~D., {Garofali}, K., {Binder}, B.~A., {Fornasini}, F., {Vulic}, N.,
  {Zezas}, A., et~al. (2023).
\newblock {submitted}.
\newblock \emph{Frontiers in Astronomy and Space Sciences}
\bibAnnoteFile{Lehmer23}

\bibitem[{{Levenson} et~al.(2002){Levenson}, {Krolik}, {{\.Z}ycki}, {Heckman},
  {Weaver}, {Awaki} et~al.}]{Levenson02}
{Levenson}, N.~A., {Krolik}, J.~H., {{\.Z}ycki}, P.~T., {Heckman}, T.~M.,
  {Weaver}, K.~A., {Awaki}, H., et~al. (2002).
\newblock {Extreme X-Ray Iron Lines in Active Galactic Nuclei}.
\newblock \emph{\apjl} 573, L81--L84.
\newblock \doi{10.1086/342092}
\bibAnnoteFile{Levenson02}

\bibitem[{{Liu} et~al.(2019){Liu}, {H{\"o}nig}, {Ricci}, and {Paltani}}]{Liu19}
{Liu}, J., {H{\"o}nig}, S.~F., {Ricci}, C., and {Paltani}, S. (2019).
\newblock {X-ray signatures of the polar dusty gas in AGN}.
\newblock \emph{\mnras} 490, 4344--4352.
\newblock \doi{10.1093/mnras/stz2908}
\bibAnnoteFile{Liu19}

\bibitem[{{Liu} and {Li}(2014)}]{Liu14}
{Liu}, Y. and {Li}, X. (2014).
\newblock {An X-Ray Spectral Model for Clumpy Tori in Active Galactic Nuclei}.
\newblock \emph{\apj} 787, 52.
\newblock \doi{10.1088/0004-637X/787/1/52}
\bibAnnoteFile{Liu14}

\bibitem[{{Maccagni} et~al.(2021){Maccagni}, {Serra}, {Gaspari}, {Kleiner},
  {Morokuma-Matsui}, {Oosterloo} et~al.}]{Maccagni21}
{Maccagni}, F.~M., {Serra}, P., {Gaspari}, M., {Kleiner}, D.,
  {Morokuma-Matsui}, K., {Oosterloo}, T.~A., et~al. (2021).
\newblock {AGN feeding and feedback in Fornax A. Kinematical analysis of the
  multi-phase ISM}.
\newblock \emph{\aap} 656, A45.
\newblock \doi{10.1051/0004-6361/202141143}
\bibAnnoteFile{Maccagni21}

\bibitem[{{Madsen} et~al.(2023){Madsen}, {Garc\'{i}a}, {Grefenstette}, {Stern},
  and {The HEX-P Team}}]{Madsen23}
{Madsen}, K., {Garc\'{i}a}, J., {Grefenstette}, B., {Stern}, D., and {The HEX-P
  Team} (2023).
\newblock {in prep.}
\newblock \emph{Frontiers in Astronomy and Space Sciences}
\bibAnnoteFile{Madsen23}

\bibitem[{{Madsen} et~al.(2015){Madsen}, {Harrison}, {Markwardt}, {An},
  {Grefenstette}, {Bachetti} et~al.}]{Madsen15}
{Madsen}, K.~K., {Harrison}, F.~A., {Markwardt}, C.~B., {An}, H.,
  {Grefenstette}, B.~W., {Bachetti}, M., et~al. (2015).
\newblock {Calibration of the NuSTAR High-energy Focusing X-ray Telescope.}
\newblock \emph{\apjs} 220, 8.
\newblock \doi{10.1088/0067-0049/220/1/8}
\bibAnnoteFile{Madsen15}

\bibitem[{Maoz et~al.(2005)Maoz, Nagar, Falcke, and Wilson}]{Maoz05}
Maoz, D., Nagar, N.~M., Falcke, H., and Wilson, A.~S. (2005).
\newblock Truncated accretion disks in active galactic nuclei.
\newblock \emph{The Astrophysical Journal} 625, 699--707.
\newblock \doi{10.1086/429625}
\bibAnnoteFile{Maoz05}

\bibitem[{{Marchesi} et~al.(2018){Marchesi}, {Ajello}, {Marcotulli},
  {Comastri}, {Lanzuisi}, and {Vignali}}]{Marchesi18}
{Marchesi}, S., {Ajello}, M., {Marcotulli}, L., {Comastri}, A., {Lanzuisi}, G.,
  and {Vignali}, C. (2018).
\newblock {Compton-thick AGNs in the NuSTAR Era}.
\newblock \emph{\apj} 854, 49.
\newblock \doi{10.3847/1538-4357/aaa410}
\bibAnnoteFile{Marchesi18}

\bibitem[{{Marchesi} et~al.(2019{\natexlab{a}}){Marchesi}, {Ajello}, {Zhao},
  {Comastri}, {La Parola}, and {Segreto}}]{Marchesi19b}
{Marchesi}, S., {Ajello}, M., {Zhao}, X., {Comastri}, A., {La Parola}, V., and
  {Segreto}, A. (2019{\natexlab{a}}).
\newblock {Compton-thick AGNs in the NuSTAR Era. V. Joint NuSTAR and XMM-Newton
  Spectral Analysis of Three {\textquotedblleft}Soft-gamma{\textquotedblright}
  Candidate CT-AGNs in the Swift/BAT 100-month Catalog}.
\newblock \emph{\apj} 882, 162.
\newblock \doi{10.3847/1538-4357/ab340a}
\bibAnnoteFile{Marchesi19b}

\bibitem[{{Marchesi} et~al.(2019{\natexlab{b}}){Marchesi}, {Ajello}, {Zhao},
  {Marcotulli}, {Balokovi{\'c}}, {Brightman} et~al.}]{Marchesi19a}
{Marchesi}, S., {Ajello}, M., {Zhao}, X., {Marcotulli}, L., {Balokovi{\'c}},
  M., {Brightman}, M., et~al. (2019{\natexlab{b}}).
\newblock {Compton-thick AGNs in the NuSTAR Era. III. A Systematic Study of the
  Torus Covering Factor}.
\newblock \emph{\apj} 872, 8.
\newblock \doi{10.3847/1538-4357/aafbeb}
\bibAnnoteFile{Marchesi19a}

\bibitem[{{Marchesi} et~al.(2017){Marchesi}, {Tremblay}, {Ajello},
  {Marcotulli}, {Paggi}, {Cusumano} et~al.}]{Marchesi17}
{Marchesi}, S., {Tremblay}, L., {Ajello}, M., {Marcotulli}, L., {Paggi}, A.,
  {Cusumano}, G., et~al. (2017).
\newblock {Chandra and NuSTAR Follow-up Observations of Swift-BAT-selected
  AGNs}.
\newblock \emph{\apj} 848, 53.
\newblock \doi{10.3847/1538-4357/aa8ee6}
\bibAnnoteFile{Marchesi17}

\bibitem[{{Marchesi} et~al.(2022){Marchesi}, {Zhao}, {Torres-Alb{\`a}},
  {Ajello}, {Gaspari}, {Pizzetti} et~al.}]{Marchesi22}
{Marchesi}, S., {Zhao}, X., {Torres-Alb{\`a}}, N., {Ajello}, M., {Gaspari}, M.,
  {Pizzetti}, A., et~al. (2022).
\newblock {Compton-thick AGN in the NuSTAR Era. VIII. A joint NuSTAR-XMM-Newton
  Monitoring of the Changing-look Compton-thick AGN NGC 1358}.
\newblock \emph{\apj} 935, 114.
\newblock \doi{10.3847/1538-4357/ac80be}
\bibAnnoteFile{Marchesi22}

\bibitem[{{Marcotulli} et~al.(2023){Marcotulli}, {Ajello}, {Boettcher},
  {Coppi}, {Costamante}, {Di Gesu} et~al.}]{Marcotulli23}
{Marcotulli}, L., {Ajello}, M., {Boettcher}, M., {Coppi}, P., {Costamante}, L.,
  {Di Gesu}, L., et~al. (2023).
\newblock {submitted}.
\newblock \emph{Frontiers in Astronomy and Space Sciences}
\bibAnnoteFile{Marcotulli23}

\bibitem[{{Marinucci} et~al.(2016){Marinucci}, {Bianchi}, {Matt}, {Alexander},
  {Balokovi{\'c}}, {Bauer} et~al.}]{Marinucci16}
{Marinucci}, A., {Bianchi}, S., {Matt}, G., {Alexander}, D.~M.,
  {Balokovi{\'c}}, M., {Bauer}, F.~E., et~al. (2016).
\newblock {NuSTAR catches the unveiling nucleus of NGC 1068}.
\newblock \emph{\mnras} 456, L94--L98.
\newblock \doi{10.1093/mnrasl/slv178}
\bibAnnoteFile{Marinucci16}

\bibitem[{{Markowitz} et~al.(2014){Markowitz}, {Krumpe}, and
  {Nikutta}}]{Markowitz14}
{Markowitz}, A.~G., {Krumpe}, M., and {Nikutta}, R. (2014).
\newblock {First X-ray-based statistical tests for clumpy-torus models: eclipse
  events from 230 years of monitoring of Seyfert AGN}.
\newblock \emph{\mnras} 439, 1403--1458.
\newblock \doi{10.1093/mnras/stt2492}
\bibAnnoteFile{Markowitz14}

\bibitem[{{Masini} et~al.(2016){Masini}, {Comastri}, {Balokovi{\'c}}, {Zaw},
  {Puccetti}, {Ballantyne} et~al.}]{Masini16}
{Masini}, A., {Comastri}, A., {Balokovi{\'c}}, M., {Zaw}, I., {Puccetti}, S.,
  {Ballantyne}, D.~R., et~al. (2016).
\newblock {NuSTAR observations of water megamaser AGN}.
\newblock \emph{\aap} 589, A59.
\newblock \doi{10.1051/0004-6361/201527689}
\bibAnnoteFile{Masini16}

\bibitem[{{Masini} et~al.(2017){Masini}, {Comastri}, {Puccetti},
  {Balokovi{\'c}}, {Gandhi}, {Guainazzi} et~al.}]{Masini17}
{Masini}, A., {Comastri}, A., {Puccetti}, S., {Balokovi{\'c}}, M., {Gandhi},
  P., {Guainazzi}, M., et~al. (2017).
\newblock {The Phoenix galaxy as seen by NuSTAR}.
\newblock \emph{\aap} 597, A100.
\newblock \doi{10.1051/0004-6361/201629444}
\bibAnnoteFile{Masini17}

\bibitem[{{Matt} et~al.(2000){Matt}, {Fabian}, {Guainazzi}, {Iwasawa},
  {Bassani}, and {Malaguti}}]{Matt00}
{Matt}, G., {Fabian}, A.~C., {Guainazzi}, M., {Iwasawa}, K., {Bassani}, L., and
  {Malaguti}, G. (2000).
\newblock {The X-ray spectra of Compton-thick Seyfert 2 galaxies as seen by
  BeppoSAX}.
\newblock \emph{\mnras} 318, 173--179.
\newblock \doi{10.1046/j.1365-8711.2000.03721.x}
\bibAnnoteFile{Matt00}

\bibitem[{{Matzeu} et~al.(2019){Matzeu}, {Braito}, {Reeves}, {Severgnini},
  {Ballo}, {Caccianiga} et~al.}]{Matzeu19}
{Matzeu}, G.~A., {Braito}, V., {Reeves}, J.~N., {Severgnini}, P., {Ballo}, L.,
  {Caccianiga}, A., et~al. (2019).
\newblock {Evidence for a clumpy disc-wind in the star-forming Seyfert 2 galaxy
  MCG-03-58-007}.
\newblock \emph{\mnras} 483, 2836--2850.
\newblock \doi{10.1093/mnras/sty3327}
\bibAnnoteFile{Matzeu19}

\bibitem[{{Matzeu} et~al.(2022){Matzeu}, {Lieu}, {Costa}, {Reeves}, {Braito},
  {Dadina} et~al.}]{Matzeu22}
{Matzeu}, G.~A., {Lieu}, M., {Costa}, M.~T., {Reeves}, J.~N., {Braito}, V.,
  {Dadina}, M., et~al. (2022).
\newblock {A new emulated Monte Carlo radiative transfer disc-wind model: X-Ray
  Accretion Disc-wind Emulator - XRADE}.
\newblock \emph{\mnras} 515, 6172--6190.
\newblock \doi{10.1093/mnras/stac2155}
\bibAnnoteFile{Matzeu22}

\bibitem[{{McConnell} and {Ma}(2013)}]{McConnell13}
{McConnell}, N.~J. and {Ma}, C.-P. (2013).
\newblock {Revisiting the Scaling Relations of Black Hole Masses and Host
  Galaxy Properties}.
\newblock \emph{\apj} 764, 184.
\newblock \doi{10.1088/0004-637X/764/2/184}
\bibAnnoteFile{McConnell13}

\bibitem[{{McKaig} et~al.(2022){McKaig}, {Ricci}, {Paltani}, and
  {Satyapal}}]{Mckaig22}
{McKaig}, J., {Ricci}, C., {Paltani}, S., and {Satyapal}, S. (2022).
\newblock {X-ray simulations of polar gas in accreting supermassive black
  holes}.
\newblock \emph{\mnras} 512, 2961--2971.
\newblock \doi{10.1093/mnras/stab3178}
\bibAnnoteFile{Mckaig22}

\bibitem[{{Meidinger} et~al.(2020){Meidinger}, {Albrecht}, {Beitler},
  {Bonholzer}, {Emberger}, {Frank} et~al.}]{Meidinger20}
{Meidinger}, N., {Albrecht}, S., {Beitler}, C., {Bonholzer}, M., {Emberger},
  V., {Frank}, J., et~al. (2020).
\newblock {Development status of the wide field imager instrument for Athena}.
\newblock In \emph{Society of Photo-Optical Instrumentation Engineers (SPIE)
  Conference Series}. vol. 11444 of \emph{Society of Photo-Optical
  Instrumentation Engineers (SPIE) Conference Series}, 114440T.
\newblock \doi{10.1117/12.2560507}
\bibAnnoteFile{Meidinger20}

\bibitem[{{Mohanadas} and {Annuar}(2023)}]{Mohanadas23}
{Mohanadas}, P. and {Annuar}, A. (2023).
\newblock {NGC 4117: A New Compton-thick AGN Revealed by Broadband X-Ray
  Spectral Analysis}.
\newblock \emph{Research in Astronomy and Astrophysics} 23, 055002.
\newblock \doi{10.1088/1674-4527/acc151}
\bibAnnoteFile{Mohanadas23}

\bibitem[{{Moran} et~al.(2014){Moran}, {Shahinyan}, {Sugarman}, {V{\'e}lez},
  and {Eracleous}}]{Moran14}
{Moran}, E.~C., {Shahinyan}, K., {Sugarman}, H.~R., {V{\'e}lez}, D.~O., and
  {Eracleous}, M. (2014).
\newblock {Black Holes At the Centers of Nearby Dwarf Galaxies}.
\newblock \emph{\aj} 148, 136.
\newblock \doi{10.1088/0004-6256/148/6/136}
\bibAnnoteFile{Moran14}

\bibitem[{{Moravec} et~al.(2022){Moravec}, {Svoboda}, {Borkar}, {Boorman},
  {Kynoch}, {Panessa} et~al.}]{Moravec22}
{Moravec}, E., {Svoboda}, J., {Borkar}, A., {Boorman}, P., {Kynoch}, D.,
  {Panessa}, F., et~al. (2022).
\newblock {Do radio active galactic nuclei reflect X-ray binary spectral
  states?}
\newblock \emph{\aap} 662, A28.
\newblock \doi{10.1051/0004-6361/202142870}
\bibAnnoteFile{Moravec22}

\bibitem[{{Murphy} and {Yaqoob}(2009)}]{Murphy09}
{Murphy}, K.~D. and {Yaqoob}, T. (2009).
\newblock {An X-ray spectral model for Compton-thick toroidal reprocessors}.
\newblock \emph{\mnras} 397, 1549--1562.
\newblock \doi{10.1111/j.1365-2966.2009.15025.x}
\bibAnnoteFile{Murphy09}

\bibitem[{Nagar et~al.(2005)Nagar, Falcke, Wilson, and Ho}]{Nagar05}
Nagar, N.~M., Falcke, H., Wilson, A.~S., and Ho, L.~C. (2005).
\newblock Accretion disk model for low-luminosity active galactic nuclei.
\newblock \emph{The Astrophysical Journal} 620, 83--89.
\newblock \doi{10.1086/427175}
\bibAnnoteFile{Nagar05}

\bibitem[{{Nandra} et~al.(2013){Nandra}, {Barret}, {Barcons}, {Fabian}, {den
  Herder}, {Piro} et~al.}]{Nandra13}
{Nandra}, K., {Barret}, D., {Barcons}, X., {Fabian}, A., {den Herder}, J.-W.,
  {Piro}, L., et~al. (2013).
\newblock {The Hot and Energetic Universe: A White Paper presenting the science
  theme motivating the Athena+ mission}.
\newblock \emph{arXiv e-prints} , arXiv:1306.2307\doi{10.48550/arXiv.1306.2307}
\bibAnnoteFile{Nandra13}

\bibitem[{{Nardini}(2017)}]{Nardini17}
{Nardini}, E. (2017).
\newblock {Nuclear absorption and emission in the AGN merger NGC 6240 : the
  hard X-ray view}.
\newblock \emph{\mnras} 471, 3483--3493.
\newblock \doi{10.1093/mnras/stx1878}
\bibAnnoteFile{Nardini17}

\bibitem[{{Nemmen} et~al.(2014){Nemmen}, {Storchi-Bergmann}, and
  {Eracleous}}]{Nemmen14}
{Nemmen}, R.~S., {Storchi-Bergmann}, T., and {Eracleous}, M. (2014).
\newblock {Spectral models for low-luminosity active galactic nuclei in LINERs:
  the role of advection-dominated accretion and jets}.
\newblock \emph{\mnras} 438, 2804--2827.
\newblock \doi{10.1093/mnras/stt2388}
\bibAnnoteFile{Nemmen14}

\bibitem[{Nemmen et~al.(2006)Nemmen, Storchi-Bergmann, Yuan, Eracleous,
  Terashima, and Wilson}]{Nemmen06}
Nemmen, R.~S., Storchi-Bergmann, T., Yuan, F., Eracleous, M., Terashima, Y.,
  and Wilson, A.~S. (2006).
\newblock The relationship between luminosity and broad-line region size in
  active galactic nuclei.
\newblock \emph{The Astrophysical Journal} 643, 652--669.
\newblock \doi{10.1086/503344}
\bibAnnoteFile{Nemmen06}

\bibitem[{{Nenkova} et~al.(2008){Nenkova}, {Sirocky}, {Ivezi{\'c}}, and
  {Elitzur}}]{Nenkova08a}
{Nenkova}, M., {Sirocky}, M.~M., {Ivezi{\'c}}, {\v Z}., and {Elitzur}, M.
  (2008).
\newblock {AGN Dusty Tori. I. Handling of Clumpy Media}.
\newblock \emph{\apj} 685, 147-159.
\newblock \doi{10.1086/590482}
\bibAnnoteFile{Nenkova08a}

\bibitem[{{Netzer}(2015)}]{Netzer15}
{Netzer}, H. (2015).
\newblock {Revisiting the Unified Model of Active Galactic Nuclei}.
\newblock \emph{\araa} 53, 365--408.
\newblock \doi{10.1146/annurev-astro-082214-122302}
\bibAnnoteFile{Netzer15}

\bibitem[{{Odaka} et~al.(2011){Odaka}, {Aharonian}, {Watanabe}, {Tanaka},
  {Khangulyan}, and {Takahashi}}]{Odaka11}
{Odaka}, H., {Aharonian}, F., {Watanabe}, S., {Tanaka}, Y., {Khangulyan}, D.,
  and {Takahashi}, T. (2011).
\newblock {X-Ray Diagnostics of Giant Molecular Clouds in the Galactic Center
  Region and Past Activity of Sgr A*}.
\newblock \emph{\apj} 740, 103.
\newblock \doi{10.1088/0004-637X/740/2/103}
\bibAnnoteFile{Odaka11}

\bibitem[{{Odaka} et~al.(2016){Odaka}, {Yoneda}, {Takahashi}, and
  {Fabian}}]{Odaka16}
{Odaka}, H., {Yoneda}, H., {Takahashi}, T., and {Fabian}, A. (2016).
\newblock {Sensitivity of the Fe K{\ensuremath{\alpha}} Compton shoulder to the
  geometry and variability of the X-ray illumination of cosmic objects}.
\newblock \emph{\mnras} 462, 2366--2381.
\newblock \doi{10.1093/mnras/stw1764}
\bibAnnoteFile{Odaka16}

\bibitem[{{Osorio-Clavijo} et~al.(2022){Osorio-Clavijo},
  {Gonz{\'a}lez-Mart{\'\i}n}, {S{\'a}nchez}, {Esparza-Arredondo}, {Masegosa},
  {Victoria-Ceballos} et~al.}]{OsorioClavijo22}
{Osorio-Clavijo}, N., {Gonz{\'a}lez-Mart{\'\i}n}, O., {S{\'a}nchez}, S.~F.,
  {Esparza-Arredondo}, D., {Masegosa}, J., {Victoria-Ceballos}, C., et~al.
  (2022).
\newblock {Observational hints on the torus obscuring gas behaviour through
  X-rays with NuSTAR data}.
\newblock \emph{\mnras} 510, 5102--5118.
\newblock \doi{10.1093/mnras/stab3752}
\bibAnnoteFile{OsorioClavijo22}

\bibitem[{{Paltani} and {Ricci}(2017)}]{Paltani17}
{Paltani}, S. and {Ricci}, C. (2017).
\newblock {RefleX: X-ray absorption and reflection in active galactic nuclei
  for arbitrary geometries}.
\newblock \emph{\aap} 607, A31.
\newblock \doi{10.1051/0004-6361/201629623}
\bibAnnoteFile{Paltani17}

\bibitem[{{Panessa} et~al.(2020){Panessa}, {Castangia}, {Malizia}, {Bassani},
  {Tarchi}, {Bazzano} et~al.}]{Panessa20}
{Panessa}, F., {Castangia}, P., {Malizia}, A., {Bassani}, L., {Tarchi}, A.,
  {Bazzano}, A., et~al. (2020).
\newblock {Water megamaser emission in hard X-ray selected AGN}.
\newblock \emph{\aap} 641, A162.
\newblock \doi{10.1051/0004-6361/201937407}
\bibAnnoteFile{Panessa20}

\bibitem[{{Pfeifle} et~al.(2023){Pfeifle}, {Boorman}, {Weaver}, {Civano},
  {Nardini}, {Ricci} et~al.}]{Pfeifle23_hexp}
{Pfeifle}, R., {Boorman}, P., {Weaver}, K., {Civano}, F., {Nardini}, E.,
  {Ricci}, C., et~al. (2023).
\newblock {submitted}.
\newblock \emph{Frontiers in Astronomy and Space Sciences}
\bibAnnoteFile{Pfeifle23_hexp}

\bibitem[{{Piotrowska} et~al.(2023){Piotrowska}, {Garc\'{i}a}, {Walton},
  {Beckmann}, {Stern}, {Ballantyne} et~al.}]{Piotrowska23}
{Piotrowska}, J.~M., {Garc\'{i}a}, J.~A., {Walton}, D.~J., {Beckmann}, R.~S.,
  {Stern}, D., {Ballantyne}, D.~R., et~al. (2023).
\newblock {submitted}.
\newblock \emph{Frontiers in Astronomy and Space Sciences}
\bibAnnoteFile{Piotrowska23}

\bibitem[{{Pizzetti} et~al.(2022){Pizzetti}, {Torres-Alb{\`a}}, {Marchesi},
  {Ajello}, {Silver}, and {Zhao}}]{Pizzetti22}
{Pizzetti}, A., {Torres-Alb{\`a}}, N., {Marchesi}, S., {Ajello}, M., {Silver},
  R., and {Zhao}, X. (2022).
\newblock {A Multiepoch X-Ray Study of the Nearby Seyfert 2 Galaxy NGC 7479:
  Linking Column Density Variability to the Torus Geometry}.
\newblock \emph{\apj} 936, 149.
\newblock \doi{10.3847/1538-4357/ac86c6}
\bibAnnoteFile{Pizzetti22}

\bibitem[{{Ptak} et~al.(2015){Ptak}, {Hornschemeier}, {Zezas}, {Lehmer},
  {Yukita}, {Wik} et~al.}]{Ptak15}
{Ptak}, A., {Hornschemeier}, A., {Zezas}, A., {Lehmer}, B., {Yukita}, M.,
  {Wik}, D., et~al. (2015).
\newblock {A Focused, Hard X-Ray Look at Arp 299 with NuSTAR}.
\newblock \emph{\apj} 800, 104.
\newblock \doi{10.1088/0004-637X/800/2/104}
\bibAnnoteFile{Ptak15}

\bibitem[{{Puccetti} et~al.(2016){Puccetti}, {Comastri}, {Bauer}, {Brandt},
  {Fiore}, {Harrison} et~al.}]{Puccetti16}
{Puccetti}, S., {Comastri}, A., {Bauer}, F.~E., {Brandt}, W.~N., {Fiore}, F.,
  {Harrison}, F.~A., et~al. (2016).
\newblock {Hard X-ray emission of the luminous infrared galaxy NGC 6240 as
  observed by NuSTAR}.
\newblock \emph{\aap} 585, A157.
\newblock \doi{10.1051/0004-6361/201527189}
\bibAnnoteFile{Puccetti16}

\bibitem[{{Puccetti} et~al.(2014){Puccetti}, {Comastri}, {Fiore},
  {Ar{\'e}valo}, {Risaliti}, {Bauer} et~al.}]{Puccetti14}
{Puccetti}, S., {Comastri}, A., {Fiore}, F., {Ar{\'e}valo}, P., {Risaliti}, G.,
  {Bauer}, F.~E., et~al. (2014).
\newblock {The Variable Hard X-Ray Emission of NGC 4945 as Observed by NuSTAR}.
\newblock \emph{\apj} 793, 26.
\newblock \doi{10.1088/0004-637X/793/1/26}
\bibAnnoteFile{Puccetti14}

\bibitem[{{Ramos Almeida} and {Ricci}(2017)}]{RamosAlmeida17}
{Ramos Almeida}, C. and {Ricci}, C. (2017).
\newblock {Nuclear obscuration in active galactic nuclei}.
\newblock \emph{Nature Astronomy} 1, 679--689.
\newblock \doi{10.1038/s41550-017-0232-z}
\bibAnnoteFile{RamosAlmeida17}

\bibitem[{{Reines}(2022)}]{Reines22}
{Reines}, A.~E. (2022).
\newblock {Hunting for massive black holes in dwarf galaxies}.
\newblock \emph{Nature Astronomy} 6, 26--34.
\newblock \doi{10.1038/s41550-021-01556-0}
\bibAnnoteFile{Reines22}

\bibitem[{{Reines} et~al.(2013){Reines}, {Greene}, and {Geha}}]{Reines13}
{Reines}, A.~E., {Greene}, J.~E., and {Geha}, M. (2013).
\newblock {Dwarf Galaxies with Optical Signatures of Active Massive Black
  Holes}.
\newblock \emph{\apj} 775, 116.
\newblock \doi{10.1088/0004-637X/775/2/116}
\bibAnnoteFile{Reines13}

\bibitem[{{Reynolds} et~al.(2015){Reynolds}, {Lohfink}, {Ogle}, {Harrison},
  {Madsen}, {Fabian} et~al.}]{Reynolds15}
{Reynolds}, C.~S., {Lohfink}, A.~M., {Ogle}, P.~M., {Harrison}, F.~A.,
  {Madsen}, K.~K., {Fabian}, A.~C., et~al. (2015).
\newblock {NuSTAR Observations of the Powerful Radio Galaxy Cygnus A}.
\newblock \emph{\apj} 808, 154.
\newblock \doi{10.1088/0004-637X/808/2/154}
\bibAnnoteFile{Reynolds15}

\bibitem[{{Ricci} et~al.(2022){Ricci}, {Ananna}, {Temple}, {Urry}, {Koss},
  {Trakhtenbrot} et~al.}]{Ricci22}
{Ricci}, C., {Ananna}, T.~T., {Temple}, M.~J., {Urry}, C.~M., {Koss}, M.~J.,
  {Trakhtenbrot}, B., et~al. (2022).
\newblock {BASS XXXVII: The Role of Radiative Feedback in the Growth and
  Obscuration Properties of Nearby Supermassive Black Holes}.
\newblock \emph{\apj} 938, 67.
\newblock \doi{10.3847/1538-4357/ac8e67}
\bibAnnoteFile{Ricci22}

\bibitem[{{Ricci} et~al.(2016){Ricci}, {Bauer}, {Arevalo}, {Boggs}, {Brandt},
  {Christensen} et~al.}]{Ricci16}
{Ricci}, C., {Bauer}, F.~E., {Arevalo}, P., {Boggs}, S., {Brandt}, W.~N.,
  {Christensen}, F.~E., et~al. (2016).
\newblock {IC 751: A New Changing Look AGN Discovered by NuSTAR}.
\newblock \emph{\apj} 820, 5.
\newblock \doi{10.3847/0004-637X/820/1/5}
\bibAnnoteFile{Ricci16}

\bibitem[{{Ricci} et~al.(2017{\natexlab{a}}){Ricci}, {Bauer}, {Treister},
  {Schawinski}, {Privon}, {Blecha} et~al.}]{Ricci17b}
{Ricci}, C., {Bauer}, F.~E., {Treister}, E., {Schawinski}, K., {Privon}, G.~C.,
  {Blecha}, L., et~al. (2017{\natexlab{a}}).
\newblock {Growing supermassive black holes in the late stages of galaxy
  mergers are heavily obscured}.
\newblock \emph{\mnras} 468, 1273--1299.
\newblock \doi{10.1093/mnras/stx173}
\bibAnnoteFile{Ricci17b}

\bibitem[{{Ricci} and {Paltani}(2023)}]{Ricci23}
{Ricci}, C. and {Paltani}, S. (2023).
\newblock {Ray-tracing Simulations and Spectral Models of X-Ray Radiation in
  Dusty Media}.
\newblock \emph{\apj} 945, 55.
\newblock \doi{10.3847/1538-4357/acb5a6}
\bibAnnoteFile{Ricci23}

\bibitem[{{Ricci} et~al.(2021){Ricci}, {Privon}, {Pfeifle}, {Armus}, {Iwasawa},
  {Torres-Alb{\`a}} et~al.}]{Ricci21}
{Ricci}, C., {Privon}, G.~C., {Pfeifle}, R.~W., {Armus}, L., {Iwasawa}, K.,
  {Torres-Alb{\`a}}, N., et~al. (2021).
\newblock {A hard X-ray view of luminous and ultra-luminous infrared galaxies
  in GOALS - I. AGN obscuration along the merger sequence}.
\newblock \emph{\mnras} 506, 5935--5950.
\newblock \doi{10.1093/mnras/stab2052}
\bibAnnoteFile{Ricci21}

\bibitem[{{Ricci} and {Trakhtenbrot}(2022)}]{Ricci22_variability}
{Ricci}, C. and {Trakhtenbrot}, B. (2022).
\newblock {Changing-look Active Galactic Nuclei}.
\newblock \emph{arXiv e-prints} ,
  arXiv:2211.05132\doi{10.48550/arXiv.2211.05132}
\bibAnnoteFile{Ricci22_variability}

\bibitem[{{Ricci} et~al.(2017{\natexlab{b}}){Ricci}, {Trakhtenbrot}, {Koss},
  {Ueda}, {Del Vecchio}, {Treister} et~al.}]{Ricci17_bassV}
{Ricci}, C., {Trakhtenbrot}, B., {Koss}, M.~J., {Ueda}, Y., {Del Vecchio}, I.,
  {Treister}, E., et~al. (2017{\natexlab{b}}).
\newblock {BAT AGN Spectroscopic Survey. V. X-Ray Properties of the Swift/BAT
  70-month AGN Catalog}.
\newblock \emph{\apjs} 233, 17.
\newblock \doi{10.3847/1538-4365/aa96ad}
\bibAnnoteFile{Ricci17_bassV}

\bibitem[{{Ricci} et~al.(2017{\natexlab{c}}){Ricci}, {Trakhtenbrot}, {Koss},
  {Ueda}, {Schawinski}, {Oh} et~al.}]{Ricci17_eddrat}
{Ricci}, C., {Trakhtenbrot}, B., {Koss}, M.~J., {Ueda}, Y., {Schawinski}, K.,
  {Oh}, K., et~al. (2017{\natexlab{c}}).
\newblock {The close environments of accreting massive black holes are shaped
  by radiative feedback}.
\newblock \emph{\nat} 549, 488--491.
\newblock \doi{10.1038/nature23906}
\bibAnnoteFile{Ricci17_eddrat}

\bibitem[{{Ricci} et~al.(2015){Ricci}, {Ueda}, {Koss}, {Trakhtenbrot}, {Bauer},
  and {Gandhi}}]{Ricci15}
{Ricci}, C., {Ueda}, Y., {Koss}, M.~J., {Trakhtenbrot}, B., {Bauer}, F.~E., and
  {Gandhi}, P. (2015).
\newblock {Compton-thick Accretion in the Local Universe}.
\newblock \emph{\apjl} 815, L13.
\newblock \doi{10.1088/2041-8205/815/1/L13}
\bibAnnoteFile{Ricci15}

\bibitem[{{Rino-Silvestre} et~al.(2022){Rino-Silvestre},
  {Gonz{\'a}lez-Gait{\'a}n}, {Stalevski}, {Smole}, {Guilherme-Garcia},
  {Carvalho} et~al.}]{RinoSilvestre22}
{Rino-Silvestre}, J., {Gonz{\'a}lez-Gait{\'a}n}, S., {Stalevski}, M., {Smole},
  M., {Guilherme-Garcia}, P., {Carvalho}, J.~P., et~al. (2022).
\newblock {EmulART: Emulating Radiative Transfer -- A pilot study on
  autoencoder based dimensionality reduction for radiative transfer models}.
\newblock \emph{arXiv e-prints} ,
  arXiv:2210.15400\doi{10.48550/arXiv.2210.15400}
\bibAnnoteFile{RinoSilvestre22}

\bibitem[{{Risaliti} et~al.(2005){Risaliti}, {Elvis}, {Fabbiano}, {Baldi}, and
  {Zezas}}]{Risaliti05}
{Risaliti}, G., {Elvis}, M., {Fabbiano}, G., {Baldi}, A., and {Zezas}, A.
  (2005).
\newblock {Rapid Compton-thick/Compton-thin Transitions in the Seyfert 2 Galaxy
  NGC 1365}.
\newblock \emph{\apjl} 623, L93--L96.
\newblock \doi{10.1086/430252}
\bibAnnoteFile{Risaliti05}

\bibitem[{{Risaliti} et~al.(2007){Risaliti}, {Elvis}, {Fabbiano}, {Baldi},
  {Zezas}, and {Salvati}}]{Risaliti07}
{Risaliti}, G., {Elvis}, M., {Fabbiano}, G., {Baldi}, A., {Zezas}, A., and
  {Salvati}, M. (2007).
\newblock {Occultation Measurement of the Size of the X-Ray-emitting Region in
  the Active Galactic Nucleus of NGC 1365}.
\newblock \emph{\apjl} 659, L111--L114.
\newblock \doi{10.1086/517884}
\bibAnnoteFile{Risaliti07}

\bibitem[{{Risaliti} et~al.(2002){Risaliti}, {Elvis}, and
  {Nicastro}}]{Risaliti02}
{Risaliti}, G., {Elvis}, M., and {Nicastro}, F. (2002).
\newblock {Ubiquitous Variability of X-Ray-absorbing Column Densities in
  Seyfert 2 Galaxies}.
\newblock \emph{\apj} 571, 234--246.
\newblock \doi{10.1086/324146}
\bibAnnoteFile{Risaliti02}

\bibitem[{{Risaliti} et~al.(1999){Risaliti}, {Maiolino}, and
  {Salvati}}]{Risaliti99}
{Risaliti}, G., {Maiolino}, R., and {Salvati}, M. (1999).
\newblock {The Distribution of Absorbing Column Densities among Seyfert 2
  Galaxies}.
\newblock \emph{\apj} 522, 157--164.
\newblock \doi{10.1086/307623}
\bibAnnoteFile{Risaliti99}

\bibitem[{{Risaliti} et~al.(2009){Risaliti}, {Salvati}, {Elvis}, {Fabbiano},
  {Baldi}, {Bianchi} et~al.}]{Risaliti09}
{Risaliti}, G., {Salvati}, M., {Elvis}, M., {Fabbiano}, G., {Baldi}, A.,
  {Bianchi}, S., et~al. (2009).
\newblock {The XMM-Newton long look of NGC 1365: uncovering of the obscured
  X-ray source}.
\newblock \emph{\mnras} 393, L1--L5.
\newblock \doi{10.1111/j.1745-3933.2008.00580.x}
\bibAnnoteFile{Risaliti09}

\bibitem[{{Rivers} et~al.(2015){Rivers}, {Balokovi{\'c}}, {Ar{\'e}valo},
  {Bauer}, {Boggs}, {Brandt} et~al.}]{Rivers15}
{Rivers}, E., {Balokovi{\'c}}, M., {Ar{\'e}valo}, P., {Bauer}, F.~E., {Boggs},
  S.~E., {Brandt}, W.~N., et~al. (2015).
\newblock {The NuSTAR View of Reflection and Absorption in NGC 7582}.
\newblock \emph{\apj} 815, 55.
\newblock \doi{10.1088/0004-637X/815/1/55}
\bibAnnoteFile{Rivers15}

\bibitem[{{Rose} et~al.(2019){Rose}, {Edge}, {Combes}, {Gaspari}, {Hamer},
  {Nesvadba} et~al.}]{Rose19}
{Rose}, T., {Edge}, A.~C., {Combes}, F., {Gaspari}, M., {Hamer}, S.,
  {Nesvadba}, N., et~al. (2019).
\newblock {Constraining cold accretion on to supermassive black holes:
  molecular gas in the cores of eight brightest cluster galaxies revealed by
  joint CO and CN absorption}.
\newblock \emph{\mnras} 489, 349--365.
\newblock \doi{10.1093/mnras/stz2138}
\bibAnnoteFile{Rose19}

\bibitem[{{Saha} et~al.(2022){Saha}, {Markowitz}, and {Buchner}}]{Saha22}
{Saha}, T., {Markowitz}, A.~G., and {Buchner}, J. (2022).
\newblock {Inferring the morphology of AGN torus using X-ray spectra: a
  reliability study}.
\newblock \emph{\mnras} 509, 5485--5510.
\newblock \doi{10.1093/mnras/stab3250}
\bibAnnoteFile{Saha22}

\bibitem[{{Shakura} and {Sunyaev}(1973)}]{Shakura73}
{Shakura}, N.~I. and {Sunyaev}, R.~A. (1973).
\newblock {Black holes in binary systems. Observational appearance.}
\newblock \emph{\aap} 24, 337--355
\bibAnnoteFile{Shakura73}

\bibitem[{{Silver} et~al.(2022){Silver}, {Torres-Alb{\`a}}, {Zhao}, {Marchesi},
  {Pizzetti}, {Cox} et~al.}]{Silver22}
{Silver}, R., {Torres-Alb{\`a}}, N., {Zhao}, X., {Marchesi}, S., {Pizzetti},
  A., {Cox}, I., et~al. (2022).
\newblock {Compton-thick AGN in the NuSTAR Era. IX. A Joint NuSTAR and
  XMM-Newton Analysis of Four Local AGN}.
\newblock \emph{\apj} 940, 148.
\newblock \doi{10.3847/1538-4357/ac9bf8}
\bibAnnoteFile{Silver22}

\bibitem[{{Springel} et~al.(2005){Springel}, {Di Matteo}, and
  {Hernquist}}]{Springel05}
{Springel}, V., {Di Matteo}, T., and {Hernquist}, L. (2005).
\newblock {Modelling feedback from stars and black holes in galaxy mergers}.
\newblock \emph{\mnras} 361, 776--794.
\newblock \doi{10.1111/j.1365-2966.2005.09238.x}
\bibAnnoteFile{Springel05}

\bibitem[{{Storchi-Bergmann} and
  {Schnorr-M{\"u}ller}(2019)}]{StorchiBergmann19}
{Storchi-Bergmann}, T. and {Schnorr-M{\"u}ller}, A. (2019).
\newblock {Observational constraints on the feeding of supermassive black
  holes}.
\newblock \emph{Nature Astronomy} 3, 48--61.
\newblock \doi{10.1038/s41550-018-0611-0}
\bibAnnoteFile{StorchiBergmann19}

\bibitem[{{Svoboda} et~al.(2017){Svoboda}, {Guainazzi}, and
  {Merloni}}]{Svoboda17}
{Svoboda}, J., {Guainazzi}, M., and {Merloni}, A. (2017).
\newblock {AGN spectral states from simultaneous UV and X-ray observations by
  XMM-Newton}.
\newblock \emph{\aap} 603, A127.
\newblock \doi{10.1051/0004-6361/201630181}
\bibAnnoteFile{Svoboda17}

\bibitem[{{Tanimoto} et~al.(2019){Tanimoto}, {Ueda}, {Odaka}, {Kawaguchi},
  {Fukazawa}, and {Kawamuro}}]{Tanimoto19}
{Tanimoto}, A., {Ueda}, Y., {Odaka}, H., {Kawaguchi}, T., {Fukazawa}, Y., and
  {Kawamuro}, T. (2019).
\newblock {XCLUMPY: X-Ray Spectral Model from Clumpy Torus and Its Application
  to the Circinus Galaxy}.
\newblock \emph{\apj} 877, 95.
\newblock \doi{10.3847/1538-4357/ab1b20}
\bibAnnoteFile{Tanimoto19}

\bibitem[{{Tanimoto} et~al.(2022){Tanimoto}, {Ueda}, {Odaka}, {Yamada}, and
  {Ricci}}]{Tanimoto22}
{Tanimoto}, A., {Ueda}, Y., {Odaka}, H., {Yamada}, S., and {Ricci}, C. (2022).
\newblock {NuSTAR Observations of 52 Compton-thick Active Galactic Nuclei
  Selected by the Swift/Burst Alert Telescope All-sky Hard X-Ray Survey}.
\newblock \emph{\apjs} 260, 30.
\newblock \doi{10.3847/1538-4365/ac5f59}
\bibAnnoteFile{Tanimoto22}

\bibitem[{{Temi} et~al.(2022){Temi}, {Gaspari}, {Brighenti}, {Werner},
  {Grossova}, {Gitti} et~al.}]{Temi22}
{Temi}, P., {Gaspari}, M., {Brighenti}, F., {Werner}, N., {Grossova}, R.,
  {Gitti}, M., et~al. (2022).
\newblock {Probing Multiphase Gas in Local Massive Elliptical Galaxies via
  Multiwavelength Observations}.
\newblock \emph{\apj} 928, 150.
\newblock \doi{10.3847/1538-4357/ac5036}
\bibAnnoteFile{Temi22}

\bibitem[{Terashima(2002)}]{Terashima02}
Terashima, Y. (2002).
\newblock Asca observations of "type 2" liners: Evidence for a stellar source
  of ionization.
\newblock \emph{The Astrophysical Journal} 576, 653--666.
\newblock \doi{10.1086/341772}
\bibAnnoteFile{Terashima02}

\bibitem[{{Torres-Alb{\`a}} et~al.(2018){Torres-Alb{\`a}}, {Iwasawa},
  {D{\'\i}az-Santos}, {Charmandaris}, {Ricci}, {Chu} et~al.}]{TorresAlba18}
{Torres-Alb{\`a}}, N., {Iwasawa}, K., {D{\'\i}az-Santos}, T., {Charmandaris},
  V., {Ricci}, C., {Chu}, J.~K., et~al. (2018).
\newblock {C-GOALS. II. Chandra observations of the lower luminosity sample of
  nearby luminous infrared galaxies in GOALS}.
\newblock \emph{\aap} 620, A140.
\newblock \doi{10.1051/0004-6361/201834105}
\bibAnnoteFile{TorresAlba18}

\bibitem[{{Torres-Alb{\`a}} et~al.(2021){Torres-Alb{\`a}}, {Marchesi}, {Zhao},
  {Ajello}, {Silver}, {Ananna} et~al.}]{TorresAlba21}
{Torres-Alb{\`a}}, N., {Marchesi}, S., {Zhao}, X., {Ajello}, M., {Silver}, R.,
  {Ananna}, T.~T., et~al. (2021).
\newblock {Compton-thick AGN in the NuSTAR Era VI: The Observed Compton-thick
  Fraction in the Local Universe}.
\newblock \emph{\apj} 922, 252.
\newblock \doi{10.3847/1538-4357/ac1c73}
\bibAnnoteFile{TorresAlba21}

\bibitem[{{Torres-Alb{\`a}} et~al.(2023){Torres-Alb{\`a}}, {Marchesi}, {Zhao},
  {Cox}, {Pizzetti}, {Ajello} et~al.}]{TorresAlba23}
{Torres-Alb{\`a}}, N., {Marchesi}, S., {Zhao}, X., {Cox}, I., {Pizzetti}, A.,
  {Ajello}, M., et~al. (2023).
\newblock {Hydrogen Column Density Variability in a Sample of Local
  Compton-Thin AGN}.
\newblock \emph{arXiv e-prints} ,
  arXiv:2301.07138\doi{10.48550/arXiv.2301.07138}
\bibAnnoteFile{TorresAlba23}

\bibitem[{{Traina} et~al.(2021){Traina}, {Marchesi}, {Vignali},
  {Torres-Alb{\`a}}, {Ajello}, {Pizzetti} et~al.}]{Traina21}
{Traina}, A., {Marchesi}, S., {Vignali}, C., {Torres-Alb{\`a}}, N., {Ajello},
  M., {Pizzetti}, A., et~al. (2021).
\newblock {Compton-Thick AGN in the NuSTAR ERA VII. A joint NuSTAR, Chandra,
  and XMM-Newton Analysis of Two Nearby, Heavily Obscured Sources}.
\newblock \emph{\apj} 922, 159.
\newblock \doi{10.3847/1538-4357/ac1fee}
\bibAnnoteFile{Traina21}

\bibitem[{{Treister} et~al.(2009){Treister}, {Urry}, and {Virani}}]{Treister09}
{Treister}, E., {Urry}, C.~M., and {Virani}, S. (2009).
\newblock {The Space Density of Compton-Thick Active Galactic Nucleus and the
  X-Ray Background}.
\newblock \emph{\apj} 696, 110--120.
\newblock \doi{10.1088/0004-637X/696/1/110}
\bibAnnoteFile{Treister09}

\bibitem[{Trump et~al.(2011)Trump, Impey, Kelly, Civano, Gabor, Diamond-Stanic
  et~al.}]{Trump11}
Trump, J.~R., Impey, C.~D., Kelly, B.~C., Civano, F., Gabor, J.~M.,
  Diamond-Stanic, A.~M., et~al. (2011).
\newblock Mid-infrared selection of active galactic nuclei with the wide-field
  infrared survey explorer. i. characterizing wise-selected agns in cosmos.
\newblock \emph{The Astrophysical Journal} 733, 60.
\newblock \doi{10.1088/0004-637X/733/1/60}
\bibAnnoteFile{Trump11}

\bibitem[{{Trump} et~al.(2015){Trump}, {Sun}, {Zeimann}, {Luck}, {Bridge},
  {Grier} et~al.}]{Trump15}
{Trump}, J.~R., {Sun}, M., {Zeimann}, G.~R., {Luck}, C., {Bridge}, J.~S.,
  {Grier}, C.~J., et~al. (2015).
\newblock {The Biases of Optical Line-Ratio Selection for Active Galactic
  Nuclei and the Intrinsic Relationship between Black Hole Accretion and Galaxy
  Star Formation}.
\newblock \emph{\apj} 811, 26.
\newblock \doi{10.1088/0004-637X/811/1/26}
\bibAnnoteFile{Trump15}

\bibitem[{{Ueda} et~al.(2014){Ueda}, {Akiyama}, {Hasinger}, {Miyaji}, and
  {Watson}}]{Ueda14}
{Ueda}, Y., {Akiyama}, M., {Hasinger}, G., {Miyaji}, T., and {Watson}, M.~G.
  (2014).
\newblock {Toward the Standard Population Synthesis Model of the X-Ray
  Background: Evolution of X-Ray Luminosity and Absorption Functions of Active
  Galactic Nuclei Including Compton-thick Populations}.
\newblock \emph{\apj} 786, 104.
\newblock \doi{10.1088/0004-637X/786/2/104}
\bibAnnoteFile{Ueda14}

\bibitem[{{Urry} and {Padovani}(1995)}]{Urry95}
{Urry}, C.~M. and {Padovani}, P. (1995).
\newblock {Unified Schemes for Radio-Loud Active Galactic Nuclei}.
\newblock \emph{\pasp} 107, 803.
\newblock \doi{10.1086/133630}
\bibAnnoteFile{Urry95}

\bibitem[{Ursini et~al.(2015)Ursini, Boissay-Malaquin, Marinucci, Matt,
  Bianchi, Capalbi et~al.}]{Ursini15}
Ursini, F., Boissay-Malaquin, R., Marinucci, A., Matt, G., Bianchi, S.,
  Capalbi, M., et~al. (2015).
\newblock Simultaneous x-ray/optical/uv snapshots of active galactic nuclei
  from xmm-newton: spectral energy distributions for the reverberation mapped
  sample.
\newblock \emph{Monthly Notices of the Royal Astronomical Society} 452,
  3266--3291.
\newblock \doi{10.1093/mnras/stv1441}
\bibAnnoteFile{Ursini15}

\bibitem[{{van Dyk} et~al.(2001){van Dyk}, {Connors}, {Kashyap}, and
  {Siemiginowska}}]{vanDyk01}
{van Dyk}, D.~A., {Connors}, A., {Kashyap}, V.~L., and {Siemiginowska}, A.
  (2001).
\newblock {Analysis of Energy Spectra with Low Photon Counts via Bayesian
  Posterior Simulation}.
\newblock \emph{\apj} 548, 224--243.
\newblock \doi{10.1086/318656}
\bibAnnoteFile{vanDyk01}

\bibitem[{{Vander Meulen} et~al.(2023){Vander Meulen}, {Camps}, {Stalevski},
  and {Baes}}]{VanderMeulen23}
{Vander Meulen}, B., {Camps}, P., {Stalevski}, M., and {Baes}, M. (2023).
\newblock {X-ray radiative transfer in full 3D with SKIRT}.
\newblock \emph{arXiv e-prints} ,
  arXiv:2304.10563\doi{10.48550/arXiv.2304.10563}
\bibAnnoteFile{VanderMeulen23}

\bibitem[{{Vasudevan} et~al.(2013){Vasudevan}, {Brandt}, {Mushotzky}, {Winter},
  {Baumgartner}, {Shimizu} et~al.}]{Vasudevan13}
{Vasudevan}, R.~V., {Brandt}, W.~N., {Mushotzky}, R.~F., {Winter}, L.~M.,
  {Baumgartner}, W.~H., {Shimizu}, T.~T., et~al. (2013).
\newblock {X-Ray Properties of the Northern Galactic Cap Sources in the 58
  Month Swift/BAT Catalog}.
\newblock \emph{\apj} 763, 111.
\newblock \doi{10.1088/0004-637X/763/2/111}
\bibAnnoteFile{Vasudevan13}

\bibitem[{{Vasudevan} et~al.(2016){Vasudevan}, {Fabian}, {Reynolds}, {Aird},
  {Dauser}, and {Gallo}}]{Vasudevan16}
{Vasudevan}, R.~V., {Fabian}, A.~C., {Reynolds}, C.~S., {Aird}, J., {Dauser},
  T., and {Gallo}, L.~C. (2016).
\newblock {A selection effect boosting the contribution from rapidly spinning
  black holes to the cosmic X-ray background}.
\newblock \emph{\mnras} 458, 2012--2023.
\newblock \doi{10.1093/mnras/stw363}
\bibAnnoteFile{Vasudevan16}

\bibitem[{{Vasylenko} et~al.(2013){Vasylenko}, {Fedorova}, and
  {Zhdanov}}]{Vasylenko13}
{Vasylenko}, A.~A., {Fedorova}, E., and {Zhdanov}, V.~I. (2013).
\newblock {Observations of Sy2 galaxy NGC 3281 by XMM-Newton and INTEGRAL
  satellites}.
\newblock \emph{Advances in Astronomy and Space Physics} 3, 120--125.
\newblock \doi{10.48550/arXiv.1311.1691}
\bibAnnoteFile{Vasylenko13}

\bibitem[{Virtanen et~al.(2020)Virtanen, Gommers, Oliphant, Haberland, Reddy,
  Cournapeau et~al.}]{2020SciPy-NMeth}
Virtanen, P., Gommers, R., Oliphant, T.~E., Haberland, M., Reddy, T.,
  Cournapeau, D., et~al. (2020).
\newblock {{SciPy} 1.0: Fundamental Algorithms for Scientific Computing in
  Python}.
\newblock \emph{Nature Methods} 17, 261--272.
\newblock \doi{10.1038/s41592-019-0686-2}
\bibAnnoteFile{2020SciPy-NMeth}

\bibitem[{{Vollmer} and {Duschl}(2002)}]{Vollmer02}
{Vollmer}, B. and {Duschl}, W.~J. (2002).
\newblock {The Dynamics of the Circumnuclear Disk and its environment in the
  Galactic Centre}.
\newblock \emph{\aap} 388, 128--148.
\newblock \doi{10.1051/0004-6361:20020422}
\bibAnnoteFile{Vollmer02}

\bibitem[{{Volonteri}(2010)}]{Volonteri10}
{Volonteri}, M. (2010).
\newblock {Formation of supermassive black holes}.
\newblock \emph{\aapr} 18, 279--315.
\newblock \doi{10.1007/s00159-010-0029-x}
\bibAnnoteFile{Volonteri10}

\bibitem[{{Volonteri} et~al.(2008){Volonteri}, {Lodato}, and
  {Natarajan}}]{Volonteri08}
{Volonteri}, M., {Lodato}, G., and {Natarajan}, P. (2008).
\newblock {The evolution of massive black hole seeds}.
\newblock \emph{\mnras} 383, 1079--1088.
\newblock \doi{10.1111/j.1365-2966.2007.12589.x}
\bibAnnoteFile{Volonteri08}

\bibitem[{{Wada}(2012)}]{Wada12}
{Wada}, K. (2012).
\newblock {Radiation-driven Fountain and Origin of Torus around Active Galactic
  Nuclei}.
\newblock \emph{\apj} 758, 66.
\newblock \doi{10.1088/0004-637X/758/1/66}
\bibAnnoteFile{Wada12}

\bibitem[{{Walton} et~al.(2014){Walton}, {Risaliti}, {Harrison}, {Fabian},
  {Miller}, {Arevalo} et~al.}]{Walton14}
{Walton}, D.~J., {Risaliti}, G., {Harrison}, F.~A., {Fabian}, A.~C., {Miller},
  J.~M., {Arevalo}, P., et~al. (2014).
\newblock {NuSTAR and XMM-NEWTON Observations of NGC 1365: Extreme Absorption
  Variability and a Constant Inner Accretion Disk}.
\newblock \emph{\apj} 788, 76.
\newblock \doi{10.1088/0004-637X/788/1/76}
\bibAnnoteFile{Walton14}

\bibitem[{{W}es {M}c{K}inney(2010)}]{mckinney-proc-scipy-2010}
{W}es {M}c{K}inney (2010).
\newblock {D}ata {S}tructures for {S}tatistical {C}omputing in {P}ython.
\newblock In \emph{{P}roceedings of the 9th {P}ython in {S}cience
  {C}onference}, eds. {S}t\'efan van~der {W}alt and {J}arrod {M}illman. 56 --
  61.
\newblock \doi{10.25080/Majora-92bf1922-00a}
\bibAnnoteFile{mckinney-proc-scipy-2010}

\bibitem[{{Williams} et~al.(2022){Williams}, {Pahari}, {Baldi}, {McHardy},
  {Mathur}, {Beswick} et~al.}]{Williams22}
{Williams}, D.~R.~A., {Pahari}, M., {Baldi}, R.~D., {McHardy}, I.~M., {Mathur},
  S., {Beswick}, R.~J., et~al. (2022).
\newblock {LeMMINGs - IV. The X-ray properties of a statistically complete
  sample of the nuclei in active and inactive galaxies from the Palomar
  sample}.
\newblock \emph{\mnras} 510, 4909--4928.
\newblock \doi{10.1093/mnras/stab3310}
\bibAnnoteFile{Williams22}

\bibitem[{{Yamauchi} et~al.(2012){Yamauchi}, {Nakai}, {Ishihara}, {Diamond},
  and {Sato}}]{Yamauchi12}
{Yamauchi}, A., {Nakai}, N., {Ishihara}, Y., {Diamond}, P., and {Sato}, N.
  (2012).
\newblock {Water-Vapor Maser Disk at the Nucleus of the Seyfert 2 Galaxy IC
  2560 and its Distance}.
\newblock \emph{\pasj} 64, 103.
\newblock \doi{10.1093/pasj/64.5.103}
\bibAnnoteFile{Yamauchi12}

\bibitem[{{Yaqoob}(2012)}]{Yaqoob12}
{Yaqoob}, T. (2012).
\newblock {The nature of the Compton-thick X-ray reprocessor in NGC 4945}.
\newblock \emph{\mnras} 423, 3360--3396.
\newblock \doi{10.1111/j.1365-2966.2012.21129.x}
\bibAnnoteFile{Yaqoob12}

\bibitem[{Younes et~al.(2011)Younes, Porquet, Sabra, Reeves, and
  Grosso}]{Younes11}
Younes, G., Porquet, D., Sabra, B., Reeves, J.~N., and Grosso, N. (2011).
\newblock Discovery of relativistic fe k$\alpha$ emission in the seyfert 1.9
  galaxy mcg-05-23-16.
\newblock \emph{Astronomy \& Astrophysics} 530, A149.
\newblock \doi{10.1051/0004-6361/201016194}
\bibAnnoteFile{Younes11}

\bibitem[{Younes et~al.(2019)Younes, Porquet, Sabra, Reeves, and
  Grosso}]{Younes19}
Younes, G., Porquet, D., Sabra, B., Reeves, J.~N., and Grosso, N. (2019).
\newblock A systematic study of the properties of relativistic fe k$\alpha$
  lines in agn.
\newblock \emph{Astronomy \& Astrophysics} 628, A56.
\newblock \doi{10.1051/0004-6361/201834282}
\bibAnnoteFile{Younes19}

\bibitem[{Young et~al.(2018)Young, McHardy, Emmanoulopoulos, and
  Nandra}]{Young18}
Young, A.~J., McHardy, I.~M., Emmanoulopoulos, D., and Nandra, K. (2018).
\newblock Nustar and xmm-newton observations of the hard state in the seyfert
  1.5 galaxy mcg--5-23-16.
\newblock \emph{The Astrophysical Journal} 853, 56.
\newblock \doi{10.3847/1538-4357/aaa5aa}
\bibAnnoteFile{Young18}

\bibitem[{{Zaino} et~al.(2020){Zaino}, {Bianchi}, {Marinucci}, {Matt}, {Bauer},
  {Brandt} et~al.}]{Zaino20}
{Zaino}, A., {Bianchi}, S., {Marinucci}, A., {Matt}, G., {Bauer}, F.~E.,
  {Brandt}, W.~N., et~al. (2020).
\newblock {Probing the circumnuclear absorbing medium of the buried AGN in NGC
  1068 through NuSTAR observations}.
\newblock \emph{\mnras} 492, 3872--3884.
\newblock \doi{10.1093/mnras/staa107}
\bibAnnoteFile{Zaino20}

\bibitem[{{Zaw} et~al.(2020){Zaw}, {Rosenthal}, {Katkov}, {Gelfand}, {Chen},
  {Greenhill} et~al.}]{Zaw20}
{Zaw}, I., {Rosenthal}, M.~J., {Katkov}, I.~Y., {Gelfand}, J.~D., {Chen},
  Y.-P., {Greenhill}, L.~J., et~al. (2020).
\newblock {An Accreting, Anomalously Low-mass Black Hole at the Center of
  Low-mass Galaxy IC 750}.
\newblock \emph{\apj} 897, 111.
\newblock \doi{10.3847/1538-4357/ab9944}
\bibAnnoteFile{Zaw20}

\bibitem[{{Zhao} et~al.(2021){Zhao}, {Marchesi}, {Ajello}, {Cole}, {Hu},
  {Silver} et~al.}]{Zhao21}
{Zhao}, X., {Marchesi}, S., {Ajello}, M., {Cole}, D., {Hu}, Z., {Silver}, R.,
  et~al. (2021).
\newblock {The properties of the AGN torus as revealed from a set of unbiased
  NuSTAR observations}.
\newblock \emph{\aap} 650, A57.
\newblock \doi{10.1051/0004-6361/202140297}
\bibAnnoteFile{Zhao21}

\bibitem[{{ZuHone} et~al.(2023){ZuHone}, {Vikhlinin}, {Tremblay}, {Randall},
  {Andrade-Santos}, and {Bourdin}}]{Zuhone23}
[Dataset] {ZuHone}, J.~A., {Vikhlinin}, A., {Tremblay}, G.~R., {Randall},
  S.~W., {Andrade-Santos}, F., and {Bourdin}, H. (2023).
\newblock {SOXS: Simulated Observations of X-ray Sources}.
\newblock Astrophysics Source Code Library, record ascl:2301.024
\bibAnnoteFile{Zuhone23}

\end{thebibliography}

\end{document}


\onecolumn
\firstpage{1}

\title {{\helveticaitalic{Supplementary Material}}}

\maketitle

\section{Supplementary Data}
\begin{landscape}
\input{ctdb_v1.tex}
\end{landscape}
Supplementary Material should be uploaded separately on submission. Please include any supplementary data, figures and/or tables. All supplementary files are deposited to FigShare for permanent storage and receive a DOI.

Supplementary material is not typeset so please ensure that all information is clearly presented, the appropriate caption is included in the file and not in the manuscript, and that the style conforms to the rest of the article. To avoid discrepancies between the published article and the supplementary material, please do not add the title, author list, affiliations or correspondence in the supplementary files.

\section{Supplementary Tables and Figures}

For more information on Supplementary Material and for details on the different file types accepted, please see \href{https://www.frontiersin.org/about/author-guidelines#supplementary-material}{the Supplementary Material section} of the Author Guidelines.

Figures, tables, and images will be published under a Creative Commons CC-BY licence and permission must be obtained for use of copyrighted material from other sources (including re-published/adapted/modified/partial figures and images from the internet). It is the responsibility of the authors to acquire the licenses, to follow any citation instructions requested by third-party rights holders, and cover any supplementary charges.


\subsection{Figures}


\begin{figure}[htbp]
\begin{center}
\includegraphics[width=9cm]{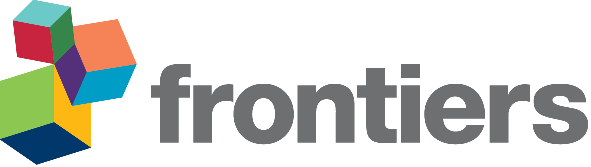}
\end{center}
\caption{ Enter the caption for your figure here.  Repeat as  necessary for each of your figures}\label{fig:1}
\end{figure}

\begin{figure}[htbp]
\begin{center}
\includegraphics[width=10cm]{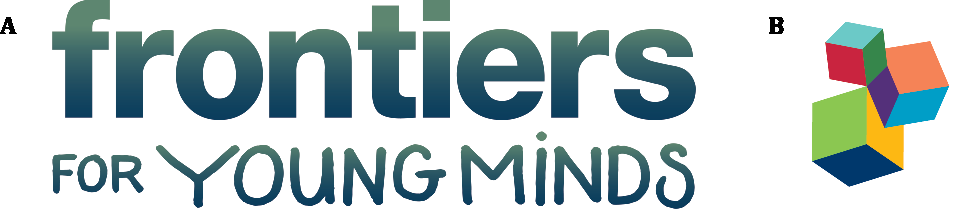}
\end{center}
\caption{This is a figure with sub figures, \textbf{(A)} is one logo, \textbf{(B)} is a different logo.}\label{fig:2}
\end{figure}

